\documentclass[fleqn,usenatbib]{mnras}

\usepackage{newtxtext,newtxmath}
\usepackage[T1]{fontenc}
\usepackage{ae,aecompl}
\usepackage{hyperref}
\usepackage{verbatim}
\usepackage{natbib}
\usepackage{graphicx}	
\usepackage{amsmath}	
\usepackage{amssymb}	
\usepackage{subfig}
\usepackage[export]{adjustbox}
\usepackage{caption,booktabs}
\usepackage{multirow}
\usepackage{enumitem}

\newcommand{\halpha}{H$\alpha$ }
\newcommand{\mstar}{M_{\rm *} }
\newcommand{\htwohone}{${\rm H}_2/{\rm H{\sc I}}$}
\newcommand{\kms}{\rm km\,s^{-1}} 
\newcommand{\msun}{\rm M_{\odot}} 

\usepackage{xcolor}
\usepackage{soul}

\newcommand{\cedit}[1]{\textcolor{black}{#1}}

\graphicspath{{./figs/}}

\title[Low-Redshift Blue Nuggets]{Linking Compact Dwarf Starburst Galaxies in the RESOLVE Survey to Downsized Blue Nuggets}

\author[Palumbo et al.]{Michael L. Palumbo III,$^{1,2}$\thanks{E-mail: palumbom@live.unc.edu, mlp95@psu.edu}
Sheila J. Kannappan,$^{1}$
Elaine M. Frazer,$^{3}$
\newauthor 
Kathleen D. Eckert,$^{4}$
Dara J. Norman,$^{5}$
Luciano Fraga,$^{6}$
Bruno C. Quint,$^{7}$
\newauthor 
Philippe Amram,$^{8}$
Claudia Mendes de Oliveira,$^{9}$
Ashley S. Bittner,$^{10}$
\newauthor 
Amanda J. Moffett,$^{11}$
David V. Stark,$^{12}$
Mark A. Norris,$^{13}$
\newauthor
Nathaniel T. Cleaves,$^{1}$
and Derrick S. Carr$^{1}$
\\
$^{1}$Department of Physics \& Astronomy, University of North Carolina at Chapel Hill, Phillips Hall, Chapel Hill, NC 27514, USA\\
$^{2}$Department of Astronomy \& Astrophysics, 525 Davey Laboratory, The Pennsylvania State University, University Park, PA, 16802, USA\\
$^{3}$Space Telescope Science Institute, 3700 San Martin Drive, Baltimore, MD 21218, USA\\
$^{4}$Department of Physics \& Astronomy, University of Pennsylvania, 209 South 33rd Street, Philadelphia, PA 19104, USA\\
$^{5}$National Optical Astronomy Observatory, 950N Cherry Avenue, Tucson, AZ 85750\\
$^{6}$Laborat\'{o}rio Nacional de Astrof\'{i}sica LNA/MCTIC, 37504-364, Itajub\'{a}, MG, Brazil\\
$^{7}$Southern Observatory for Astrophysical Research, Casilla 603, La Serena, Chile\\
$^{8}$Aix Marseille University, CNRS, CNES, LAM, 38 rue Fr\'ed\'eric Joliot Curie, 13248 Marseille Cedex 13, France\\
$^{9}$Instituto de Astronomia, Geof\'isica e Ci\^encias Atmosf\'ericas da U. de S\~{a}o Paulo, Cidade Universit\'aria,
05508-900, S\~{a}o Paulo, SP, Brazil\\
$^{10}$Department of Civil, Construction, and Environmental Engineering, North Carolina State University, Raleigh, NC 27605, USA\\
$^{11}$Department of Physics and Astronomy, University of North Georgia, 3820 Mundy Mill Road, Oakwood, GA 30566, USA \\
$^{12}$Kavli IPMU (WPI), UTIAS, The University of Tokyo, Kashiwa, Chiba 277-8583, Japan\\
$^{13}$Jeremiah Horrocks Institute, University of Central Lancashire, Preston, Lancashire, PR1 2HE, United Kingdom\\
}

\date{Accepted 27 March 2020.}
\pubyear{2020}

\begin{document}
\label{firstpage}
\pagerange{\pageref{firstpage}--\pageref{lastpage}}
\maketitle

\begin{abstract}
We identify and characterize compact dwarf starburst (CDS) galaxies in the RESOLVE survey, a volume-limited census of galaxies in the local universe, to probe whether this population contains any residual ``blue nuggets,'' a class of intensely star-forming compact galaxies first identified at high redshift $z$. Our 50 low-$z$ CDS galaxies are defined by dwarf masses (stellar mass $M_* < 10^{9.5}$ M$_{\odot}$), compact bulged-disk or spheroid-dominated morphologies (using a quantitative criterion, $\mu_\Delta > 8.6$), and specific star formation rates above the defining threshold for high-$z$ blue nuggets ($\log$ SSFR [Gyr$^{-1}] > -0.5$). Across redshifts, blue nuggets exhibit three defining properties: compactness relative to contemporaneous galaxies, abundant cold gas, and formation via compaction in mergers or colliding streams. Those with halo mass below $M_{\rm halo} \sim 10^{11.5}$ M$_{\odot}$ may in theory evade permanent quenching and cyclically refuel until the present day. Selected only for compactness and starburst activity, our CDS galaxies generally have $M_{\rm halo} \lesssim 10^{11.5}$ M$_{\odot}$ and gas-to-stellar mass ratio $\gtrsim$1. Moreover, analysis of archival DECaLS photometry and new 3D spectroscopic observations for CDS galaxies reveals a high rate of photometric and kinematic disturbances suggestive of dwarf mergers. The SSFRs, surface mass densities, and number counts of CDS galaxies are compatible with theoretical and observational expectations for redshift evolution in blue nuggets. We argue that CDS galaxies represent a maximally-starbursting subset of traditional compact dwarf classes such as blue compact dwarfs and blue E/S0s. We conclude that CDS galaxies represent a low-$z$ tail of the blue nugget phenomenon formed via a moderated compaction channel that leaves open the possibility of disk regrowth and evolution into normal disk galaxies.
\end{abstract}


\section{Introduction} \label{intro}

The first observations of a class of compact massive ($\mstar \gtrsim 10^{11} \,\msun$) elliptical galaxies with oblate morphologies and suppressed star formation rates at redshift $z \sim 2 - 3$ \citep{2008ApJ...677L...5V, 2010sf2a.conf..355I} posed an evolutionary mystery. These galaxies, known as ``red nuggets" \citep{2009ApJ...695..101D, 2010ApJ...717L.103N}, are extraordinarily compact, with radii up to $\sim 5$ times smaller than typical galaxies of comparable mass in the low-$z$ universe \citep{2007MNRAS.382..109T, 2007ApJ...671..285T, 2014ApJ...780....1W}. Their abundance at high $z$, contrasted with the scarcity of such dense early-type galaxies today, rules out simple monolithic  models of galaxy formation \citep[e.g.,][]{1962ApJ...136..748E}, since such models do not allow for significant structural evolution following initial galaxy assembly \citep{2008ApJ...677L...5V}. Their compactness also rules out gas-poor major merger models, since non-dissipative events preferentially scatter pre-existing stars to larger orbital radii, ``puffing up" rather than compacting the resulting object \citep{2007ApJ...658..710N, 2009ApJ...690.1452N}. The first observations of likely progenitors \citep{2013ApJ...765..104B} pointed to massive gas-rich compact starbursts capable of rapid quenching. \par

In response to these observations, \citet{2014MNRAS.438.1870D} proposed that a new class of ultra-compact, star-forming galaxies known as ``blue nuggets'' could be the evolutionary progenitors of red nuggets. Using a toy model, Dekel \& Burkert describe how blue nuggets could form via ``fast-track growth'' channels driven by extreme gas inflow from smooth cosmic streams or gas-rich mergers. The gas inflow episodes ignite a chain of events that culminates in violent disk instability, contraction, and rapid star formation, which continues as long as the rate of dissipative gas inflow exceeds the rate of gas depletion by star formation in the central regions of the galaxy. \citet{2014MNRAS.438.1870D} postulate that once blue nuggets run out of gas, they undergo inside-out quenching while maintaining their extreme compactness, making them suitable red nugget progenitors. The fraction of galaxies that become blue nuggets depends strongly on the fraction of cold gas and young stellar mass with respect to total baryonic mass, $f_{\rm c}$, a parameter that decreases dramatically over cosmic time. \citet{2014MNRAS.438.1870D} show that at $f_{\rm c} \simeq 0.5$, the blue nugget fraction is approximately $\sim 0.4$, compared to a dramatically smaller blue nugget fraction of $\sim 0.02$ at $f_{\rm c} \simeq 0.2$. \par 

Following up on the toy model of \citet{2014MNRAS.438.1870D}, cosmological simulations by \citet{2015MNRAS.450.2327Z}, \citet{2015MNRAS.453..408C}, and \citet{2016MNRAS.457.2790T} verified that gas-rich mergers favorably produce star-forming remnants with ultra-compact cores as a result of the dissipative nature of the inflowing gas, which allows for angular momentum loss \citep{2006ApJ...648L..21K, 2008ApJ...689...17H}. Indeed, star-forming galaxies consistent with the predictions of these simulations and of \citet{2014MNRAS.438.1870D} had been previously observed \citep[e.g., by][]{2013ApJ...765..104B, 2013ApJ...766...15P, 2014Natur.513..394N}. Moreover, the rapid appearance of a red sequence on a timescale of $\sim 0.7$ Gyr, as observed by \citet{2013ApJ...765..104B}, is consistent with a population of galaxies experiencing dramatic fast-track growth followed by absolute quenching. \par

Evolution from the blue nugget phase to the red nugget phase may be accompanied by evolution in morphology. Cosmological simulations by \citet{2015MNRAS.453..408C} and \citet{2016MNRAS.458.4477T} suggest that torques induced by the dark matter (DM) halo during the gas compaction phase of blue nugget formation create $\mstar < 10^{9.5} \,\msun$ objects elongated by formation along dominant large-scale filaments. These prolate objects rotate along their minor axes, then later evolve into oblate objects as part of the transition to self-gravitating, quenched red nuggets. Possibly consistent with such a morphological transition, analyses of projected ellipticities of star-forming galaxies at $z > 1$ by \citet{2012ApJ...745...85L}, \citet{2013ApJ...773..149C}, and \citet{2014ApJ...792L...6V} suggest that the fraction of oblate galaxies with $\mstar < 10^{10.5} \,\msun$ increases with cosmic time, while the fraction of prolate galaxies with $\mstar \sim 10^{8.5} - 10^{9.5} \,\msun$ steadily decreases following $z \sim 1 - 2$.  \par

The morphological evolution of blue nuggets is potentially complicated by successive cycles of compaction and temporary quenching episodes prior to absolute quenching and morphological transformation. \cedit{The high-$z$ simulations of \citet{2015MNRAS.453..408C}, as studied in \citet{2016MNRAS.457.2790T}, indicate} that this ``self-regulated evolution'' will continue so long as there is a fresh supply of gas to replenish central star formation and ultimately tends toward a red nugget when hot halo quenching dominates above halo virial mass $M_{\rm vir} \sim 10^{11.5} \,\msun$, corresponding to a stellar mass of $M_{\rm*} \sim 10^{9.5-10} \,\msun$ \citep[e.g.,][]{2016ApJ...824..124E}. Tacchella et al. suggest full quenching will likely occur above this halo mass (except in the $z>3$ regime of penetrating cold streams) as the gas replenishment time starts to exceed the depletion time. In keeping with these results, an earlier analysis of the same simulations by \citet{2015MNRAS.450.2327Z} found that lower-mass blue nuggets are not fully quenched and may continue cyclic blue nugget phases to intermediate and low redshifts. Thus, starbursts are permanently quenched earlier for more massive galaxies and later for less massive galaxies --- a.k.a. \cedit{star-formation} "downsizing" \citep{1996AJ....112..839C} --- because massive galaxy halos more quickly reach a threshold halo mass for suppression of a fresh gas supply by stable shock heating of halo gas. Although lower-mass blue nuggets exhibit lower star formation rates and mass densities than higher-mass blue nuggets, \citet{2015MNRAS.450.2327Z} still classify them as blue nuggets because of their formation via gas-rich compaction events (fast-track growth). \par

The possibility of a \cedit{residual} ``tail'' of the high-$z$ blue nugget phenomenon persisting to lower redshifts is also suggested by observations. \citet{2013ApJ...776...63F} describe compact star-forming and quenched galaxies with $\mstar \gtrsim 10^{9.75} \ \msun$ at low redshifts $z < 0.075$, and both \citet{2014MNRAS.438.1870D} and \citet{2015MNRAS.450.2327Z} discuss these objects as low-$z$ blue and red nuggets based on their compactness and star formation. Relative to the original high-$z$ blue nuggets reported in \citet[][]{2013ApJ...765..104B}, the \citet{2013ApJ...776...63F} objects are generally less massive, less compact, and less vigorously star-forming. More recently, \citet{2018ApJ...865...49W} have described a population of compact star-forming galaxies with $\mstar \gtrsim 10^{9.5} \ \msun$ at $z\lesssim0.05$ that appear to be in the midst of quenching to form red nuggets, based on evidence of depressed gas content. At still lower stellar masses, the observational analysis of \citet{2014ApJ...792L...6V} suggests that the fraction of prolate galaxies with $\mstar \lesssim 10^{9.5} \, {\rm \msun}$, i.e., potential blue nuggets in the regime of self-regulated evolution by cyclic fueling, decreases dramatically from $z = 2$ to $z = 0$ but is non-vanishing at $z = 0$.  If \cedit{blue nuggets} truly \cedit{persist as a residual population} all the way to $z\sim0$ as these observations suggest, then per \citet{2014MNRAS.438.1870D} and \citet{2015MNRAS.450.2327Z}, they are defined by the gas-rich compaction formation process and are not necessarily expected to produce red nuggets in the cyclic refueling regime below $\mstar \sim 10^{9.5} \ \msun$. Additionally, \citet{2014MNRAS.438.1870D} and \citet{2015MNRAS.450.2327Z} find that \cedit{as the cosmic gas inventory evolves,} the fast-track growth channel \cedit{ becomes less intense}. Thus, \cedit{while} three properties should characterize low-$z$ blue nuggets by definition --- compactness relative to contemporaneous galaxies, an upper halo mass limit of  $M_{\rm vir} \sim 10^{11.5} \,\msun$ \citep{2016MNRAS.457.2790T}, and formation via gas-rich mergers or other intense gas-compaction channels ---  their future evolution may not always involve red nugget formation. \par

With this context, the $z\sim0$ ``Fueling Diagram'' of \citet{2013ApJ...769...82S} presents some interesting parallels with the blue nugget story. The Fueling Diagram is a plot of global ${\rm H}_2/{\rm HI}$ gas mass ratio versus ``blue centeredness,'' a metric of recent central starburst activity. Building on the prior identification of blue centeredness with merger- or interaction-driven gas inflows by \citet{2004AJ....127.1371K}, \citet{2013ApJ...769...82S} argue that the triangular locus of galaxies in the Fueling Diagram reflects cycles of gas-rich merging, gas depletion/feedback, and fresh gas accretion. In particular, two of the sides of the triangle, representing post-merger gas depletion and refueling, are primarily populated by blue E/S0s and blue compact dwarfs (BCDs). Patterns of total gas content and stellar population age along these two sides suggest an evolutionary route characterized by rapid, gas-rich merger-driven star formation, gas depletion, and disk regrowth. Even with the reduced fresh gas inventory of the present-day universe, this pattern naturally prompts comparison between the evolutionary cycles of blue nuggets and low-$z$ blue E/S0s and BCDs. \par

To our knowledge, there has been no dedicated investigation of blue nuggets in the abundant gas fueling regime below $\mstar \sim 10^{9.5} \, {\rm \msun}$ at $z\sim0$. This paper seeks to identify candidate blue nuggets among $z\sim0$ compact, starbursting dwarf galaxies and determine:

\begin{enumerate}[leftmargin=*]
\item whether they in fact qualify as blue nuggets based on their compactness, halo masses/gas richness, and formation mechanisms,
\item how their numbers and properties align with redshift trends predicted by \citet{2014MNRAS.438.1870D} and \citet{2015MNRAS.450.2327Z},
\item to what extent their morphological and kinematic properties favor formation via gas-rich mergers vs.\ colliding gas streams, with implications for how evolutionary trajectories of low-$z$ blue nuggets compare to those of high-$z$ blue nuggets, and 
\item how these compact starburst galaxies, and especially any \cedit{residual} blue nuggets among them, relate to other classes of compact star-forming dwarf galaxies at $z\sim0$.
\end{enumerate}

To answer these questions, we employ the REsolved Spectroscopy Of a Local VolumE (RESOLVE) Survey\footnote{\url{https://resolve.astro.unc.edu/}}, a volume-limited census of galaxies in the local universe whose high completeness down to \textit{baryonic} masses $\sim$$10^{9.2} \, {\rm \msun}$ is ideal for identifying and characterizing \cedit{low-$z$} blue nuggets over a dex lower in stellar mass than those in the \citet{2018ApJ...865...49W} sample. In \S\ref{data} we describe the RESOLVE survey and its data, identify \cedit{50} compact dwarf starburst (CDS) galaxies within RESOLVE by criteria suitable for finding low-$z$ blue nuggets, and introduce additional supporting data (archival DECaLS imaging and new 3D spectroscopy) used in our analysis. In \S\ref{sec:methods} we discuss our methods for identifying \cedit{evidence of recent mergers} with DECaLS imaging as well as our analysis of the 3D spectroscopic data, including reduction of velocity fields, continuum maps, and H$\alpha$ line flux maps. In \S\ref{sec:results} we first assess whether our low-$z$ CDS galaxies may in fact be \cedit{residual low-$z$} blue nuggets based on their numbers, environments, gas richness, compactness, and starburst activity, then examine morphological and kinematic signatures of formation by either gas-rich mergers or colliding streams to constrain the roles of these two mechanisms of compaction at $z\sim0$. In \S\ref{sec:discussion} we discuss what our results imply about evolution in the fast-track growth channel itself, and we relate CDS galaxies to alternately-selected classes of low-$z$ compact star-forming galaxies, such as BCDs, blue E/S0s, and green peas, to put their evolution in context within the broader story of \cedit{star-formation} downsizing and fast-track growth. Finally, we summarize our findings in \S\ref{sec:conclusions}. \par

We assume a standard $\Lambda$CDM cosmology with $\Omega_{\rm m} = 0.3$, $\Omega_\Lambda = 0.7$, and $H_0 = 70$ km s$^{-1}$ Mpc$^{-1}$ for distance measurements and other derived quantities in this work. \par

\section{Data} \label{data}

\begin{figure}
\includegraphics[width=\columnwidth]{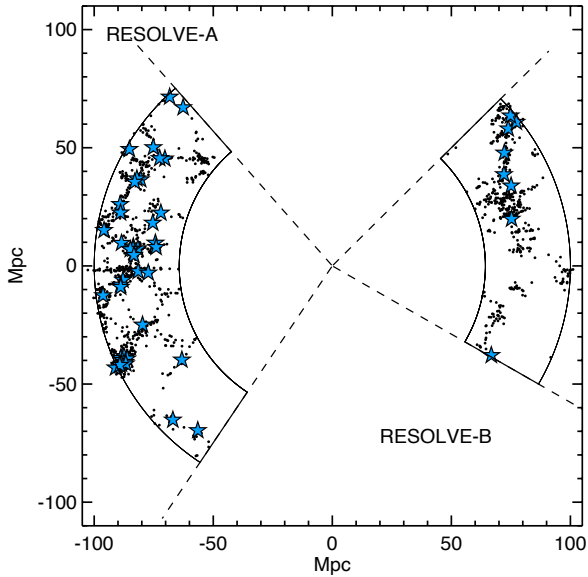}
\caption{RESOLVE survey footprint with low-$z$ compact dwarf starburst (CDS) galaxies highlighted as blue stars. RESOLVE is volume-limited and highly complete down to baryonic masses $M_{\rm bary} \sim 10^{9.2} \, \msun$.}
\label{fig:allresolve}
\end{figure}

\subsection{The RESOLVE Survey} \label{subsec:RESOLVE}

For this work, we use the RESOLVE (REsolved Spectroscopy of a Local VolumE) survey, a volume-limited census of mass in stars, cold gas, and dark matter for a statistically complete subset of the $z \sim 0$ galaxy population \citep{2008AIPC.1035..163K, 2015ApJ...810..166E, 2016ApJ...824..124E, 2017ApJ...849...20E, 2016ApJ...832..126S}. As a result of its volume-limited nature (see Figure~\ref{fig:allresolve}), RESOLVE is ideal for examining the properties of the full low-$z$ CDS population without the statistical completeness corrections necessary for flux-limited surveys. Moreover, the low mass floor of RESOLVE allows us to probe the stellar mass regime below $\mstar < 10^{9.5} \,\msun$, predicted to characterize low-$z$ blue nuggets in the gas-rich refueling regime. \par

\subsubsection{Full Survey Definition and Ancillary Data} \label{subsubsec:survey_def}

RESOLVE covers two equatorial strips, denoted RESOLVE-A and RESOLVE-B, bounded in Local Group-corrected velocity $V_{\rm LG} = 4500 - 7000 \ \kms$ and together enclosing $\sim 52,100\, {\rm Mpc}^3$ of the $z \sim 0$ universe \citep{2015ApJ...810..166E}. To avoid situations where peculiar velocities may affect survey membership, group (rather than individual) redshifts are used to decide final survey membership. RESOLVE is contained within the SDSS footprint and uses the SDSS redshift survey to build survey membership as well as additional redshifts from archival sources and new observations \citep[see][and references therein]{2015ApJ...810..166E}. \par

As shown in \citet{2013ApJ...777...42K} and \citet{2016ApJ...824..124E}, RESOLVE-A is complete down to absolute \textit{r}-band magnitude $M_{\rm r, tot} = -17.33$, which enables a baryonic mass completeness limit of $M_{\rm bary} \sim 10^{9.3} \, \msun$ and a stellar mass completeness limit of  $\mstar \sim 10^{8.9} \, \msun$, where $M_{\rm bary}$ is the combined mass of cold gas and stars. For the slightly deeper RESOLVE-B, the luminosity completeness limit of $M_{\rm r, tot} = -17.0$ enables a baryonic mass completeness limit of  $M_{\rm bary} \sim 10^{9.1} \, \msun$ and a stellar mass completeness limit of $\mstar \sim 10^{8.7} \, \msun$. These stellar and baryonic mass completeness limits are defined to include the full natural scatter in mass-to-light ratios \citep{2016ApJ...824..124E}. We define the floor of the RESOLVE survey by its $r$-band absolute magnitude limits but add in the handful of galaxies fainter than these limits with estimated $M_{\rm bary}$ above the baryonic mass completeness limits, consistent with the 21 cm targeting strategy for the RESOLVE H{\sc I} census \citep{2016ApJ...832..126S}. \par

RESOLVE employs custom reprocessed photometry from the UV through NIR as well as stellar masses calculated by spectral energy distribution modeling \citep{2015ApJ...810..166E}. Star formation rates are estimated from NUV and NIR observations from GALEX and WISE, respectively, using the calibrations of \citet{2011A&A...529A..22B}, \citet{2012MNRAS.424.1522W}, and \citet{2013AJ....145....6J}. $\rm HI$ masses and upper limits are obtained from deep, pointed observations with the GBT and Arecibo telescopes supplementing the blind 21 cm ALFALFA survey \citep{2011AJ....142..170H}, as described in \citet{2016ApJ...832..126S}. In this paper we use gas masses $M_{\rm gas}$ based on the $\rm HI$ mass corrected for helium, with missing or upper limit data filled in using photometric gas fractions as described in \citet{2015ApJ...810..166E}. The RESOLVE $\rm H{\sc I}$ census is $\sim$94\% complete \citep{2016ApJ...832..126S}. \par

The RESOLVE group catalog was defined by \citet{2017ApJ...849...20E} using the Friends-of-Friends (FoF) algorithm as described in \citet{2006ApJS..167....1B}. The brightest galaxy in the \textit{r} band in each group is labeled the central, and all other galaxies are labeled satellites. We note that both peculiar velocities and FoF group identification may induce catastrophic errors in group membership and central/satellite designation. Group halo masses are assigned using halo abundance matching on the group-integrated \textit{r}-band luminosity as described in \citet{2006ApJS..167....1B} and \citet{2017ApJ...849...20E}. Nearest neighbor distances are determined using the kd-tree algorithm of \citet{1999cs........1013M}, ``correcting'' for peculiar velocities as described in \citet{2018ApJ...857..144H} by using the group catalog to suppress apparent line-of-sight distances within, but not between, groups.  \par

\subsubsection{Compact Dwarf Starburst (CDS) Galaxy Sample} \label{subsubsec:sample}

\begin{figure}
\captionsetup[subfigure]{labelformat=empty}
\vspace{-4mm}
\subfloat[]{%
  \includegraphics[clip,width=0.93\columnwidth]{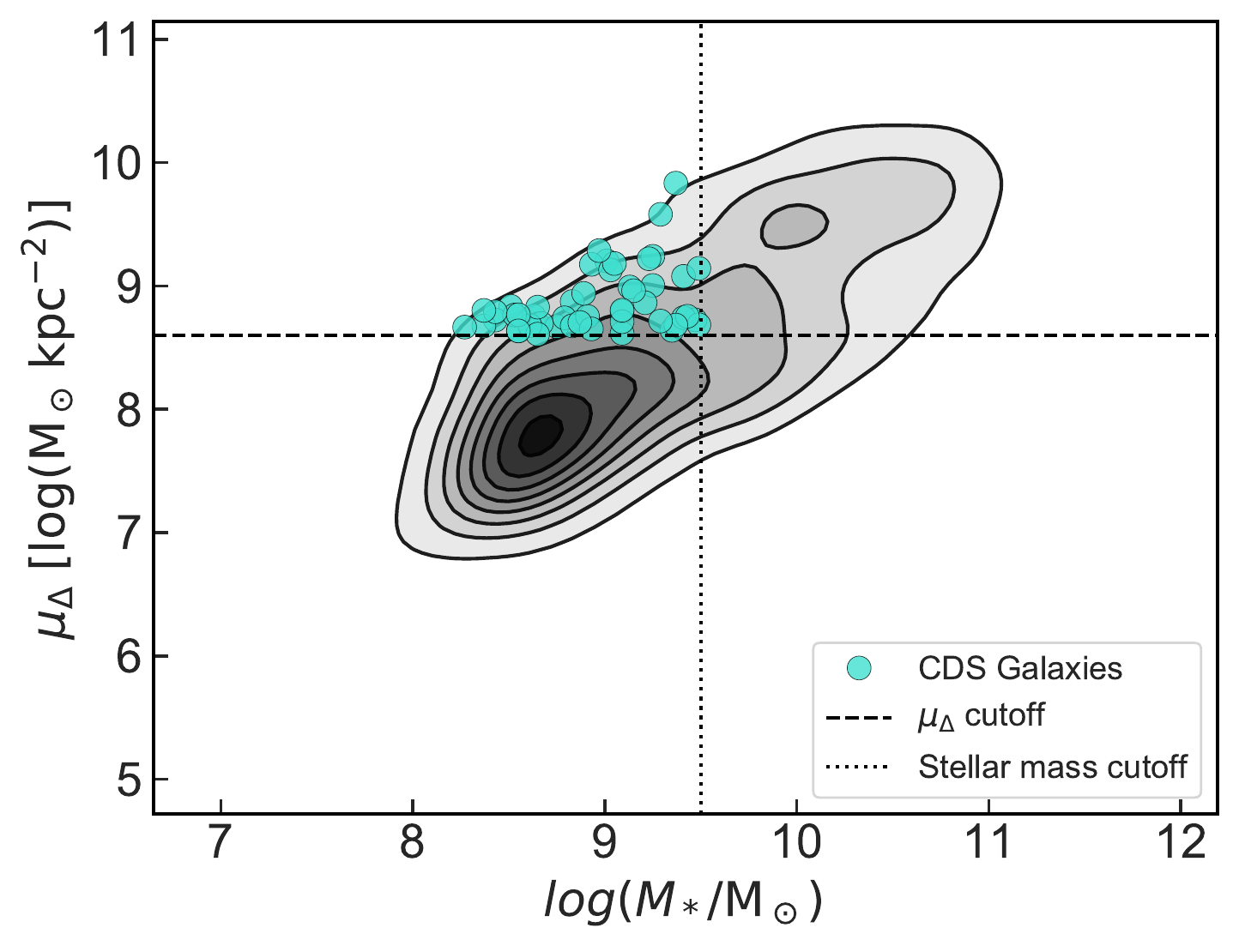}%
}\\
\vspace{-4mm}
\subfloat[]{%
  \includegraphics[clip,width=0.93\columnwidth]{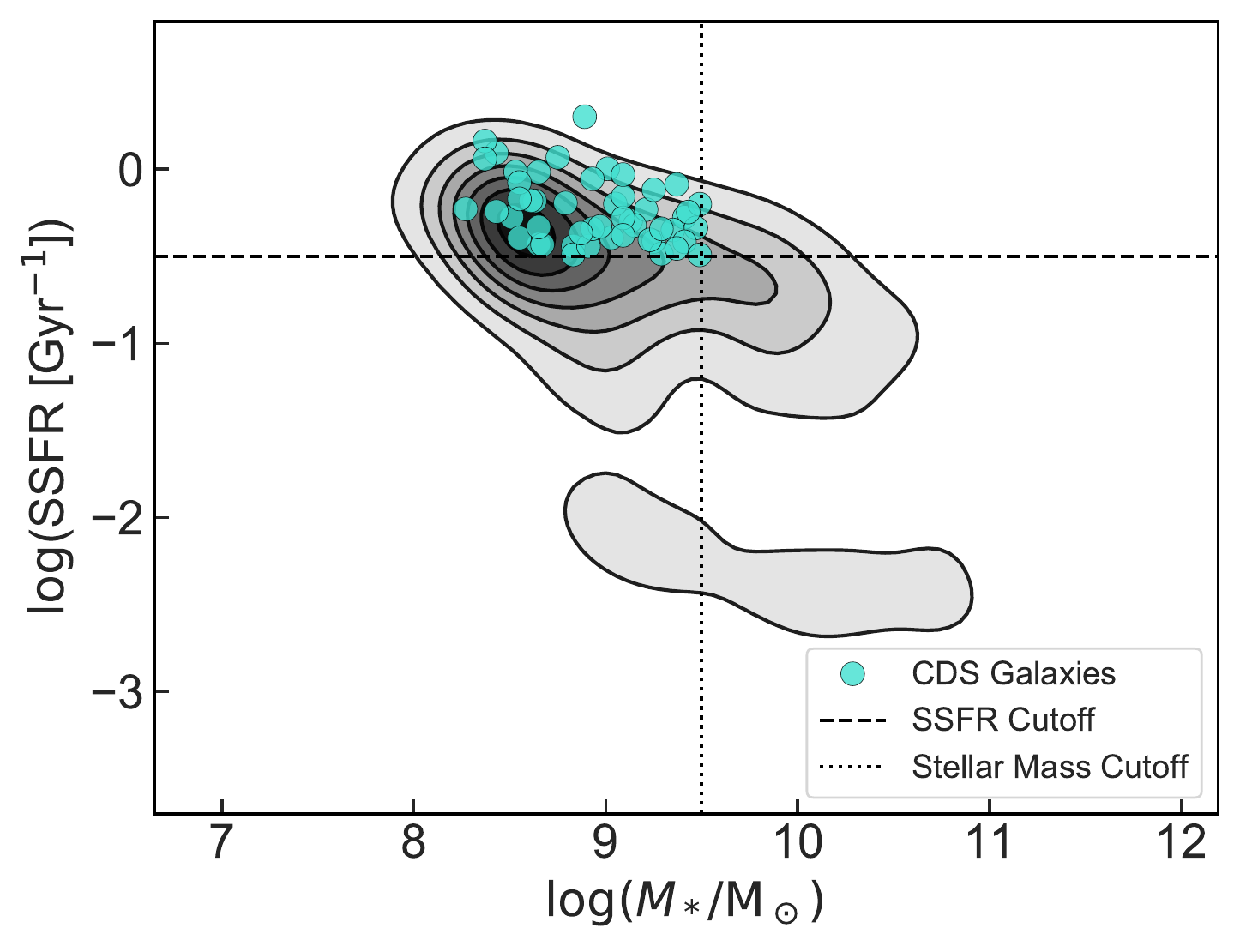}%
}\\
\vspace{-4mm}
\subfloat[]{%
  \includegraphics[clip,width=0.93\columnwidth]{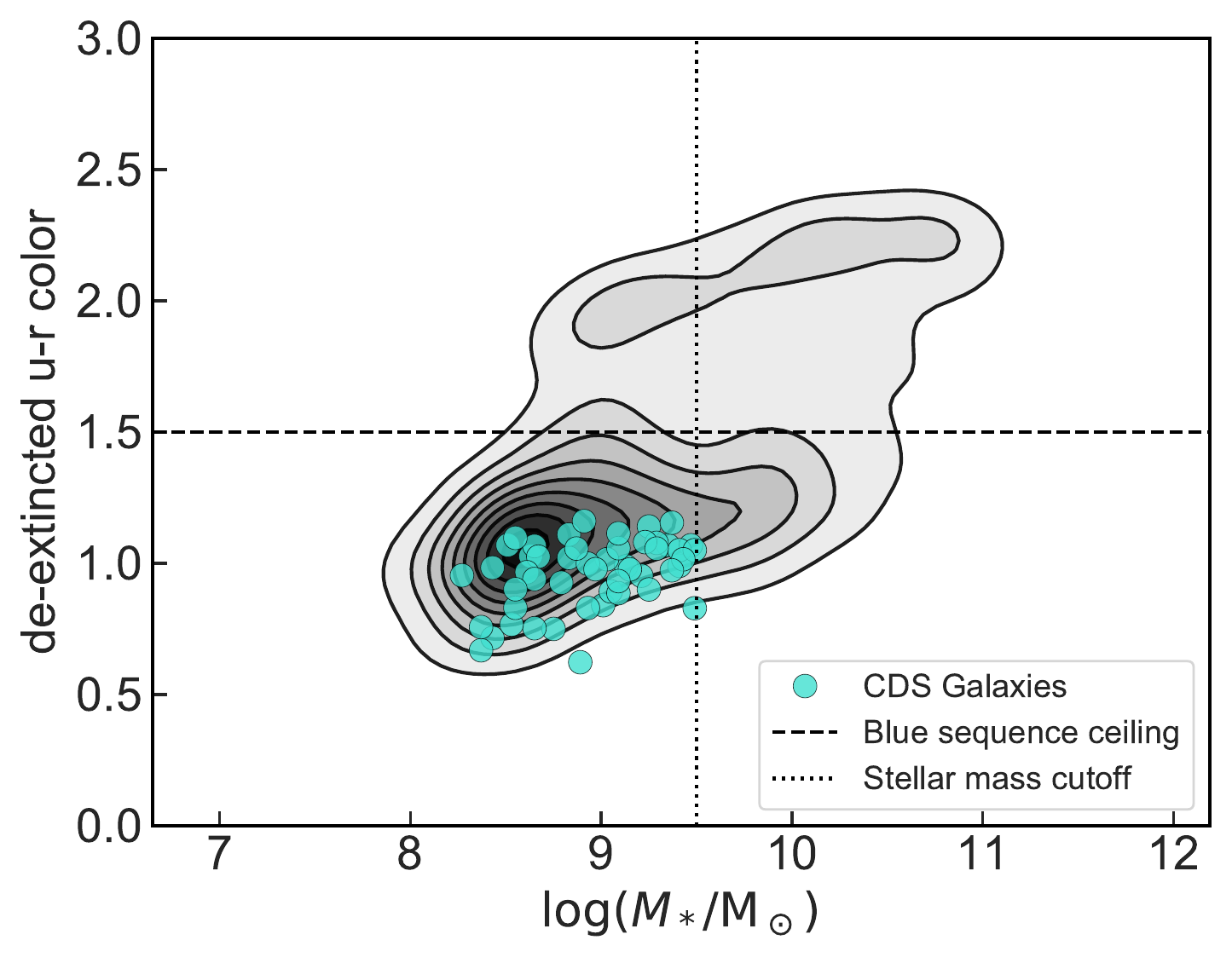}%
  }
\caption{Selection of low-$z$ CDS galaxies with $\mstar < 10^{9.5} \, \msun$. \textbf{Top:} Selected using the $\mu_\Delta$ parameter as a surrogate for Hubble type, the compactness of CDS galaxies reflects the intense compaction events that have driven their starbursts. \textbf{Middle:} The selection of $\log(\textnormal{SSFR [{\rm Gyr}$^{-1}$}]) > -0.5$, as in the \citet{2013ApJ...765..104B} high-redshift sample, ensures significant starbust activity. \textbf{Bottom:} The SSFR restriction implicitly requires that galaxies fall on the blue sequence, as shown. In all figures hereafter, the grayscale contours represent the 2D kernel density estimate (KDE) of all RESOLVE galaxies above the survey floor described in \S\ref{subsubsec:sample}. The contour levels are drawn to correspond to the 10th, 20th, etc.\ percentiles of the probability density function. In the top and middle panels, the lowest level contour is suppressed for visual clarity.}
\label{fig:selection}
\end{figure}

Guided by observed and predicted \cedit{redshift evolution} trends for blue nuggets, we restrict our sample of RESOLVE galaxies to a region of parameter space containing compact starburst galaxies in the dwarf mass regime. Implementing selection criteria in stellar mass, morphology, and specific star formation rate (SSFR), we identify a sample of \cedit{50} CDS galaxies in the RESOLVE survey as follows:

\begin{itemize}[leftmargin=*]
\item Theoretical study of blue nuggets \citep{2015MNRAS.453..408C, 2016MNRAS.457.2790T} and observations of prolate galaxies \citep{2014ApJ...792L...6V} have shown that hot halo quenching should begin to shut down cosmic gas accretion in \cedit{low-redshift} blue nuggets above stellar mass $\mstar \sim 10^{9.5}\, \msun$ as this is the approximate stellar mass corresponding to a central galaxy in a halo of mass $M_{\rm halo} \sim 10^{11.5}\, \msun$. This stellar mass also corresponds to the upper limit of the dwarf galaxy regime at $z\sim0$, the ``threshold scale" of \citet[see also \citealt{2003MNRAS.341...54K} and \citealt{2012ApJ...757...85G}]{2013ApJ...777...42K}. We select galaxies below this stellar mass ceiling and above the observational survey floor described in \S\ref{subsubsec:survey_def}. In this work, all comparison samples controlled on mass use the same stellar mass ceiling and observational survey floor.

\item As shown in high-$z$ simulations, blue nuggets are expected to be compact, prolate ellipsoids. To select on morphology, we use the $\mu_\Delta$ parameter developed by \citet{2013ApJ...777...42K} as a quantitative surrogate for Hubble type\footnote{$\mu_\Delta$ combines the overall stellar surface mass density with a stellar surface mass density contrast term: $\mu_\Delta = \mu_{90} + 1.7 \Delta_\mu$. The contrast term $\Delta_\mu$ is expressed as the difference between the stellar surface mass densities within the 50\% light radius and between the 50\% and 90\% light radii. See \citet{2013ApJ...777...42K}.}. \cedit{This parameter performs better than a simple concentration index ($R_{90}/R_{50}$) at separating early- and late-type galaxies, as demonstrated by \citet{2015ApJ...812...89M}. We also prefer $\mu_\Delta$ to density metrics such as $\Sigma_{\rm e}$ or $\Sigma_{1.5}$ since its contrast term better chooses galaxies with compact cores, as shown in \S\ref{subsubsec:ssmdcomp}.} Imposing a restriction of $\mu_\Delta > 8.6$ limits our sample to bulged-disk and spheroid-dominated galaxies with high degrees of concentration (Figure~\ref{fig:selection}, top panel).

\item Blue nuggets are impressive starbursts as a result of the intense gas compaction in their centers. To select galaxies undergoing starbursts, we impose the same specific star formation rate limit as \citet{2013ApJ...765..104B}: $\log(\textnormal{SSFR [{\rm Gyr}$^{-1}$}]) > -0.5$ (Figure~\ref{fig:selection}, middle panel).
\end{itemize}

\begin{figure}
\captionsetup[subfigure]{labelformat=empty}
\includegraphics[clip,width=\columnwidth]{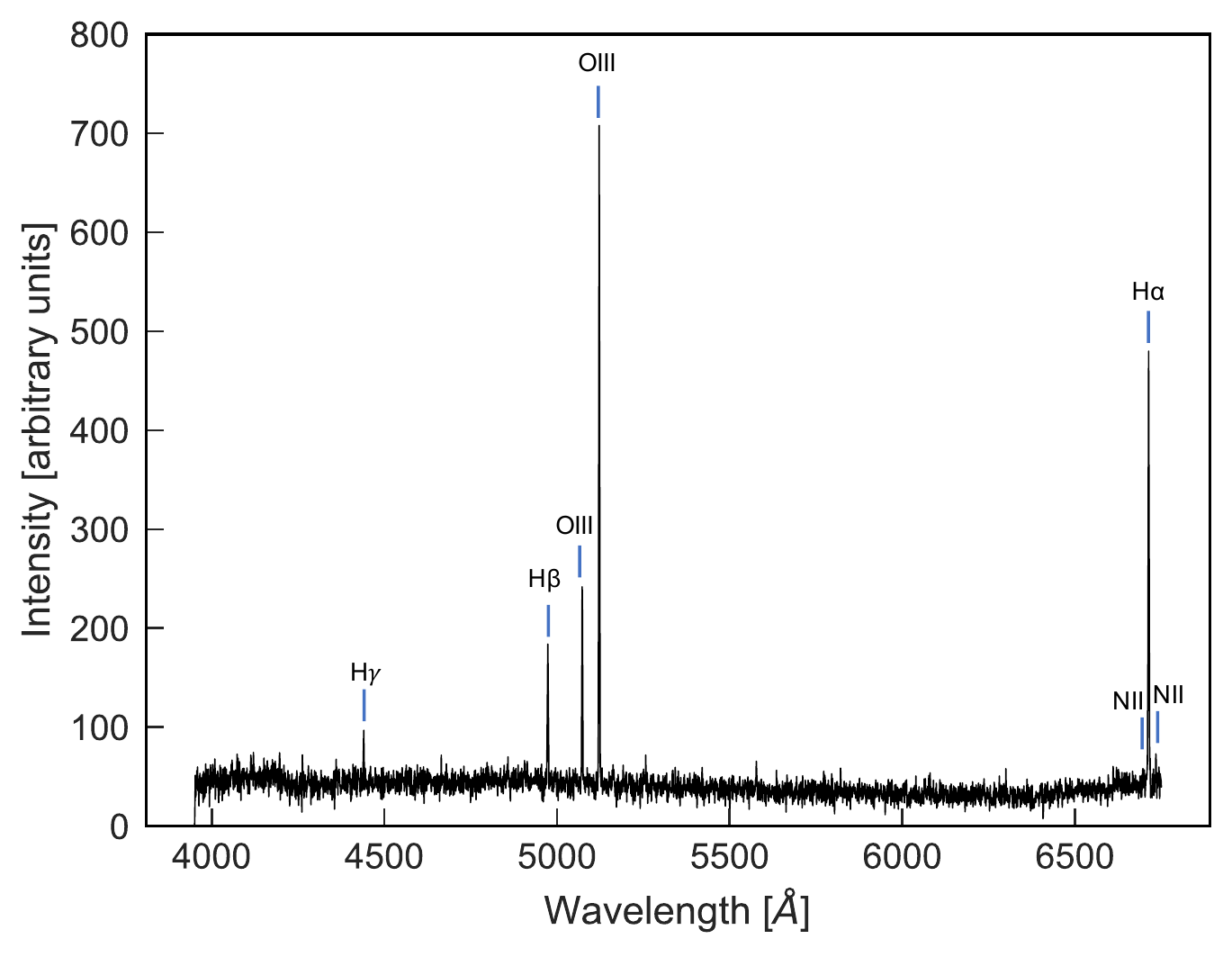}
\caption{SOAR Integral Field Spectrograph (SIFS, see \S\ref{subsec:3Dspec}) spectrum for the RESOLVE CDS galaxy rs1103 displaying strong Balmer series and [OIII] emission lines typical of CDS galaxy spectra. The strong H$\alpha$ emission signifies intense star formation, which is selected for by our UV- and IR-based SSFR restriction. The even stronger [OIII] emission lines relative to H$\beta$ and [NII] are likely indicative of metal-poor gas \citep{2004MNRAS.348L..59P}.}
\label{fig:ex_spec}
\end{figure}

These selection criteria isolate a population of \cedit{50} CDS galaxies in RESOLVE. As shown in the bottom panel of Figure~\ref{fig:selection}, our NUV-based SSFR restriction implicitly selects galaxies that fall on the blue side of the blue sequence, well below its upper edge as determined by \citet{2015ApJ...812...89M}, de-extincted $u-r<1.5$ for the mass range of interest. These galaxies' spectra are dominated by hydrogen Balmer series, doubly-ionized oxygen, and singly-ionized nitrogen emission features, with relatively low-level continuum light (Figure~\ref{fig:ex_spec}). We find that our selection criteria implicitly select RESOLVE galaxies whose de-extincted central H$\alpha$ fluxes lie in the upper $\sim 50^{\rm th}$ percentile. Moreover, the frequent presence of doubly-ionized oxygen lines comparable to or stronger than H$\beta$ likely implies low gas-phase metallicities \citep[see][]{2004MNRAS.348L..59P}. \cedit{A traditional BPT analysis does not reveal that any of these galaxies are AGN, but we note that their low metallicities and strong emission lines from star formation create a bias against AGN identification.} \par

These selection criteria explicitly seek \cedit{low-$z$} blue nuggets, whose masses, stellar densities, and SSFRs are typically less than or equal to those of high-$z$ blue nuggets. In \S\ref{sec:results} and \S\ref{sec:discussion} we will discuss blue nugget evolution with redshift, comparing our low-$z$ CDS sample to high-$z$ blue nugget samples. \par

\subsection{DECaLS Data} \label{subsec:decals}

We use photometry from Data Release\cedit{s} \cedit{7 (DR7) and 8 (DR8)} of the Dark Energy Camera Legacy Survey (DECaLS)\footnote{\url{http://legacysurvey.org/decamls/}} to probe the frequency of \cedit{likely recent mergers} in our CDS sample versus a control sample (see \S\ref{subsec:decals_clas} and \S\ref{subsubsec:doublenuc}). We also use DECaLS photometry to verify the E/S0 classifications of the blue E/S0 galaxies discussed in \S\ref{subsubsec:ssmdcomp} and \S\ref{sec:discussion}. DECaLS images have been previously calibrated by the DECam Community Pipeline\footnote{\url{https://www.noao.edu/noao/staff/fvaldes/CPDocPrelim/PL201_3.html}} before being processed by The Tractor \citep{2016ascl.soft04008L}, which creates the probabilistically motivated models and residual images we use in \S\ref{subsec:decals_clas} and \S\ref{subsubsec:doublenuc}. The Tractor approximates galaxy profiles with mixture-of-Gaussian models \citep{2013PASP..125..719H} which are fitted by $\chi^2$ minimization. We inspect residual images, in addition to the photometry and models, in the DECaLS Sky Viewer\footnote{\url{http://legacysurvey.org/viewer}}. \par

\subsection{3D Spectroscopic Data} \label{subsec:3Dspec}

For this work, we use 3D spectroscopic data obtained from the Gemini South Multi-Object Spectrograph Integral Field Unit (GMOS IFU), the SOAR Integral Field Spectrograph (SIFS), and the SOAR Adaptive Module Fabry-Perot (SAM FP). Both the GMOS IFU and SIFS use fiber-fed lenslet arrays to sample galaxy spectra at each lenslet position. Compared to the 1D velocity curves produced with traditional long-slit spectroscopy, integral field spectroscopy allows us to construct spatially-resolved velocity fields of extended objects by sampling object spectra at many discrete points. In a similar vein, the SAM FP uses an etalon to image a target at discrete wavelengths within a narrow spectral range. In comparison to IFU instruments, the SAM FP is able to sample at a higher spatial resolution at the expense of spectral range. A summary of the instrument setups used for this work is given in Table \ref{tab:setups}. Eight CDS galaxies were observed, but we do not list or analyze the SIFS data for rs0380, which were unusable due to persistent cross-talk. \par

To extract spatially-resolved velocity and continuum information, we rely primarily upon the strong H$\alpha$ emission line. Depending on the wavelength range of the instrument setup, we may also use the [NII] doublet, the [OIII] doublet, and H$\beta$. 
For full spatial coverage of the galaxy body, spatial tiling of exposures was necessary for the GMOS observations of rf0250 and rf0266. \par

From these observations, we obtain spatially resolved velocity fields, continuum maps, and H$\alpha$ flux maps, as described in \S\ref{subsec:cubes}. These data serve as our primary probe of i) minor-axis rotation or multiple misaligned rotation components and ii) kinematically-confirmed \cedit{merger evidence}. \par

For the GMOS IFU data we have developed a Gemini reduction pipeline in order to transform the two-dimensional detector data into three-dimensional data cubes. For a description of the Gemini reduction, see Appendix \ref{ap:gem}. Since we obtained the SIFS and SAM FP data in science verification (SV) time, the respective SV teams have performed the reduction for these data. The SAM FP reduction was performed as described in \citet{2017MNRAS.469.3424M}, including calibrations analogous to the GMOS IFU reduction (e.g.\ bias subtraction, flat fielding, wavelength calibration, cosmic ray rejection, etc.). We subsequently performed astrometric calibration on the SAM FP data by matching field stars in the Aladin Sky Atlas desktop program \citep{2000A&AS..143...33B}. The SIFS data reduction was performed in a process largely analogous to the GMOS IFU reduction process. However, an additional calibration step was required to extract the fiber spectra due to the dense packing of information along the spatial direction on the CCD, which would otherwise lead to severe fiber ``cross-talk" in the extracted spectra (Luciano Fraga, et al., in prep.). \par

\section{Methods} \label{sec:methods}

\begin{figure*}
\captionsetup[subfigure]{labelformat=empty}
\includegraphics[clip,width=2\columnwidth,valign=t]{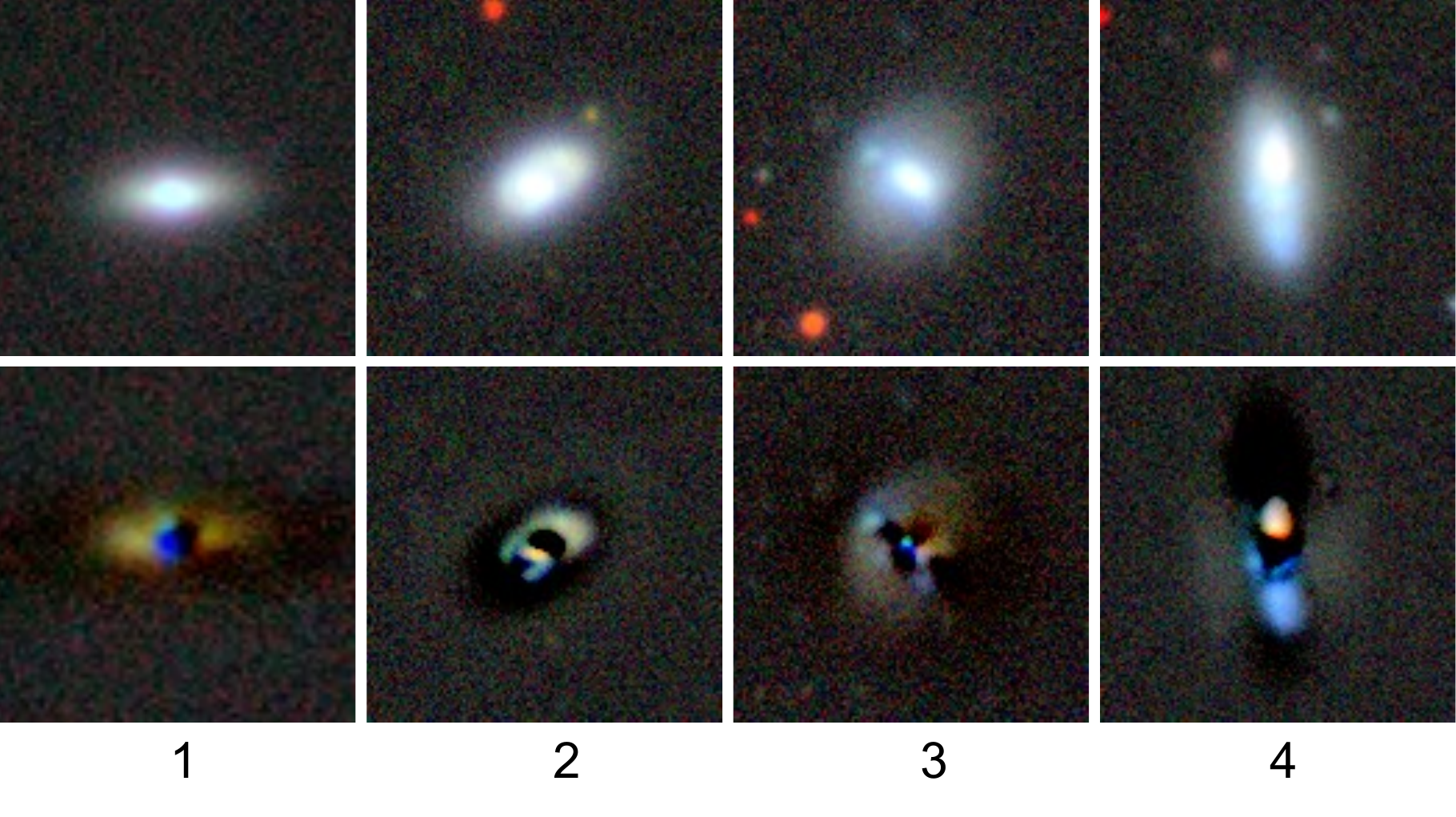}
\caption{\cedit{Example merger/non-merger classifications of CDS galaxies using DECaLS DR7 images (top panels) and model residuals (bottom panels). Galaxies are assigned categories, where 1 is an unambiguous non-merger, 2 an ambiguous non-merger, 3 an ambiguous merger, and 4 an unambiguous merger, as detailed in \S\ref{subsec:decals_clas}.}}
\label{fig:lineup}
\end{figure*}

\begin{figure*}
\captionsetup[subfigure]{labelformat=empty}
\subfloat[]{%
  \includegraphics[clip,width=0.58\columnwidth,valign=t]{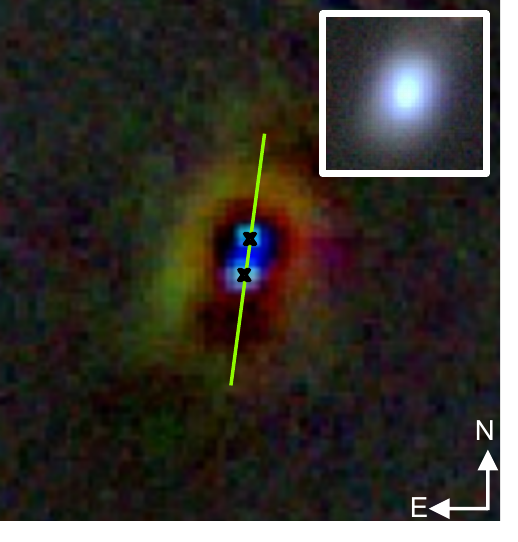}
  \includegraphics[clip,width=0.82\columnwidth,valign=t]{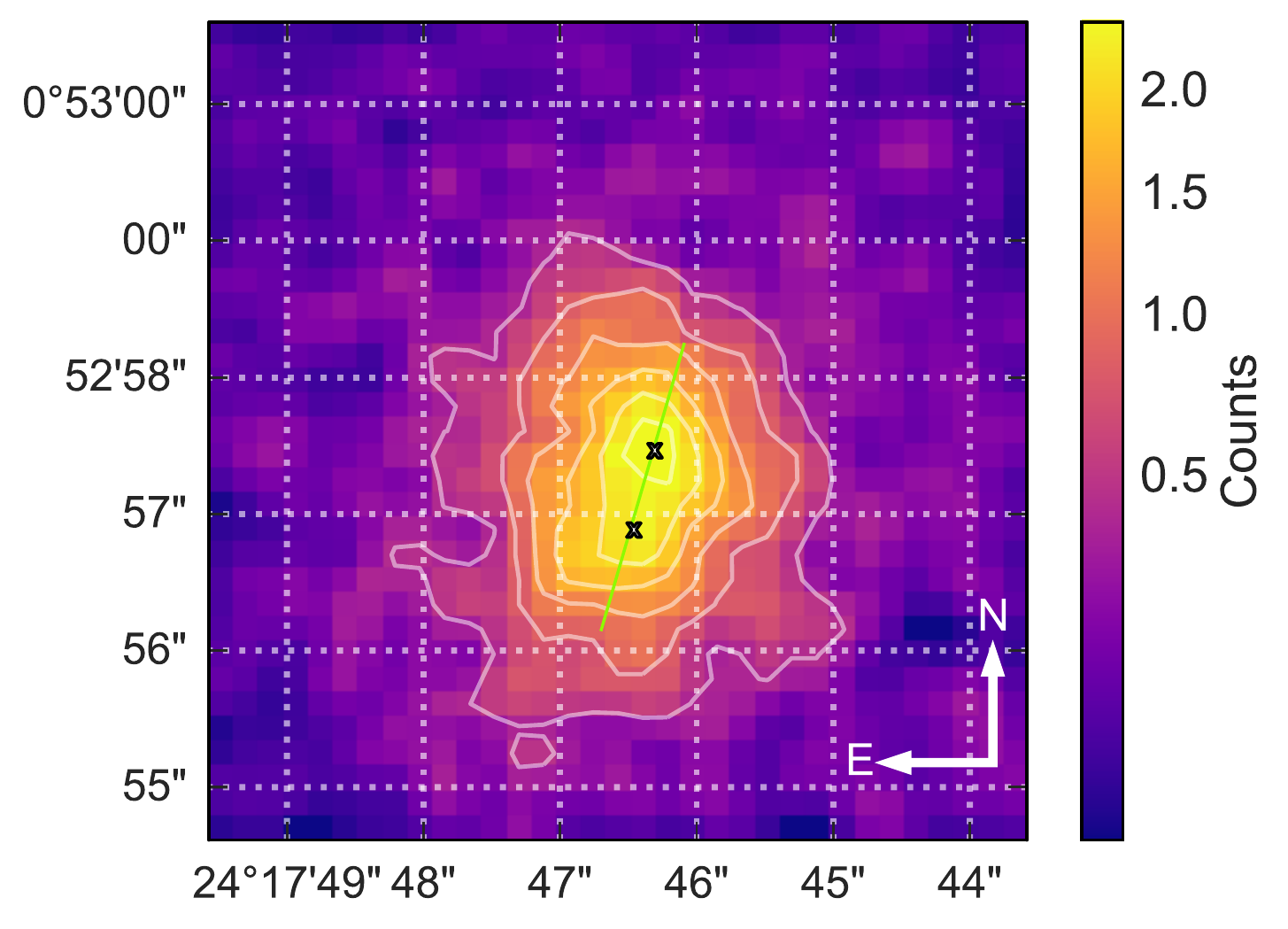}
  \includegraphics[clip,width=0.75\columnwidth,valign=t]{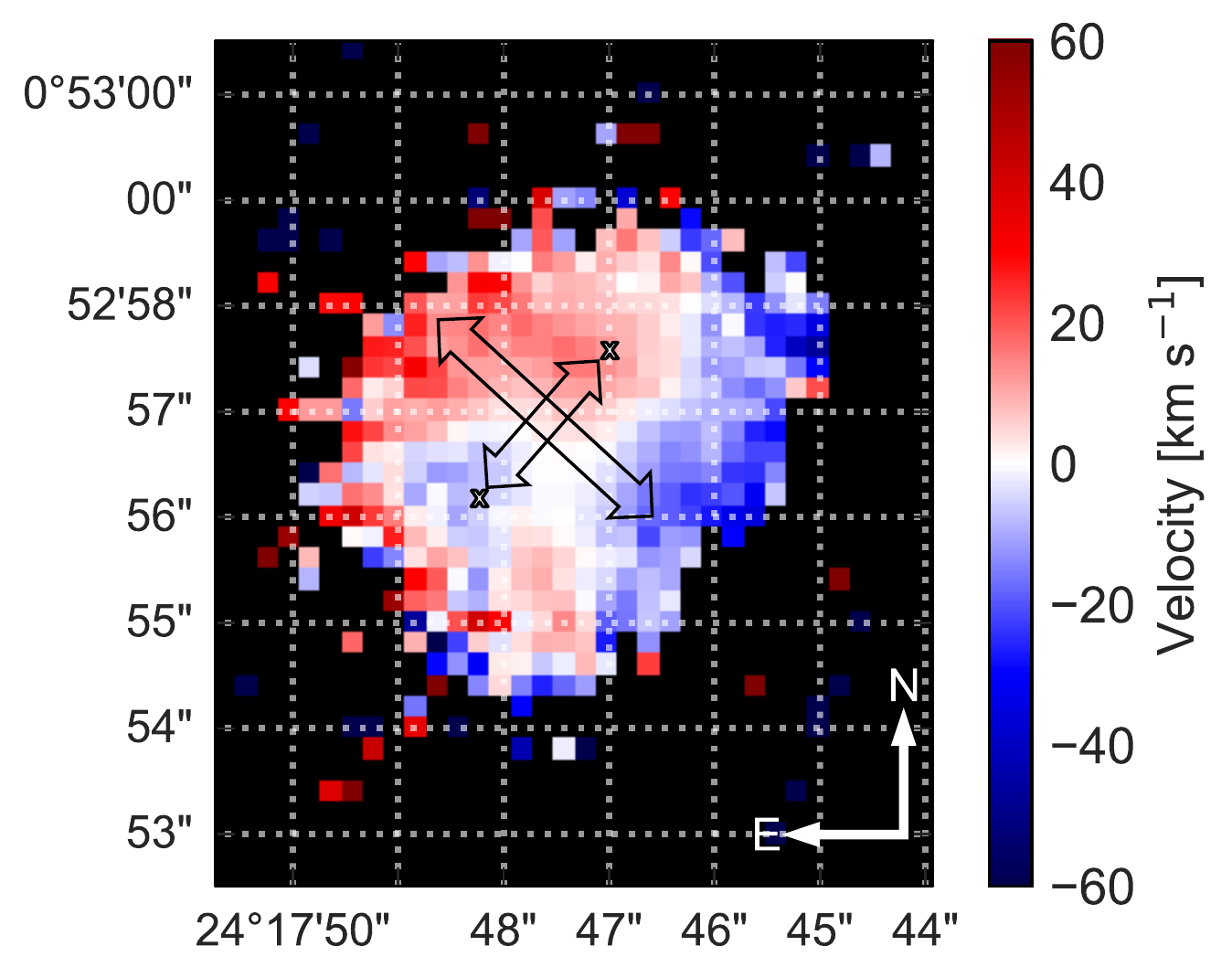}
}
\caption{\cedit{Merger} confirmation in rf0250 by three methods. \textbf{Left:} \cedit{Merger evidence, in the form of an apparent double nucleus,} found by eye in DECaLS residuals. The green line corresponds to the approximate northeast alignment axis of the double nucleus. \cedit{For reference, the DECaLS DR7 \textit{grz} photometry is inset in the upper right corner.} \textbf{Middle:} SAM FP continuum flux map for rf0250. Concentrations appear in two areas roughly matching the position of the \cedit{apparent} DECaLS nuclei. \textbf{Right:} GMOS IFU velocity field for rf0250. The inner velocity peaks (see \S\ref{subsubsec:kin}) approximately align with the positions of the DECaLS double nucleus. \cedit{Arrows in the right panel are drawn by eye to correspond to the numerically-identified positions of velocity maxima and minima, as described in \S\ref{subsec:cubes}.} The mismatch between the absolute celestial coordinates of the GMOS and SAM FP images is the result of imprecisions in the GMOS IFU World Coordinate Solution as described in Appendix \ref{ap:gem}.}
\label{fig:double_nuc}
\end{figure*}

\begin{table*}
	\centering
	\begin{tabular}{llll}
		\hline
		& GMOS IFU & SIFS & SAM FP \\
		\hline
		spectral range & 5500 - 6900 \AA & 4000 - 6800 \AA & varies$^{\rm a}$ \\
		lines of interest & H$\alpha$, [NII] & H$\beta$, [OIII], H$\alpha$, [NII] & H$\alpha$ \\
		grating & B600 & 700B & -- \\
		filter & r-G0326 & none & LAM-M13, LAM-M15 \\
		pseudoslits & 2 slits & 1 slit & -- \\
		field of view & 5" $\times$ 7" & 7.8" $\times$ 15" & $3' \times 3'$\\
		spectral resolution & 1688 & 4200 & $\sim 12000^{\rm a}$ \\ 
		projected fiber diameter & 0.2" & 0.3" & -- \\
		program ID & GS-2013B-Q-51 & -- & -- \\
		galaxies & rf0250, rf0266 & rf0363, rs0463, & rf0250, rf0266, rs0804 \\
		& & rs1103, rs1259 & \\
		\hline
	\end{tabular}
\caption{Summary of 3D spectroscopic instrument setups. $^{\rm a}$The spectral resolution of the SAM FP is dependent on the free spectral range (FSR), which is set by the etalon interference order and wavelength. The wavelength range depends upon the interference filter chosen to correspond to the mean heliocentric velocity of the observed object. See \citet{2017MNRAS.469.3424M}. \label{tab:setups}}
\end{table*}

To probe whether the gas-rich merger mechanism of blue nugget formation applies to CDS galaxies in the low-$z$ universe, we use both (1) DECaLS imaging to assess \cedit{the rate of evidence for recent mergers} in the entire low-$z$ CDS sample, and (2) follow-up 3D spectroscopy to examine the gas kinematics and continuum/H$\alpha$ morphology, as well as to cross-check \cedit{the DECaLS merger classification}, for a subset of seven CDS galaxies. In \S\ref{subsec:decals_clas}, we describe our \cedit{classification} of \cedit{galaxies} \cedit{using} DECaLS \cedit{imaging and} Tractor model residuals. In \S\ref{subsec:cubes}, we describe our extraction of spatially-resolved velocity fields, including how we assess multiple misaligned rotation axes, and our production of continuum and line flux maps. \par

\subsection{Analysis of DECaLS Imaging} \label{subsec:decals_clas}

As a probe of the formation histories of CDS galaxies, we use by-eye classification of DECaLS \cedit{DR7} images and model residuals to assign a \cedit{binary merger/non-merger} flag to RESOLVE galaxies in two samples: i) the CDS galaxy sample, and ii) a control sample consisting of the \cedit{125} RESOLVE galaxies obeying the CDS mass and morphology selection but falling short of the SSFRs required for CDS galaxies. To perform the classification, we examine \cedit{both images and model} residuals, consulting the models as needed, looking for \cedit{features indicative of recent mergers such as distinct nuclei and tidal streams.} \cedit{In many cases} in both the CDS and control samples it is somewhat unclear whether \cedit{the observed features are truly indicative of a recent merger.} In these cases, \cedit{the galaxies are} flagged as ``ambiguous" to distinguish \cedit{them} from clear \cedit{classifications.} \par

\cedit{This system of classification leads to four categories of galaxies illustrated in Figure~\ref{fig:lineup}: unambiguous mergers, ambiguous mergers, ambiguous non-mergers, and unambiguous non-mergers. To compute statistics from these classifications, we assign each category a score of 1-4, where 1 corresponds to an unambiguous non-merger and 4 an unambiguous merger. To ensure the robustness of this by-eye method, two of the authors (MP and SK) performed the DR7 classifications independently. Any galaxies with scores differing by greater than one were independently revisited; only one required discussion to achieve convergence within a score difference of one. We take the average classification score for each galaxy to compute the statistics reported in this paper \S\ref{subsubsec:doublenuc}. As an independent check on the robustness of our results, another of the authors (DC) independently classified the same galaxies using DR8 images and model residuals. These classifications yield consistent results, as described in \S\ref{subsubsec:doublenuc}.} \par

\subsection{Analysis of Spectroscopic Data Cubes} \label{subsec:cubes}

\begin{figure}
\includegraphics[width=\columnwidth]{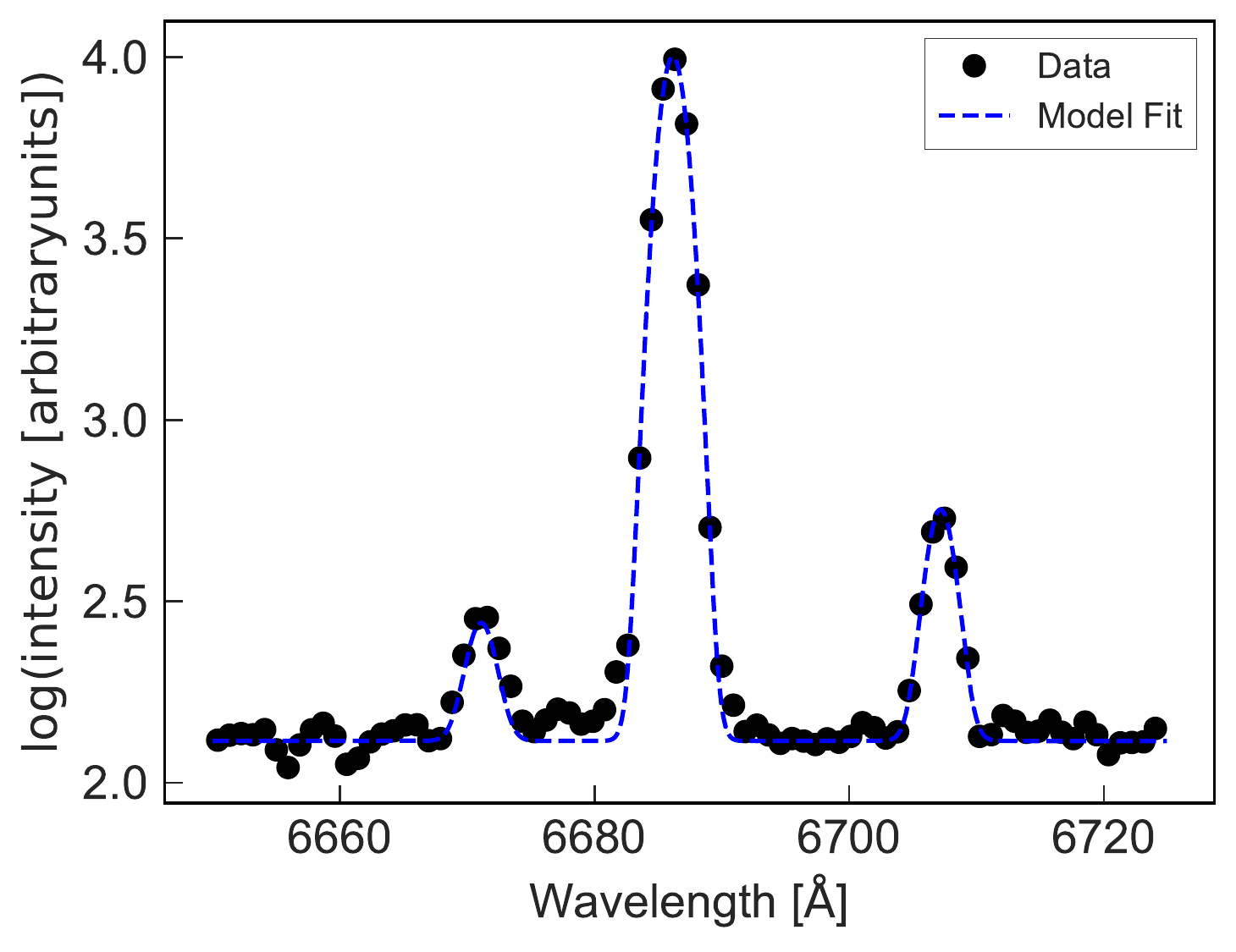}
\caption{Example spectrum extracted from GMOS IFU data for rf0266 with Gaussian fit. The \halpha and [NII] emission line data are plotted as black points, with the best-fit triple-Gaussian model overplotted as a dashed blue curve. We obtain greater centroiding accuracy by fitting to the \halpha and [NII] lines, as opposed to only the \halpha line. The $y$-axis is plotted on a log scale for visual clarity.}
\label{fig:fit_spec}
\end{figure}

\cedit{To further explore evidence for mergers and/or colliding gas streams in CDS galaxies, we analyze spectroscopic data cubes for features such as double nuclei and minor-axis rotation. A comparison of imaging and spectroscopy for a sample galaxy is shown in Figure~\ref{fig:double_nuc}.}  We extract three key pieces of information from the 3D spectroscopic data: velocity fields, continuum maps, and H$\alpha$ line flux maps. We use the \textit{mpfit} algorithm \citep{2009ASPC..411..251M}, translated into Python by Mark Rivers\footnote{\url{http://cars9.uchicago.edu/software/python/mpfit.html}}, to perform non-linear least-squares fitting of the spectra with a Gaussian line model. Wherever possible, we fit multiple emission lines to obtain increased centroiding accuracy. For the SAM FP, we fit a single Gaussian to the H$\alpha$ line. For the GMOS IFU, we fit three Gaussians to the [NII] and H$\alpha$ lines. For SIFS, we fit the [OIII] doublet, H$\beta$, the [NII] doublet, and H$\alpha$. An example Gaussian fit for the GMOS IFU data is plotted in Figure~\ref{fig:fit_spec}. \par

\subsubsection{Velocity Fields} \label{subsec:vfield}

The emission line radial velocity $z$ is the product of internal rotation, peculiar motion, and cosmological redshift components:

\begin{equation} \label{eq:red}
1 + z = \left(1 + \frac{v_{\rm rot}}{c}\right) \left(1 + \frac{v_{\rm pec}}{c}\right) \left(1 + \frac{v_{\rm cosm}}{c}\right)
\end{equation}

\noindent Since we cannot disentangle peculiar and cosmological redshifts, we assume that the peculiar and cosmological terms can be consolidated as such:

\begin{equation} \label{eq:simple_red}
1 + z = \left(1 + \frac{v_{\rm rot}}{c}\right) \left(1 + \frac{v_{\rm hel}}{c}\right),
\end{equation}

\noindent where $v_{\rm hel}$ is the recessional velocity of the galaxy in the heliocentric reference frame, taken from catalog redshift measurements as described in \S\ref{subsubsec:survey_def}. Rearranging Equation \ref{eq:simple_red}, we calculate the internal rotation velocity as:

\begin{equation}
v_{\rm rot} = c\left(\left(1+z\right)  \left(1 + \frac{v_{\rm hel}}{c}\right)^{-1} - 1 \right),
\end{equation}

\noindent where $z$ is obtained from the model Gaussian fit. We iteratively improve our measurement of $v_{\rm rot}$ by remeasuring $v_{\rm hel}$ as the average velocity in the high signal-to-noise ratio (SNR) region of the initial velocity map. \par

As shown in \citet{2015MNRAS.453..408C} and \citet{2016MNRAS.458.4477T}, blue nuggets tend to display complicated rotation patterns, which are preferentially oriented along the minor axis. We observe both minor-axis rotation and multiple misaligned rotation components in our velocity fields (\S\ref{subsubsec:kin}). We have confirmed that the complex velocity features are real physical features as opposed to erroneous detections by careful investigation of the flat-fielding (see Appendix \ref{ap:gem}). In addition, two galaxies were observed with both the GMOS IFU and SAM FP, yielding consistent results.  \par

In order to methodically and reproducibly identify rotation axis directions in the velocity fields, we numerically identify the positions of the velocity field maxima and minima. For a given velocity field, we first mask the low-SNR spaxels before smoothing the field by convolving it with a 2D median filter with a kernel size of five spaxels. The smoothing both ensures that any high-frequency noise is filtered out and also further improves the SNR of the velocity field. To specify the direction of the dominant rotation pattern, we measure the spatial positions of the velocity field maxima and minima. Where a second component may be present, as in the case of the GMOS IFU observations of rf0250 and rf0266, we then mask the spaxels beyond the galaxy half-light radius and recalculate the positions of the smaller amplitude, inner galaxy velocity peaks to specify a second rotation direction. In Figure~\ref{fig:double_nuc} the arrows indicating the galaxy rotation directions are drawn by eye, corresponding to the numerically-determined positions of the velocity peaks. The widths of the arrows are drawn arbitrarily to maximize visual clarity. \par

\subsubsection{Continuum and Line Flux Maps} \label{subsec:cont}

We use continuum maps from the follow-up 3D spectroscopic data to provide direct confirmation of some of the \cedit{merger signatures} seen in the DECaLS data. We simply determine the relative level of the continuum light from the vertical offset in our Gaussian line model. \par

The strong star formation in CDS galaxies also implies significant H$\alpha$ flux. As a result, H$\alpha$ flux maps are often more promising probes of detailed galaxy structure compared to the weaker continuum light maps. Also for data taken with the SAM FP, the instrument's narrow spectral range leads to a poor sampling of the continuum light. \par

\section{Results} \label{sec:results}

The most basic question we seek to answer is whether some or all $z\sim0$ CDS galaxies qualify as blue nuggets based on halo mass/gas richness, compactness relative to contemporary galaxies, and formation by compaction. We also wish to compare CDS galaxy numbers and properties with expected blue nugget \cedit{redshift evolution} trends, and we aim to assess the relative importance of gas-rich mergers vs.\ colliding streams in forming these potential low-$z$ blue nuggets. In \S\ref{subsec:link} below, we examine properties of CDS galaxies such as environment, gas content, compactness, frequency, and SSFR distribution as probes of blue nugget status. In \S\ref{subsec:result_formation}, we look for specific signatures of formation by either gas-rich mergers or colliding gas streams --- in particular, prolate structure, \cedit{perturbed morphology (such as tidal streams and double nuclei)}, and unusual kinematics --- both to verify formation by compaction and to assess the relative importance of its two main channels. \par

\subsection{Could Compact Dwarf Starburst Galaxies Be Low-$z$ Blue Nuggets?} \label{subsec:link}

In this section we assess CDS galaxy halo masses/gas richness, compactness, frequency in the galaxy population, and specific star formation rates. Together, these results suggest our sample of CDS galaxies largely consists of \cedit{low-$z$} blue nuggets. We assess direct evidence for formation by compaction, another defining feature of blue nuggets, in \S\ref{subsec:result_formation}. \par

\subsubsection{Halo Masses, Gas Richness, and Neighbors} \label{subsubsec:environ}

\begin{figure*}
\captionsetup[subfigure]{labelformat=empty}
\subfloat[]{%
  \includegraphics[clip,width=\columnwidth]{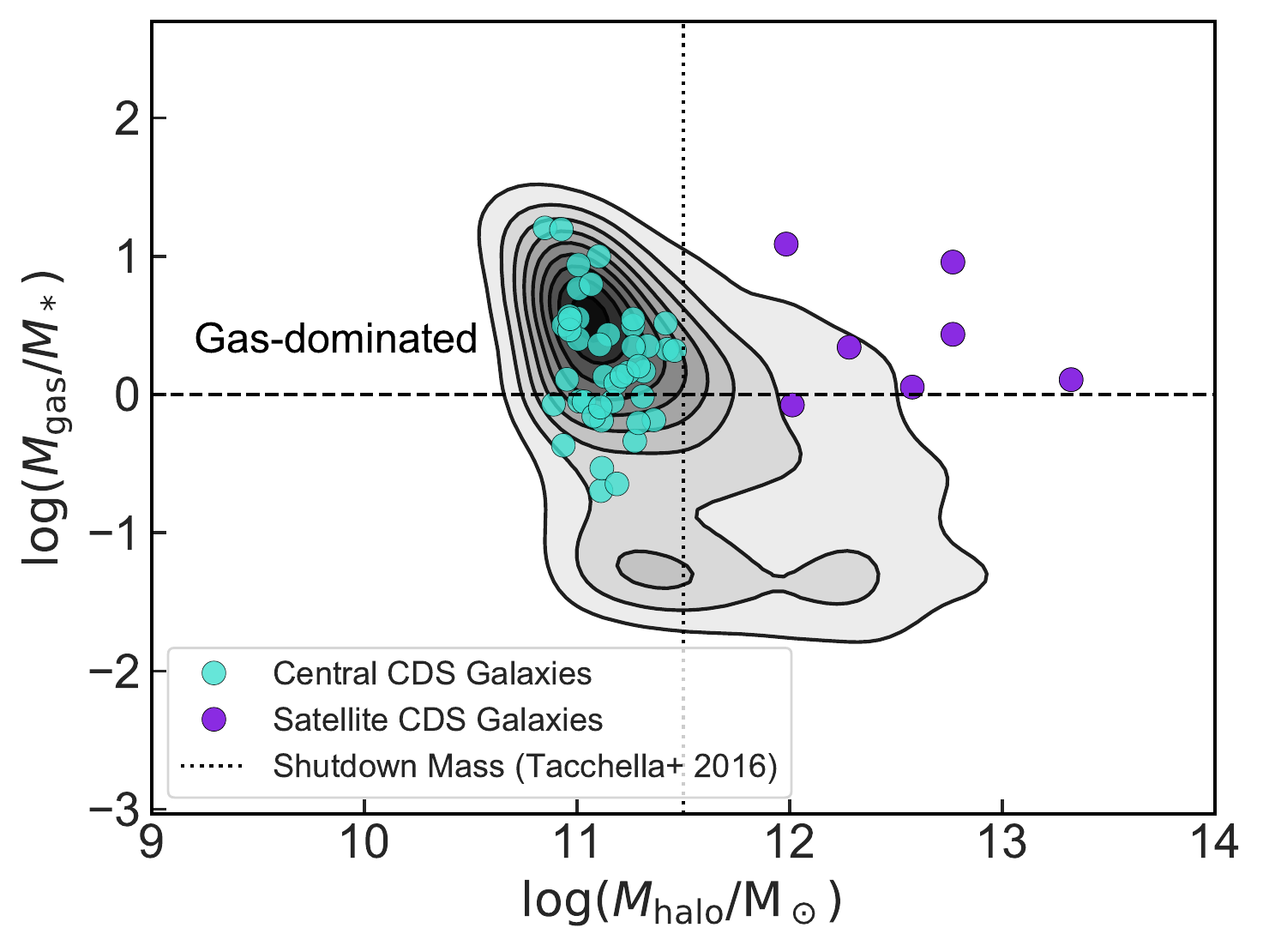}%
  \includegraphics[clip,width=\columnwidth]{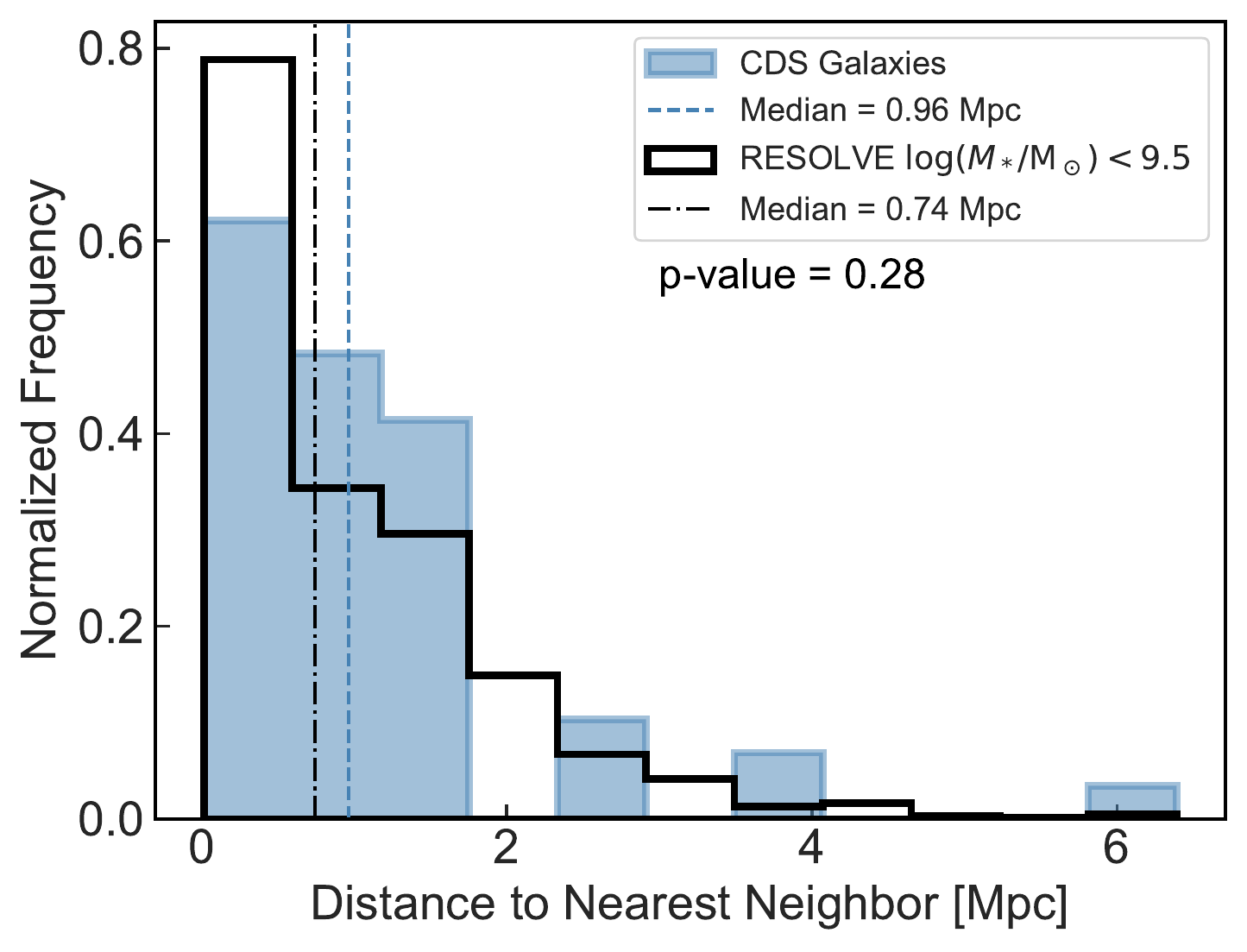}%
}
\caption{\textbf{Left:} CDS galaxies in RESOLVE generally occupy halos below the hot halo quenching threshold of $M_{\rm halo} \sim 10^{11.5}\, \msun$ and exhibit atomic gas-to-stellar mass ratios, $M_{\rm gas}/\mstar$, greater than unity. All but seven low-$z$ CDS galaxies are the central galaxies (turquoise) in their halos; the rest are satellites (purple). The satellites are all highly gas rich, which may reflect recent infall into their groups or may indicate group-finding errors (see \S\ref{subsubsec:survey_def}). Grayscale contours are drawn to correspond to the 10th, 20th, etc.\ percentiles of the probability density function. \textbf{Right:} Distribution of CDS nearest-neighbor distances revealing that they exist in environments similar to those of other dwarf galaxies, with if anything slightly larger nearest-neighbor distances.}
\label{fig:nn}
\end{figure*}

\citet{2014MNRAS.438.1870D} define blue nuggets as compact starbursts forming via fast-track growth involving extreme gas accretion. Simulations \citep{2016MNRAS.457.2790T} and observations \citep{2013ApJ...765..104B} of blue nuggets suggest that while many of the properties of blue nuggets (such as number, mass, compactness, and star formation rate) should evolve with cosmic time, the halo mass above which hot-halo quenching shuts down efficient cosmic gas accretion should remain relatively fixed following $z=2$--$3$. \cedit{According to \citet{2006MNRAS.368....2D} large-scale cold streams in cosmic filaments become too thick to pierce the hot gas in massive halos after $z\sim3$, and thereafter efficient cold gas accretion occurs solely in halos below $M_{\rm halo} \sim 10^{11.5} M_\odot$, where rapid halo gas cooling dominates over shock-heating independent of redshift. In fact, rapid gas cooling below $M_{\rm halo} \sim 10^{11.5} M_\odot$ is a general feature of theoretical models \cite{} \citep[e.g.,\ ][]{2011MNRAS.416..660L, 2013MNRAS.429.3353N} and not just the cold-mode picture, consistent with observational results suggesting the dominance of both cold HI and WHIM gas below this critical halo mass \citep[e.g.,\ ][]{2010ApJ...708L..14M, 2016ApJ...832..126S}.} \par

As shown in Figure~\ref{fig:nn} (left panel), the majority of our low-$z$ CDS galaxies occupy halos below $M_{\rm halo} \sim 10^{11.5}\, \msun$ ($\sim$86\%) and/or have gas-dominated composition with atomic gas-to-stellar mass ratio $M_{\rm gas}/\mstar > 1$ ($\sim$68\%), consistent with expectations for blue nuggets. Seven CDS galaxies appear to lie in massive halos ($M_{\rm halo} \gtrsim 10^{11.5}\, \msun$), but all exhibit high $M_{\rm gas}/\mstar \gtrsim 1$, suggesting these seven objects may have only recently fallen into their groups\cedit{, consistent with their high atomic-gas fractions,} or may reflect group-finding errors (see \S\ref{subsubsec:survey_def}). We find that $\sim$32\% of CDS galaxies that occupy low-mass halos below the quenching threshold have $M_{\rm gas}/\mstar < 1$ based on our atomic gas data. Most are still gas rich, but perhaps $\sim$10\% of CDS galaxies may be considered (atomic) gas poor. If these galaxies formed by compaction, they may be either molecular-gas dominated (see \S\ref{sec:discussion}) or in depletion phases within the cycle of compaction and depletion. \par

We note that our CDS sample follows the same distribution of nearest neighbor distances as the general dwarf galaxy population in RESOLVE (with $p=0.28$ in a two-sample K-S test, Figure~\ref{fig:nn}, right panel). The median nearest neighbor distance for CDS galaxies is $\sim$$0.96$ Mpc compared to the general dwarf population median of $\sim$$0.74$ Mpc, suggesting that if anything, CDS galaxies are preferentially found in greater isolation. \cedit{Moreover, 43 out of 50 CDS galaxies are centrals in low-mass halos.} These results are compatible with their formation by either colliding gas streams or gas-rich mergers, where in the latter case the CDS galaxies are remnants of recent dwarf-dwarf mergers. \cedit{For the seven that are satellites in massive halos, assuming they do not represent group-finding errors, similar formation scenarios may apply if they are on first infall (alternatively, they may represent ram-pressure confined starbursts as in \citealt{2019ApJ...875...58D}).} We will examine morphological and kinematic evidence to assess the gas-rich merger and colliding stream scenarios in \S\ref{subsec:result_formation}. \par 

\subsubsection{Compactness} \label{subsubsec:ssmdcomp}

Blue nuggets are expected to be less compact at later epochs (\citealt{2014MNRAS.438.1870D} and \citealt{2015MNRAS.450.2327Z}; see also \citealt{2013ApJ...776...63F}). Nevertheless, \cedit{low-$z$} blue nuggets are defined in part by extreme compactness relative to contemporaneous galaxies. Our sample selection has enforced compactness only generically, with a threshold in $\mu_\Delta$ corresponding to bulged-disk and spheroid morphologies (\S\ref{subsubsec:sample}). Thus, it is worthwhile to assess whether our final CDS sample is in fact notably compact with respect to contemporaneous RESOLVE galaxies in the mass-size relation, and to what extent our CDS galaxies have reduced compactness compared to high-$z$ blue nuggets as expected. \par

\begin{figure}
\includegraphics[width=\columnwidth]{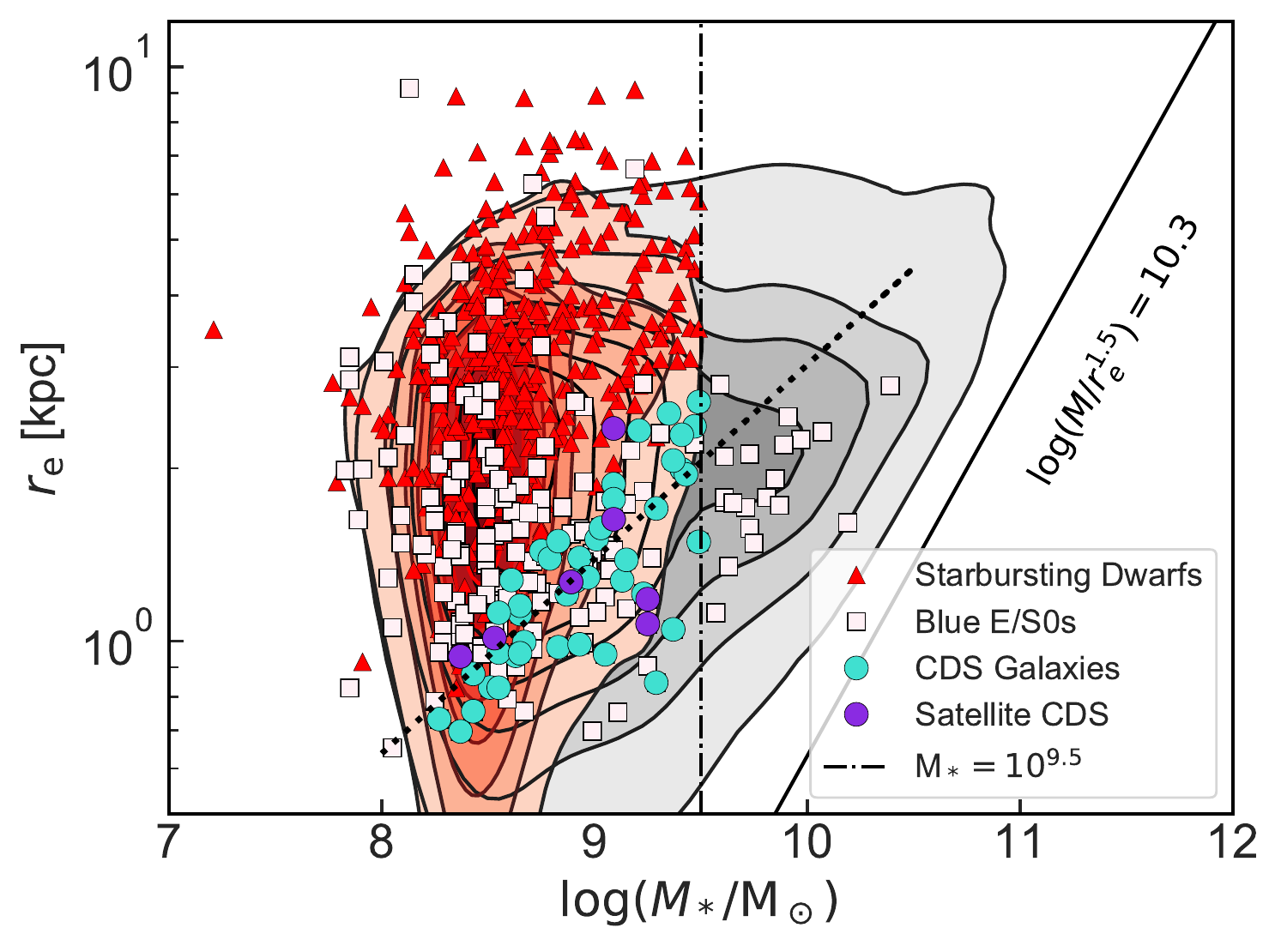}
\caption{Mass-size diagram for the RESOLVE survey (contours as in Figure~\ref{fig:nn}) with subpopulations marked, revealing the small sizes at fixed mass (high densities) of our CDS sample \cedit{(centrals as turquoise dots and satellites as purple dots)} relative to the general population of high-SSFR dwarfs selected without any $\mu_\Delta$ morphology restriction (red triangles and contours). The low-$z$ CDS mass-size relation (dotted line fit) is tighter and significantly offset toward the edge of the general sample distribution. Some subpopulations such as blue E/S0 galaxies (light pink squares; further discussed in \S\ref{sec:discussion}) overlap the CDS sample with scatter to even higher (and lower) compactness. Note however that we show blue E/S0s extending above the CDS mass cutoff (\cedit{vertical} black dash-dotted line, see \S\ref{subsubsec:ssmdcomp}), where the general RESOLVE mass-size distribution also becomes slightly more compact. Notably, no part of the $z\sim0$ galaxy population reaches the mass-size relation implied by the surface density selection of \citet{2013ApJ...765..104B}, indicated by the solid black line.}
\label{fig:mass_size}
\end{figure}

As shown in Figure~\ref{fig:mass_size}, the mass-size relation for our CDS sample (turquoise \cedit{and purple} dots and dotted line) forms a tight locus on the lower edge of the mass-size distribution for star-forming dwarf galaxies in general (red triangles and red contours). The general comparison sample is selected to have dwarf masses and high SSFRs just like the CDS sample but allows any value of $\mu_\Delta$. The tightness of the CDS mass-size relation clearly reflects the $\mu_\Delta$ selection, since the mass and SSFR criteria alone do not favor compactness.  \cedit{We find no difference in the  compactness distributions of central and satellite CDS galaxies (as distinguished in Figure~\ref{fig:nn} and Figure~\ref{fig:mass_size}), albeit the number of satellites is small.}

Anticipating \S\ref{sec:discussion}, where we will compare our candidate low-$z$ blue nuggets to other low-$z$ compact starburst populations, Figure~\ref{fig:mass_size} also highlights RESOLVE's blue E/S0 galaxies (E/S0s located on the blue sequence as defined by \citealt{2015ApJ...812...89M}), which partially overlap the CDS sample but display greater range in mass and compactness (potentially reflecting a wider range of evolutionary states; see \S\ref{sec:discussion}). At the high-mass end of their mass range, blue E/S0s are the galaxies that most closely approach the mass-size line equivalent to the surface density cut used by \citet{2013ApJ...765..104B} to select high-$z$ blue nuggets (solid black line in Figure~\ref{fig:mass_size}). However, no galaxy in the complete, volume-limited RESOLVE survey crosses this line. The failure of low-$z$ galaxies to meet this threshold is consistent with the cosmic time evolution of compactness seen by \citet{2013ApJ...765..104B}, who find that galaxies below $\mstar \sim 10^{10} \ \msun$ evolve toward lower surface densities with redshift. Moreover, \citet{2015MNRAS.450.2327Z} have shown in simulations that less massive galaxies tend to compact to lower densities than their high-mass counterparts. Although the least massive galaxy they simulate is more massive than any of our low-$z$ CDS galaxies, the lower density of RESOLVE CDS galaxies appears consistent with the general trends discussed in \citet{2015MNRAS.450.2327Z} and \citet{2014MNRAS.438.1870D} as well as the observations of \citet{2013ApJ...776...63F}. Likewise, the fact that the highest density blue E/S0s in Figure~\ref{fig:mass_size} have $\mstar$ in the giant galaxy regime is in line with the results of \citet{2013ApJ...765..104B}, who find that across redshifts, more massive galaxies are able to achieve greater compactness.  \par

\begin{figure*}
\includegraphics[width=1.9\columnwidth]{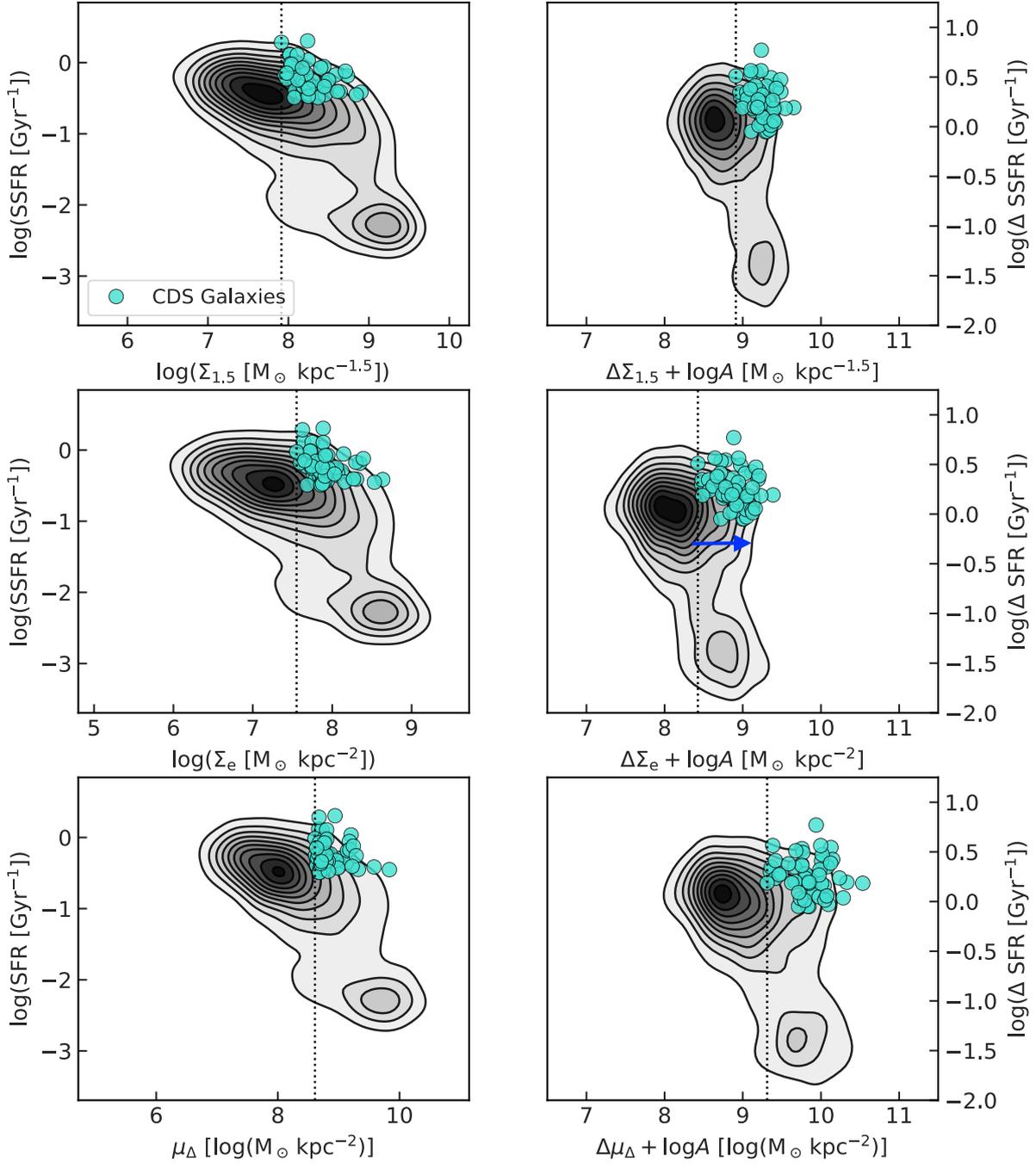}
\caption{\cedit{\textbf{Left column}: Position of low-$z$ CDS galaxies in plots of SSFR vs. $\Sigma_{1.5}$, $\Sigma_{\rm e}$, and $\mu_\Delta$. \textbf{Right column}: Plots of the ``$\Delta$''  quantities as in Figure 7 of \citet{2017ApJ...840...47B}. These $\Delta$ quantities may be thought of as residuals from the star-formation main sequence and structural relation power laws. The blue arrow in the middle right plot qualitatively represents the trajectory of redshift evolution over $z \sim 3.0 - 0.5$ in Figure 6 of \citet{2017ApJ...840...47B}. We find that our $\mu_\Delta$ cutoff from \S\ref{subsubsec:sample} corresponds to effective cutoffs in both  $\Sigma_{1.5}$ and $\Sigma_{\rm e}$ space (vertical dotted lines). However, the spread of points in the $\Delta \mu_\Delta$ plot (bottom right panel) is larger than for the other density metrics. This difference is due to the contrast term in $\mu_\Delta$, which is sensitive to galaxies with compact cores. Grayscale contours are as in Figure~\ref{fig:selection}.}}
\label{fig:six_panel}
\end{figure*}

\cedit{By design, our CDS galaxies have been selected with extreme compactness in mind. Although the low-$z$ CDS sample fails to reach the surface density threshold used to identify high-$z$ blue nuggets by \citealt{2013ApJ...765..104B} (as seen in Figure~\ref{fig:mass_size}), we find that our CDS galaxies have 0.7 dex higher surface mass densities than other RESOLVE dwarfs. This compactness relative to contemporaneous dwarfs is consistent with being possible low-$z$ blue nuggets.} \par

\cedit{To further explore the compactness of our sample with different density metrics, we carry out an analysis similar to that presented in \citet{2017ApJ...840...47B}. We compare three density metrics, as respectively defined in \citet{2013ApJ...765..104B}, \citet{2017ApJ...840...47B}, and \citet{2013ApJ...777...42K}:}

\begin{align}
\log(\Sigma_{1.5}) & = \log \left( \frac{\mstar}{r_{\rm e}^{1.5}} \right) \\
\log(\Sigma_{\rm e}) & = \log \left( \frac{\mstar}{2\pi r_{\rm e}^{2}} \right) \\
\mu_\Delta & = \mu_{90} + 1.7 \Delta_\mu
\end{align}

\cedit{The left column of Figure~\ref{fig:six_panel} shows SSFR plotted against these three metrics. Following \citet{2017ApJ...840...47B}, we also calculate and plot residual SFRs (relative to the Main Sequence) and residual density metrics (relative to quiescent galaxy density metric-mass relations). \citet{2017ApJ...840...47B} refer to these quantities as $\Delta$SFR, $\Delta \Sigma_{\rm e}$, etc. We describe the calculation of these quantities in detail in Appendix~\ref{app:delta} and plot them in the right column of Figure~\ref{fig:six_panel}.} \par

\cedit{Figure~\ref{fig:six_panel} demonstrates that our selection for $\mu_\Delta > 8.6$ corresponds to effective cutoffs in $\Sigma_{1.5}$ and $\Sigma_{\rm e}$ space. We note that $\Delta \mu_\Delta$ yields a wider distribution of points compared to the other density metrics, indicating greater sensitivity to density contrasts in galaxy cores and envelopes. Using $\Delta \Sigma_{\rm e}$ we can compare to the redshift evolution observed by \citet{2013ApJ...765..104B} and depicted in Figure 7 of  \citet{2017ApJ...840...47B}. We find that our $z\sim0$ CDS galaxies have shifted rightward to higher values of $\Delta \Sigma_{\rm e} + \log A$ roughly as expected based on these previous observations (blue arrow).}

\subsubsection{Frequency in the General Galaxy Population} \label{subsubsec:freq}

For the CDS sample in RESOLVE, we find a median stellar mass of  $\mstar = 10^{8.95} \, \msun$, albeit with an enforced stellar mass ceiling of  $M_{\rm *} = 10^{9.5} \ \msun$ based on theoretical expectations (\S\ref{subsubsec:sample}). Comparison with the mass-unrestricted blue E/S0 sample in Figure~\ref{fig:mass_size} suggests at most a small tail of compact starbursts persist to higher masses. Thus, we believe our sample contains nearly all of the \cedit{low-$z$} blue nuggets in RESOLVE, and their low stellar masses are consistent with the theoretical \cedit{expectations of blue nugget redshift evolution} \citep{2015MNRAS.450.2327Z}. \par

In their toy model for blue nugget formation and evolution, \citet{2014MNRAS.438.1870D} show that the fraction of galaxies in the blue nugget phase may be analytically estimated from the fraction of cold baryonic mass. From the equations laid out in their \S3.2, roughly $\sim$5$\%$ of $z\sim0$ galaxies with $M_{\rm halo} < 10^{11.5} \ \msun$ and $M_{\rm *}  \sim 10^{9.5} \ \msun$ should experience compaction events characteristic of blue nuggets. Our \cedit{50}-galaxy CDS sample constitutes $\sim$5.3$\%$ of the 965 dwarf galaxies (defined by $\mstar < 10^{9.5} \, \msun$) in the volume-limited RESOLVE survey, which is naturally dwarf-dominated and complete to $\mstar \sim 10^{8.8} \, \msun$ with slightly variable depth beyond that, as described in \S\ref{subsubsec:survey_def}. The 43 CDS galaxies with assigned $M_{\rm halo} < 10^{11.5} \ \msun$ constitute $\sim$6\% of the 728 dwarf galaxies in similarly low-mass halos, and of those 43, the 27 with gas-to-stellar mass ratios $>$1 constitute $\sim$4\% of the 728. Thus, we conclude that the frequency of CDS galaxies --- or alternatively, of the most blue-nugget-like galaxies among CDS galaxies --- is order-of-magnitude consistent with theoretical expectations for the frequency of blue nuggets at $z\sim0$, although with our small sample we cannot compare strictly analogous quantities. \par

\subsubsection{Specific Star Formation Rates} \label{subsubsec:sfr}

\begin{figure}
\includegraphics[width=\columnwidth]{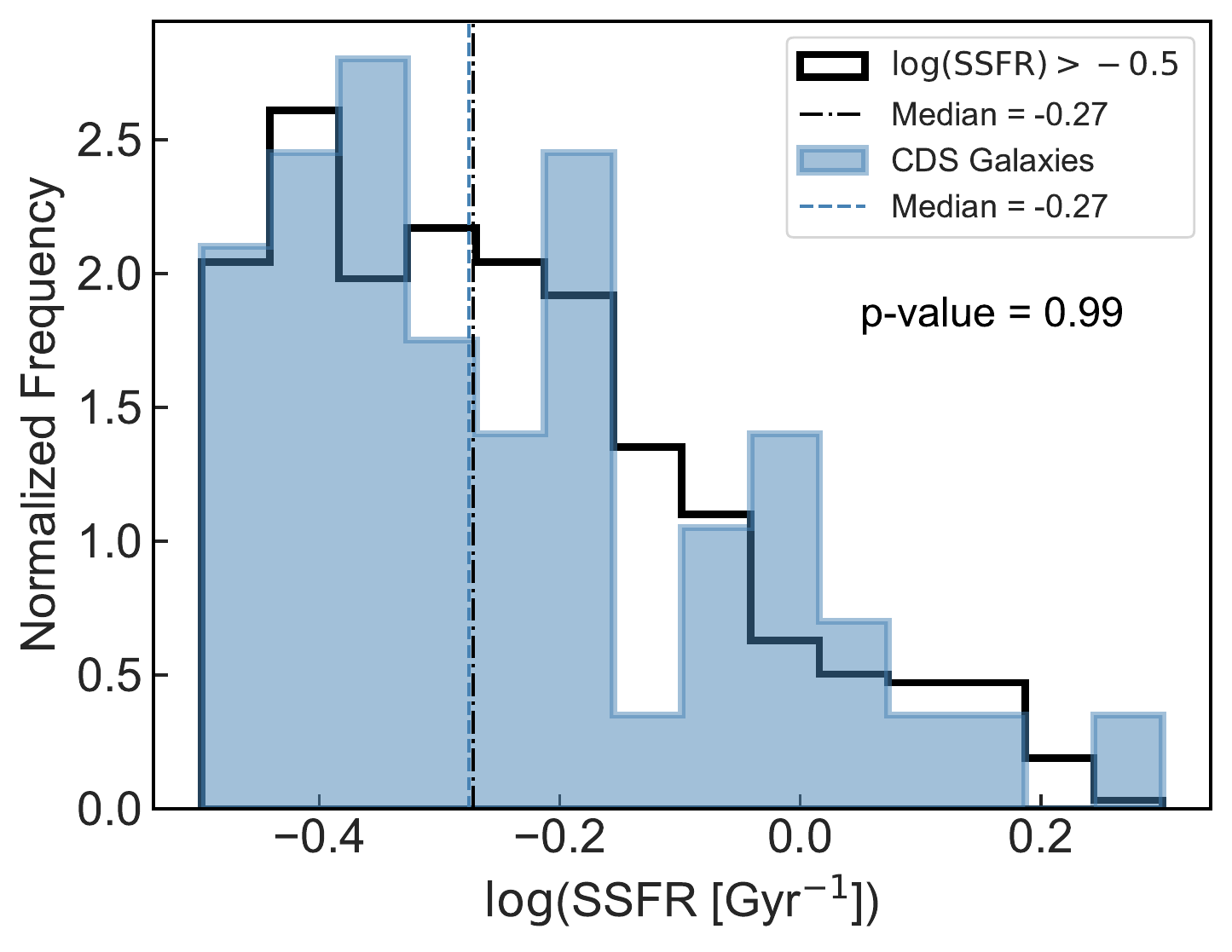}
\caption{Distribution of RESOLVE galaxy SSFRs revealing that non-compact and compact star-forming galaxies follow same distribution of specific star formation rates at $z\sim0$. In contrast, \citet{2013ApJ...765..104B} find that high-$z$ blue nuggets exhibit \textit{lower} SSFRs than high-$z$ non-compact star-forming galaxies (their Figure 4).}
\label{fig:ssfr_dist}
\end{figure}

Both theoretical \citep{2015MNRAS.450.2327Z} and observational \citep{2013ApJ...765..104B, 2013ApJ...776...63F} studies of blue nuggets have shown that the peaks of their SSFR distributions evolve downward over cosmic time, reflecting \cedit{star-formation} downsizing. Nonetheless, to select the most intensely star-forming galaxies in RESOLVE, we implemented the same SSFR floor as used previously in high-$z$ blue nugget studies: $\log(\textnormal{SSFR [{\rm Gyr}$^{-1}$}]) > -0.5$ \citep[e.g.,][]{2013ApJ...765..104B}. With this choice (which we will revisit below), the resulting RESOLVE CDS SSFR distribution overlaps the lower end of SSFRs observed for the Barro et al.\ high-$z$ sample: whereas the high-$z$ blue nugget sample has median $\log{({\rm SSFR}\ [{\rm Gyr}^{-1}])}\sim 0.1$, our low-$z$ CDS galaxies exhibit median $\log{({\rm SSFR}\ [{\rm Gyr}^{-1}])}\sim -0.27$, nearly 0.4 dex lower. The lower SSFRs of the CDS sample relative to the high-$z$ sample reflect the \cedit{star-formation} downsizing effect and suggest decreasing intensity in the fast-track growth phenomenon. \par

At the same time, the low-$z$ fast mode is somewhat more intense than might be expected: while \citet{2013ApJ...765..104B} find that high-$z$ blue nuggets exhibit, on average, SSFRs about $\sim 0.24$ dex lower than non-compact star-forming galaxies at high $z$, our low-$z$ CDS galaxies display SSFRs similar to non-compact star-forming galaxies in RESOLVE, as confirmed by a two-sample K-S test (Figure~\ref{fig:ssfr_dist}). This result may imply that while star formation has universally slowed toward the present-day, the compact dwarf starburst population has been differentially less affected. However, the very different mass distribution of the Barro et al.\ sample (selected to have $\mstar > 10^{10} \ \msun$) is an uncontrolled factor in this comparison; our CDS galaxies are in a mass regime of continual fueling, while some of the \citet{2013ApJ...765..104B} blue nuggets may have started to quench.  Analyzing a $z\sim0$ sample selected more similarly to the Barro et al.\ sample, \citet{2018ApJ...865...49W} compare the star formation rates of compact and non-compact star-forming galaxies \textit{above} $\mstar = 10^{9.5} \,\msun$ and see little effect of compactness on SSFR in the mass range between $10^{9.5} < \mstar < 10^{10.5}\, \msun$. The \citet{2018ApJ...865...49W} result appears more consistent with our low-$z$, lower-$\mstar$ result than with the \citet{2013ApJ...765..104B} high-$z$, similar-$\mstar$ result, but its relevance to the RESOLVE CDS sample is unclear: the \citet{2018ApJ...865...49W} selection above a stellar mass floor implies the low-mass end of their sample is mostly made up of satellites in massive halos (their Figure~6), whereas RESOLVE's approximately baryonic mass limited design implies the CDS sample is mostly made up of gas-dominated centrals in low-mass halos (Figure~\ref{fig:nn}). \par

Another uncontrolled factor in this comparison is our choice to impose the same SSFR restriction as \citet{2013ApJ...765..104B} in order to capture the most intense present-day starbursting galaxies. Using the top panel of Figure~4 from \citet{2013ApJ...770...57B} to extrapolate from the \citet{2013ApJ...765..104B} median log SSFR of $\sim0.1$, we might expect to see more than a dex in the downsizing of the SSFR distribution from $z\sim2$ to $z\sim0$. Although we only observe a \cedit{decrease} of about $\sim0.4$ dex, our median $z=0$ SSFR sits $\sim0.3$ dex above the limiting SSFR restriction, suggesting the possibility that we have excluded a more weakly starbursting tail of the $z\sim0$ CDS galaxy population. \par

In order to probe this possibility, we lower our SSFR restriction to $\log(\textnormal{SSFR [{\rm Gyr}$^{-1}$}]) > -1.5$, in line with the approximate SSFR expected for $M_{*}  \sim 10^9\, \msun$ as shown in \citealt{2013ApJ...770...57B} (their Figure~4). We find that this enlarged sample includes an additional 60 galaxies, for a total sample size of 110. The enlarged sample exhibits a mean log SSFR of $-0.56$, or about $0.28$ dex below the original CDS galaxy sample. We note that this sample may be contaminated by galaxies formed in scenarios other than compaction, for example group infall or tidal harassment. Indeed, we observe that 23 of the 60 additional galaxies ($\sim$38\%) in the enlarged sample occupy groups with $N>1$, and 18 of the 60 ($\sim$30\%) are satellites. Additionally, only 19 of the 60 ($\sim$31\%) have gas-to-stellar mass ratios exceeding unity, compared to 34 of the \cedit{50} ($\sim$68\%) of the original CDS galaxies (see \S\ref{subsubsec:environ}). Thus, the frequency of compaction-formed CDS galaxies may be higher than estimated in \S\ref{subsubsec:freq} by up to a factor of $\sim $1.8, but using the same high SSFR cutoff adopted at high $z$ by Barro et al.\ ensures a sample dominated by compaction-formed objects. \par

\subsection{Morphological and Kinematic Constraints on Possible Formation by Compaction} \label{subsec:result_formation}

By definition, blue nuggets form as the result of fast-track gas-compaction events, as described in \citet{2014MNRAS.438.1870D}. In the high-$z$ universe, when cold gas mass fractions were much higher than at the present-day, either colliding streams or gas-rich mergers could provide sufficient gas to drive compaction. However, it is unclear whether both of these blue nugget formation mechanisms are still prevalent in the low-$z$ universe. To probe the formation mechanisms of low-$z$ CDS galaxies, we examine three indicators of the two modes of compaction: prolateness, which is associated with formation in filamentary streams; \cedit{structural disturbances, such as double nuclei and tidal streams, which are associated with formation in galaxy mergers}; and abnormal kinematics, which may reveal evidence of either prolate structure (reflected in minor axis rotation) or recent mergers (reflected in multi-component velocity fields). \par

\subsubsection{Prolateness and Blue Nugget Morphology} \label{subsubsec:prolate}

As shown by \citet{2015MNRAS.453..408C} and \citet{2016MNRAS.458.4477T}, simulated blue nuggets display elongation aligned with the cosmic web structure in which they form, linking the colliding stream formation scenario to prolate morphology, which becomes oblate as the simulated nuggets quench and dynamically relax into red nuggets (\S\ref{intro}). Observations of $z > 1.5$ red nuggets do show fairly high median projected axial ratios (median $b/a \sim 0.67$ in \citealt{2011ApJ...730...38V} and $\sim$0.54 in \citealt{2013ApJ...765..104B}), but two-component fits suggest most are flattened disks \citep{2011ApJ...730...38V}. Also contrary to the prolate-to-oblate narrative for blue-to-red nugget evolution, \citet{2013ApJ...765..104B} find higher (rounder) median projected axial ratios for blue nuggets ($\sim$0.65), which they interpret as possible evidence of disk regrowth as red nuggets form after gas-rich mergers. Nonetheless, \citet{2014ApJ...792L...6V} do find observational evidence for the predicted prolate morphologies, using the statistics of projected axial ratios to infer 3D shape distributions in survey data ranging over redshifts $0<z<2.5$. The prolate galaxy fraction increases toward higher redshifts and lower masses, with the highest fractions found in the dwarf regime below $\mstar \sim 10^{9.5} \,\msun$, where a non-zero fraction persists down to $z\sim 0$.\par

\begin{figure}
\includegraphics[width=\columnwidth]{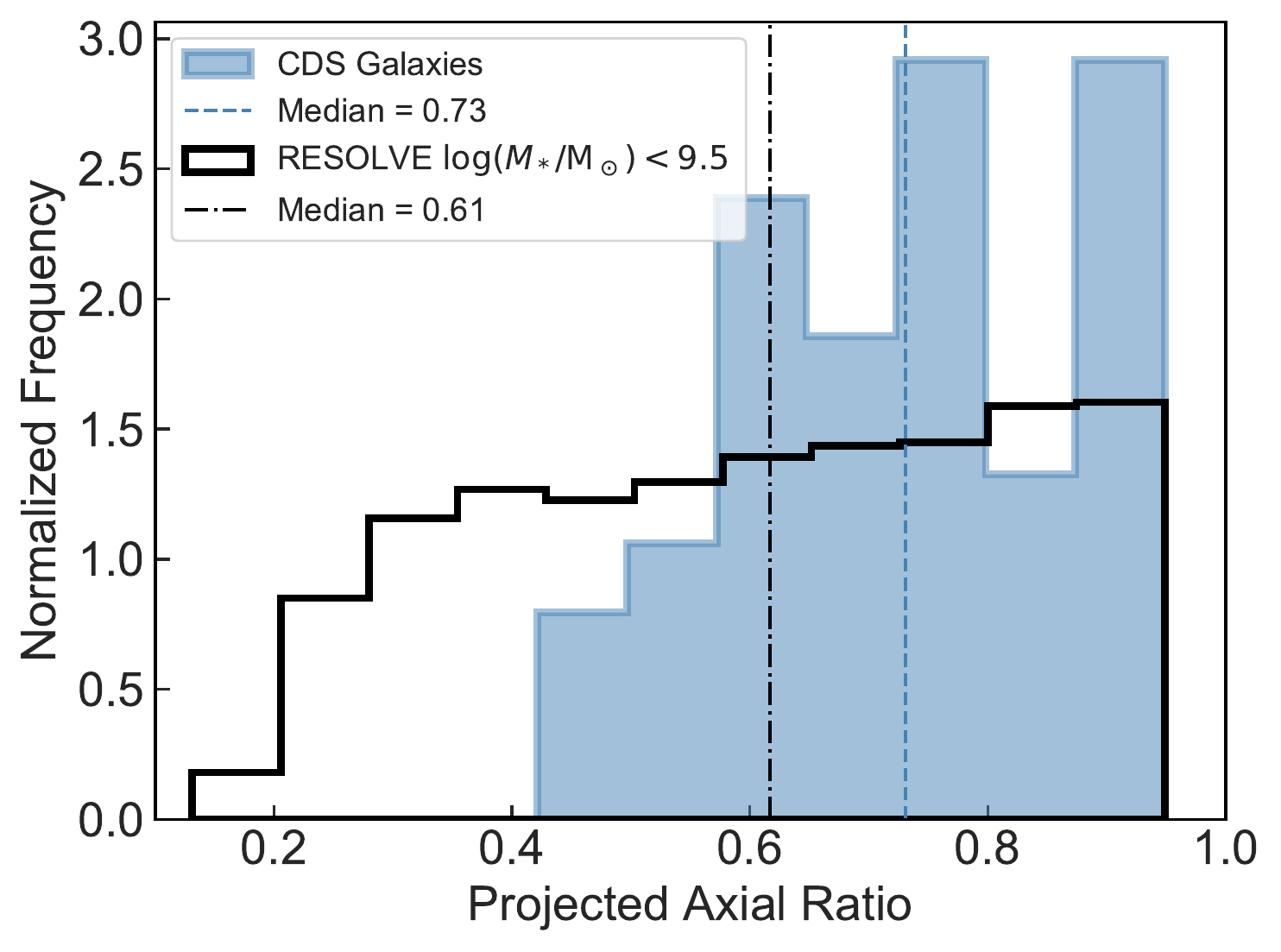}
\caption{Distribution of RESOLVE CDS projected axial ratios. The median axial ratio of $b/a \sim 0.73$ is slightly more spheroidal than the median axial ratio of $b/a \sim 0.65$ for high-$z$ blue nuggets \citep{2013ApJ...765..104B}. The steep drop-off in the distribution below $b/a \sim 0.5$ suggests CDS galaxies are less disky than the general dwarf population; few if any seem to be prolate.}
\label{fig:ba}
\end{figure}

The RESOLVE $z\sim0$ CDS sample is selected in this optimal $\mstar < 10^{9.5} \,\msun$ regime for prolate morphology. However, the median CDS projected axial ratio is $b/a \sim 0.73 \pm 0.03$ (adopting the standard error $\sigma_{\rm med} = 1.253 \frac{\sigma}{\sqrt[]{N}}$ where $\sigma$ is the standard deviation on $b/a$ and $N$ is sample size). For a comparison sample selected only on mass, the median axial ratio is $b/a \sim 0.61 \pm 0.01$, meaning our CDS galaxies appear rounder than the general population of dwarfs at $z\sim0$ (see Figure~\ref{fig:ba}), consistent with \citet{2013ApJ...765..104B}. \cedit{Our sample shows no correlation between projected axial ratio and central/satellite status.} \cedit{We do note that our comparison to \citet{2013ApJ...765..104B}} is somewhat biased, as our axial ratios are derived from ellipse fits to SDSS imaging with typical seeing $\sim$1$\farcs6$ \citep{2015ApJ...810..166E}, and the seeing necessarily affects compact galaxies more than the general population. The blurring is significant but not overwhelming: for the CDS sample the median $90\%$ light major-axis diameter is $\sim$16\arcsec, about 10 times the typical SDSS seeing of $\sim$1 -- $2$\arcsec, while the smallest $90\%$ light major-axis diameter is half that, $\sim$8\arcsec. These results suggest that prolate morphology is most likely rare to nonexistent in our low-$z$ CDS sample, which is consistent with the small prolate fraction inferred by \citet{2014ApJ...792L...6V} at $z\sim0$. \par

\subsubsection{\cedit{Merger Classification with DECaLS}} 
\label{subsubsec:doublenuc}

\begin{figure}
\includegraphics[width=\columnwidth]{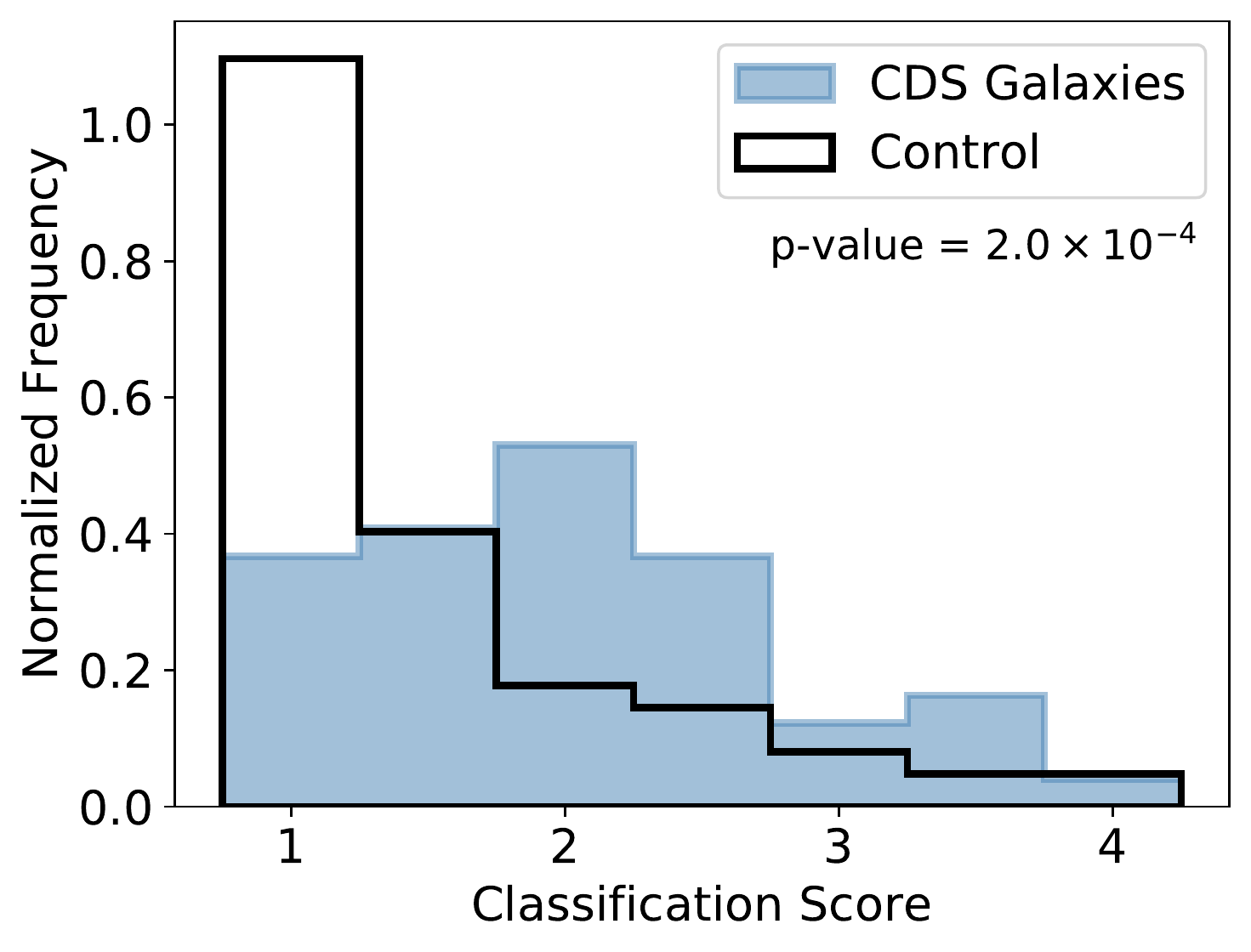}
\caption{\cedit{Distributions of merger classification scores for the CDS and control galaxy samples. A two-sample K-S test returns a p-value of $\sim$$2\times10^{-4}$, confirming that these data are unlikely to come from the same parent distribution. The median score is 1.0 for the control galaxies compared to 2.0 for the CDS galaxies. Moreover, there are roughly twice as many "likely" mergers, i.e., galaxies with average score $\geq$3.5, in the CDS sample than in the control sample ($\sim$$10\%$ versus $\sim$$5\%$, respectively).}}
\label{fig:classification}
\end{figure}

To constrain the \cedit{occurrence rate} of \cedit{recent mergers} in the RESOLVE low-$z$ CDS sample, we performed by-eye classification using DECaLS DR7 images and model residuals as described in \S\ref{subsec:decals_clas}. \cedit{The CDS galaxy sample has a median score of 2.0 (representing "ambiguous non-merger") and $\sim$$36.0^{+8.0}_{-7.3}\%$ of CDS galaxies are "possible" mergers (defined by average score $\geq$ 2.5). For a control sample consisting of \cedit{125} RESOLVE galaxies matching the mass and morphology selection of our CDS galaxies, but having less extreme SSFRs, we find a median score of 1.0 (representing "definite non-merger") and $\sim$$15.2^{+4.0}_{-3.2}\%$ of the control galaxies are "possible" mergers. Here two-sided one-sigma confidence intervals are numerically calculated using the equations set forth in \citet{1986ApJ...303..336G} for binomial upper and lower limits. "Likely" mergers, defined by average score $\geq$3.5, constitute $\sim$$10.0^{+6.2}_{-4.2}\%$ of the CDS sample and $\sim$$4.8^{+2.8}_{-1.9}\%$ of the control sample. Figure \ref{fig:classification} shows the distribution of scores for both populations. A two-sample K-S test confirms that the data are unlikely to be drawn from the same parent distribution (p-value $\sim 2\times10^{-4}$). The independent DECaLS DR8 classifications mentioned in \S\ref{subsec:decals_clas} yield medians identical to those from our DR7 classifications for both the CDS and control samples separately. Statistics computed from the DR8 classifications reveal similar frequencies of "possible" and "likely" mergers in the CDS sample ($\sim$$36\%$ and $\sim$$14\%$, respectively), but slightly lower frequencies in the control sample ($\sim$$8\%$ and $\sim$$2.4\%$, respectively).} \par

\cedit{The $\sim$\,$2\times$ higher frequency of visually-obvious merger evidence seen in CDS galaxies compared to the control sample is consistent with the merger origin proposed for blue nugget formation. A significant fraction of simulated high-$z$ nuggets are also characterized as forming via (mostly minor) mergers, while others are characterized as forming via colliding streams \citep{2015MNRAS.450.2327Z}.} To our knowledge, previous observational and theoretical works have not addressed the possibility of colliding gas streams producing \cedit{features typically associated with mergers, such as tidal streams or} double nuclei. \citet{2016MNRAS.458.4477T}, \citet{2015MNRAS.453..408C}, \citet{2016MNRAS.457.2790T}, and \citet{2015MNRAS.450.2327Z} do not discuss \cedit{such features} in their simulations. Likewise, \citet{2014ApJ...780....1W} and \citet{2013ApJ...765..104B} do not discuss the appearance of \cedit{such features} in their observations. \cedit{However, \citet{2018ApJ...857..144H} do note a higher rate of tidal features in gas-rich galaxies than can be explained by known companion interactions.} \par

\subsubsection{Kinematic Structure} \label{subsubsec:kin}

We obtained follow-up 3D spectroscopic observations for seven CDS galaxies with the GMOS IFU, SAM FP, and/or SIFS (described in \S\ref{subsec:3Dspec}). As seen in Figure~\ref{fig:3dspec1} and detailed in the individual galaxy notes in Appendix B, the internal structure and dynamics of the observed CDS galaxies can be quite varied and intricate.  The four SIFS observations are somewhat shallow, but reveal two cases of double nuclei seen in both continuum and H$\alpha$ maps (rs0463 and rs1259), plus one more double nucleus seen just in H$\alpha$ (rf0363). The higher SNR GMOS and SAM FP data reveal a fourth double nucleus (rf0250) corresponding to a distinct inner kinematic component in the velocity field. Two of the four  galaxies with IFU-detected double nuclei (rf0363 and rf0250) are also classified as \cedit{"likely" mergers} based on DECaLS DR7 residuals, \cedit{with visually discernible double nuclei} (\S\ref{subsubsec:doublenuc}). The lack of \cedit{visually obvious} double-nucleus signatures in the DECaLS residuals for the other two galaxies, rs0463 and rs1259, may simply result from inadequate seeing; these nuclei are separated by about an arcsecond in the SIFS maps (see Figure~\ref{fig:3dspec1}), comparable to the typical seeing of DECaLS. The SIFS data \cedit{for two of these four double-nucleus galaxies (rf0363 and rs1259)} suggest single-component minor-axis rotation, plausibly consistent with prolate structure, which is in both cases oddly perpendicular to the alignment axis of the apparent double nucleus. In addition, GMOS/SAM FP data for both rf0250 and a fifth galaxy (rf0266) show signatures of two distinct minor and major axis rotation components at different radii. The SAM FP velocity field for a sixth galaxy (rs0804) reveals a previously unknown merging smaller dwarf companion. Thus, in all, at least 6 of the 7 galaxies show peculiarities in their 3D spectroscopy suggestive of recent or ongoing mergers. Four galaxies have minor axis rotation, which might by itself indicate prolateness due to formation in colliding streams (\S\ref{subsubsec:prolate}). However, in two cases the minor axis rotation coexists with major axis rotation and in the other two it coexists with apparent double nuclei aligned with the major axis. We conclude that although a merger formation scenario is favored, formation in colliding streams is not ruled out in a few cases. \par

\begin{figure*}
\captionsetup[subfigure]{labelformat=empty}
  \subfloat[GMOS IFU: rf0266 velocity field]{\includegraphics[clip,width=0.7\columnwidth]{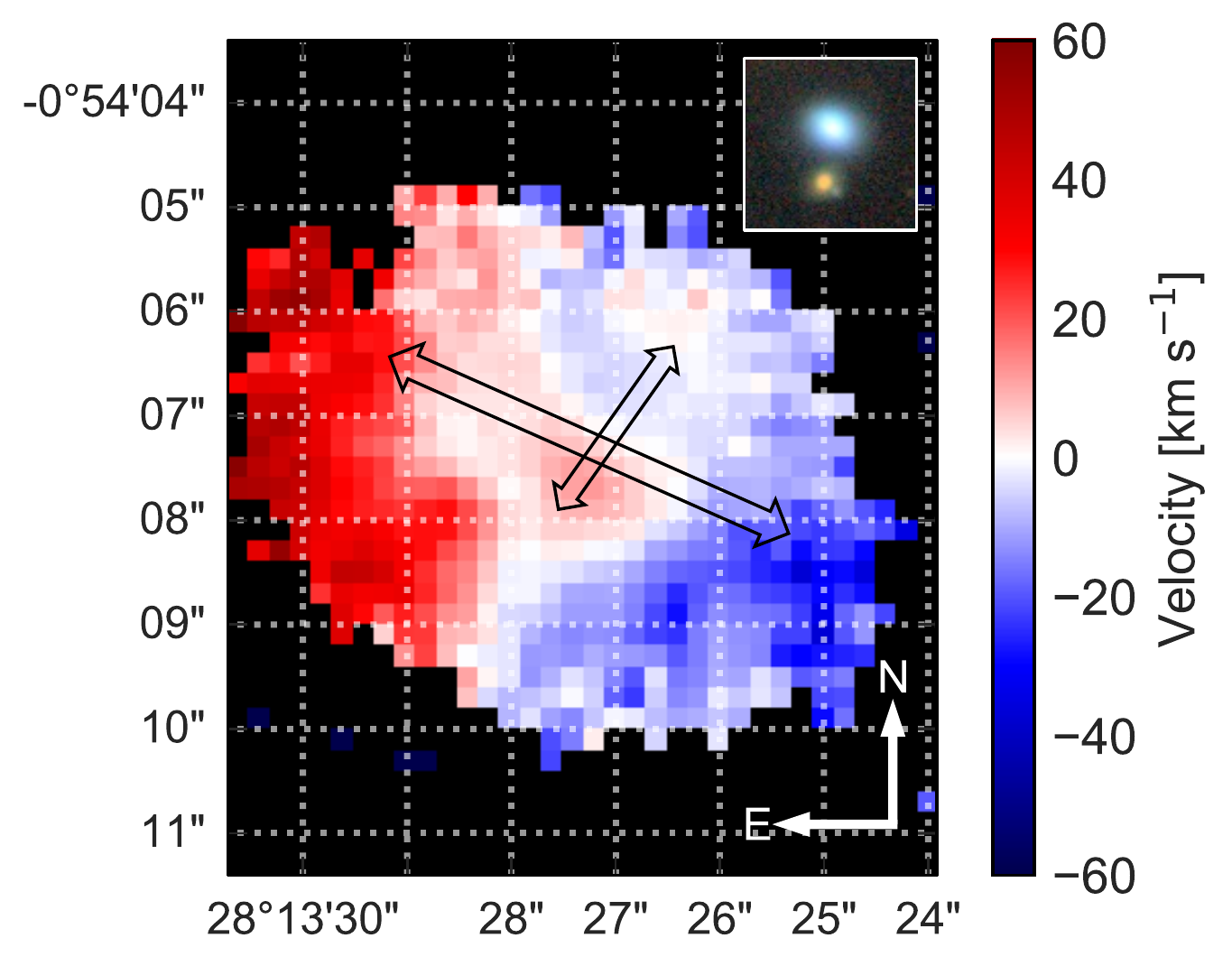}}
  \subfloat[GMOS IFU: rf0266 continuum]{ \includegraphics[clip,width=0.7\columnwidth]{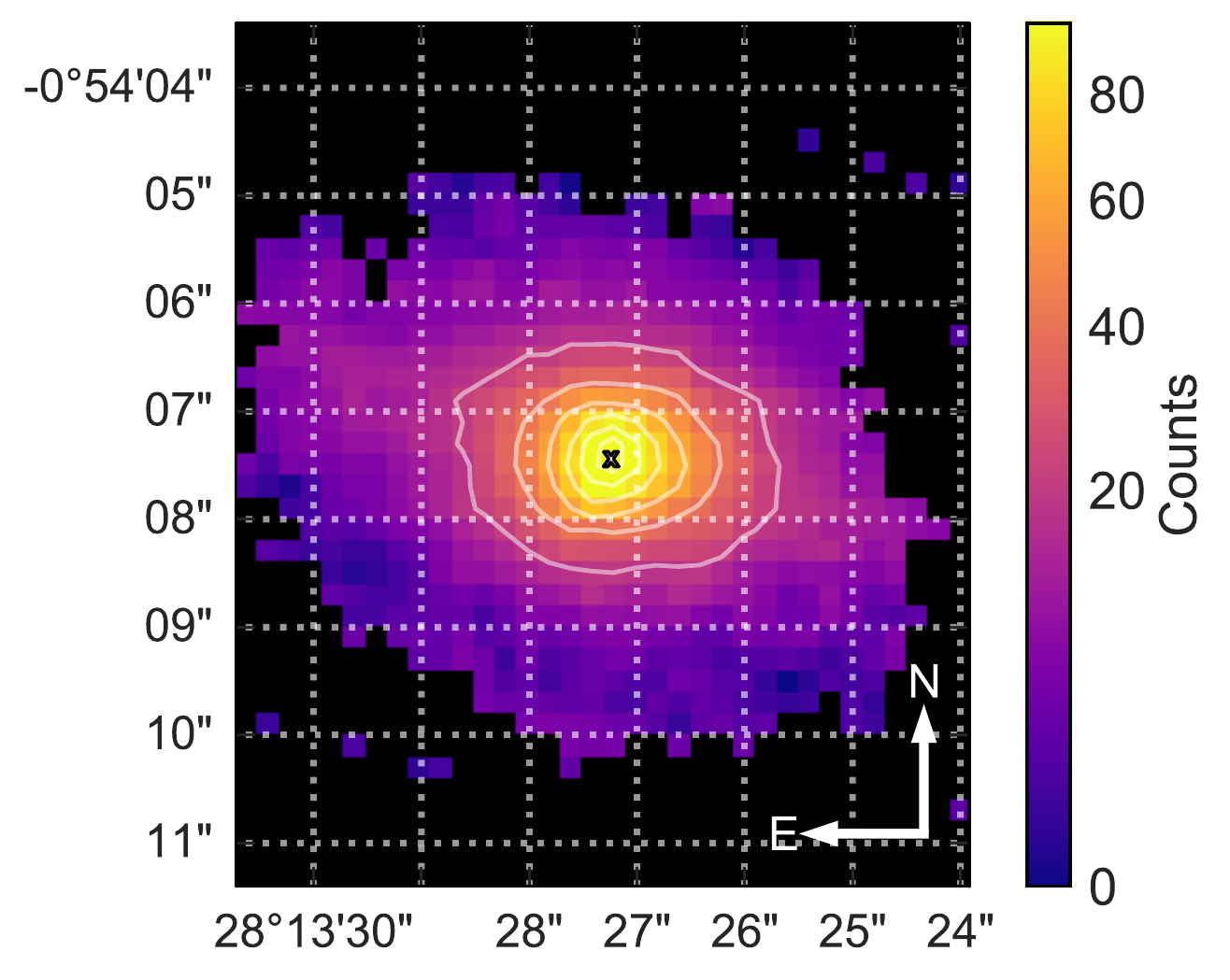}}
  \subfloat[GMOS IFU: rf0266 \halpha flux]{\includegraphics[clip,width=0.7\columnwidth]{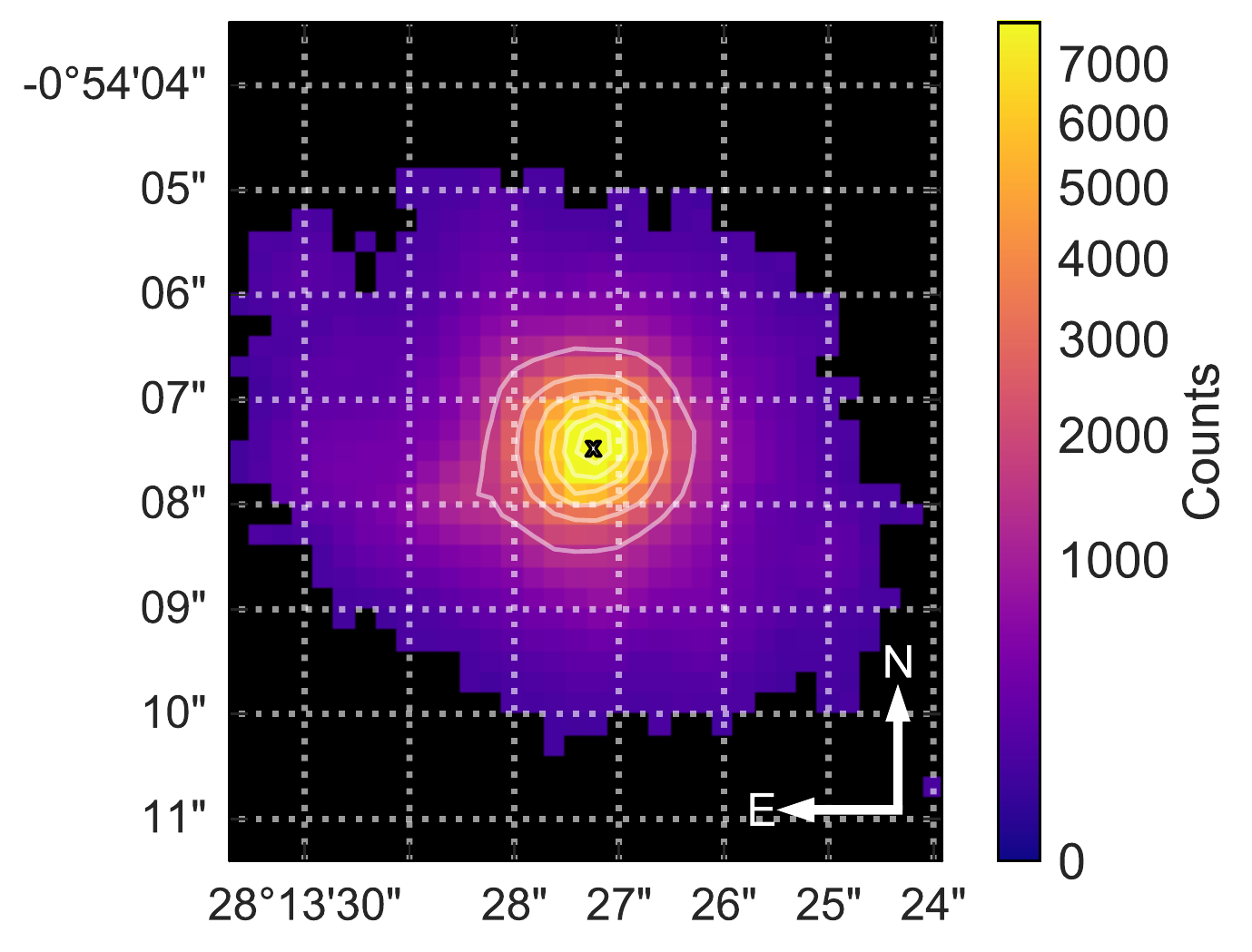}
}\\
   \subfloat[SAM FP: rf0266 velocity field]{\includegraphics[clip,width=0.7\columnwidth]{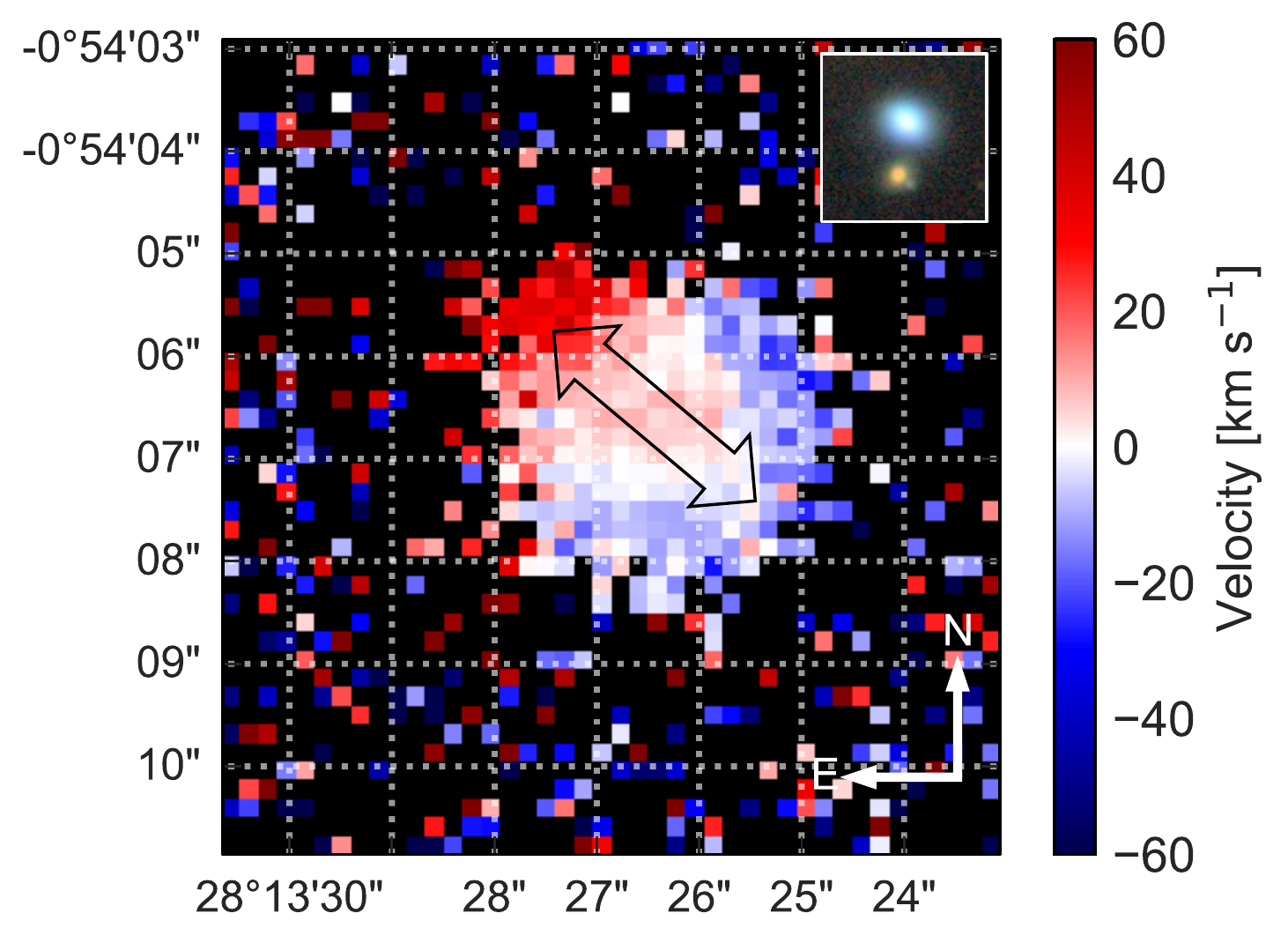}}
   \subfloat[SAM FP: rf0266 continuum]{\includegraphics[clip,width=0.7\columnwidth]{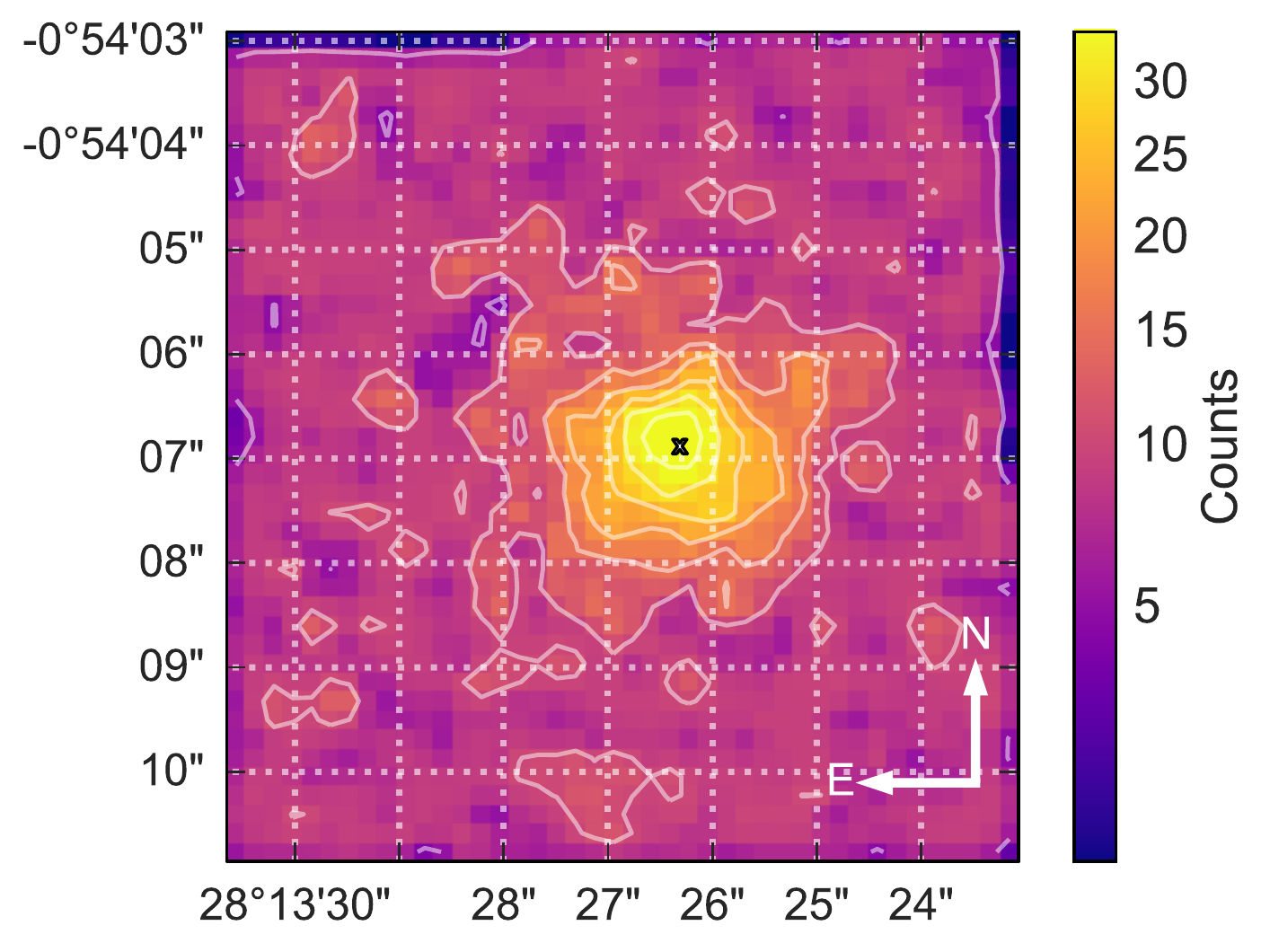}}
   \subfloat[SAM FP: rf0266 \halpha flux]{\includegraphics[clip,width=0.7\columnwidth]{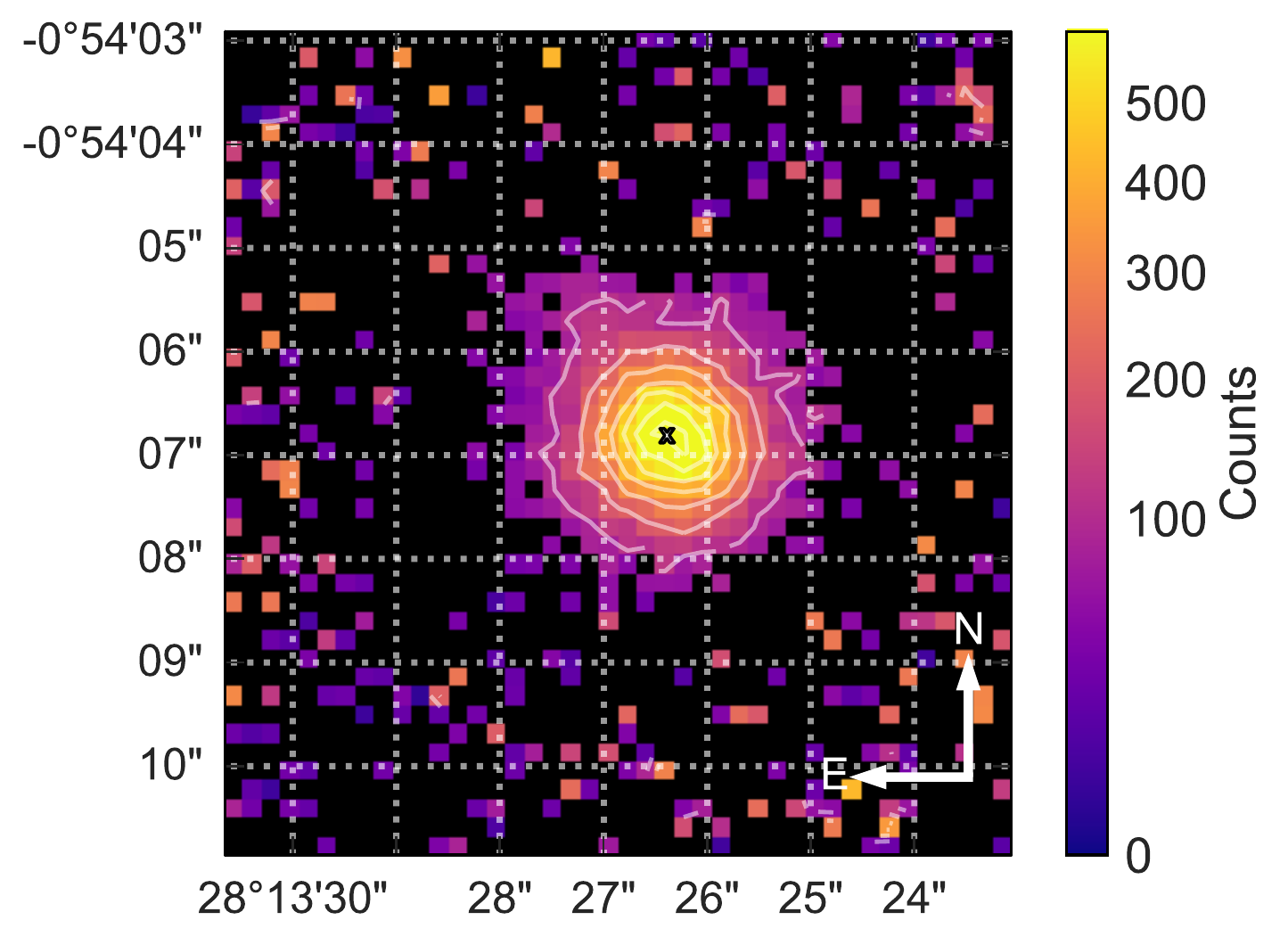}
}\\
   \subfloat[GMOS IFU: rf0250 velocity field]{\includegraphics[clip,width=0.7\columnwidth]{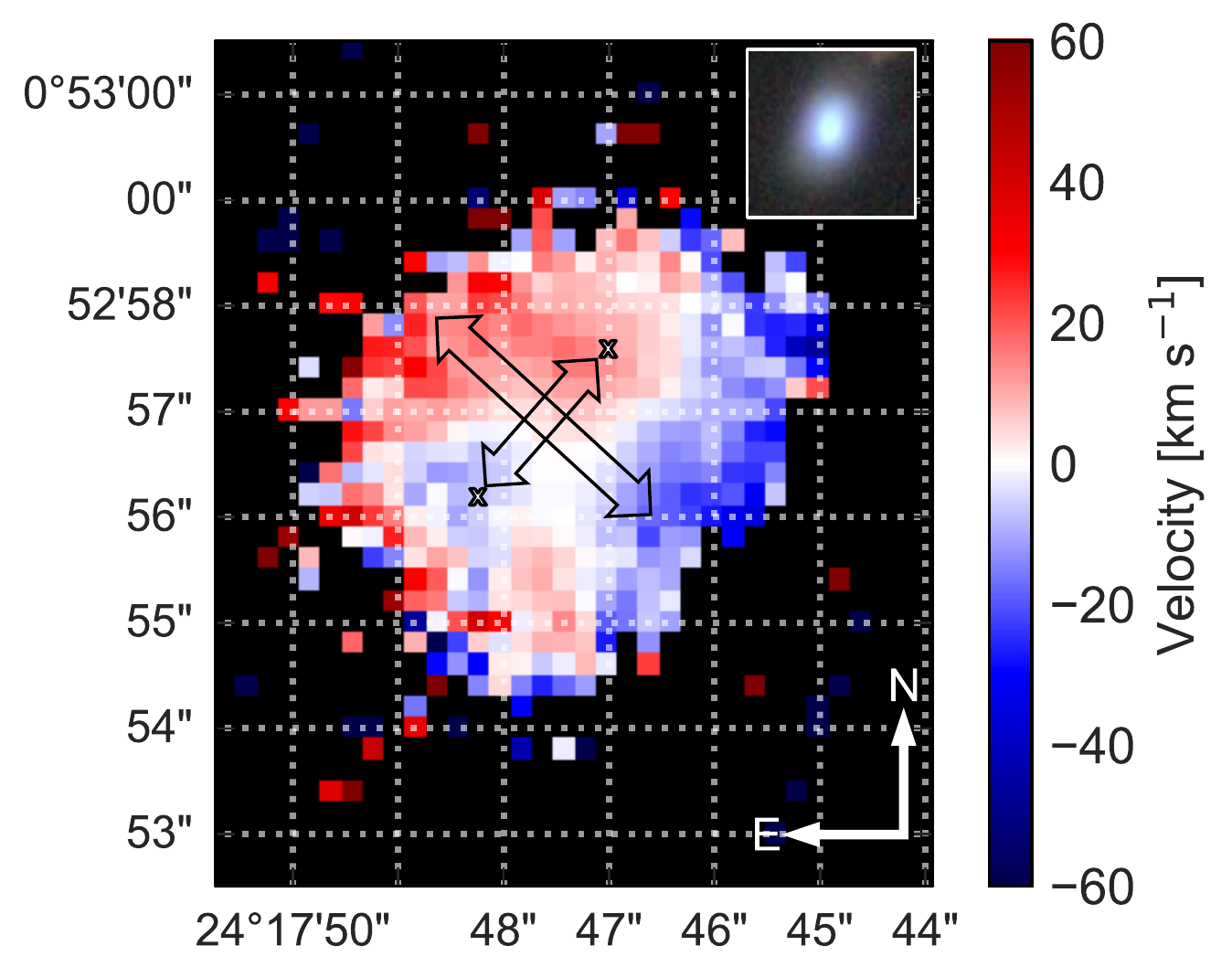}}
   \subfloat[GMOS IFU: rf0250 continuum]{\includegraphics[clip,width=0.7\columnwidth]{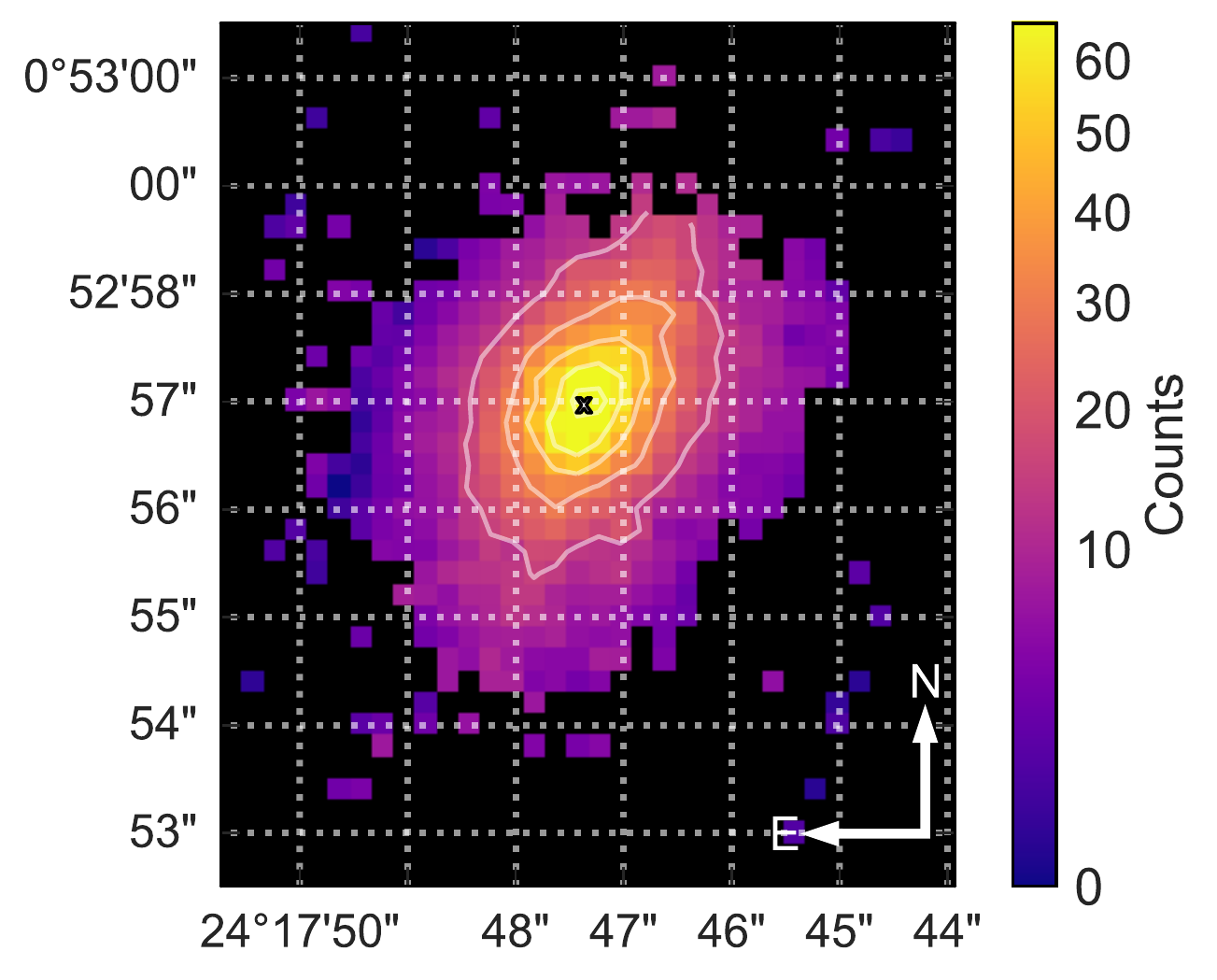}}
   \subfloat[GMOS IFU \halpha flux]{\includegraphics[clip,width=0.7\columnwidth]{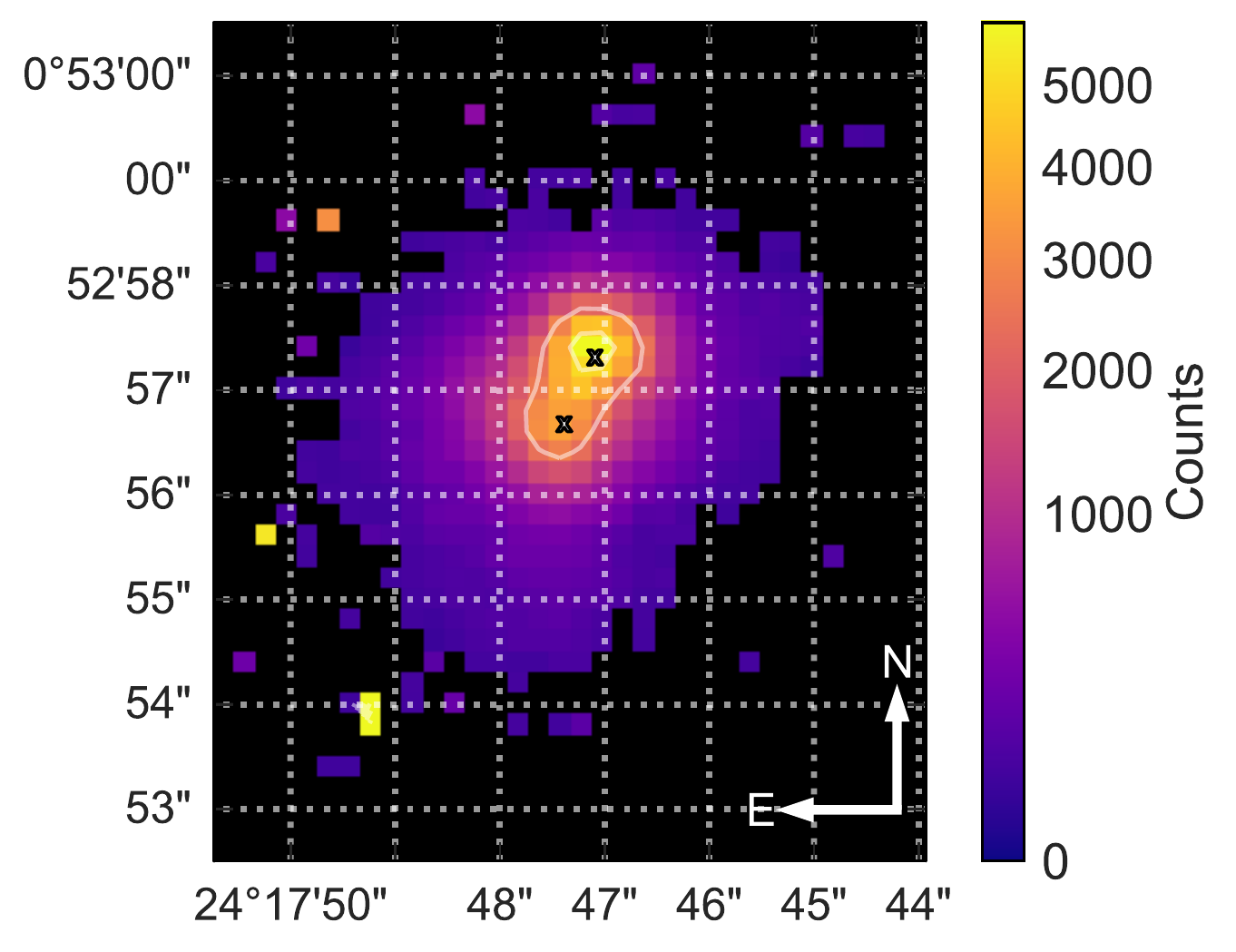}
}\\
   \subfloat[SAM FP: rf0250 velocity field]{\includegraphics[clip,width=0.7\columnwidth]{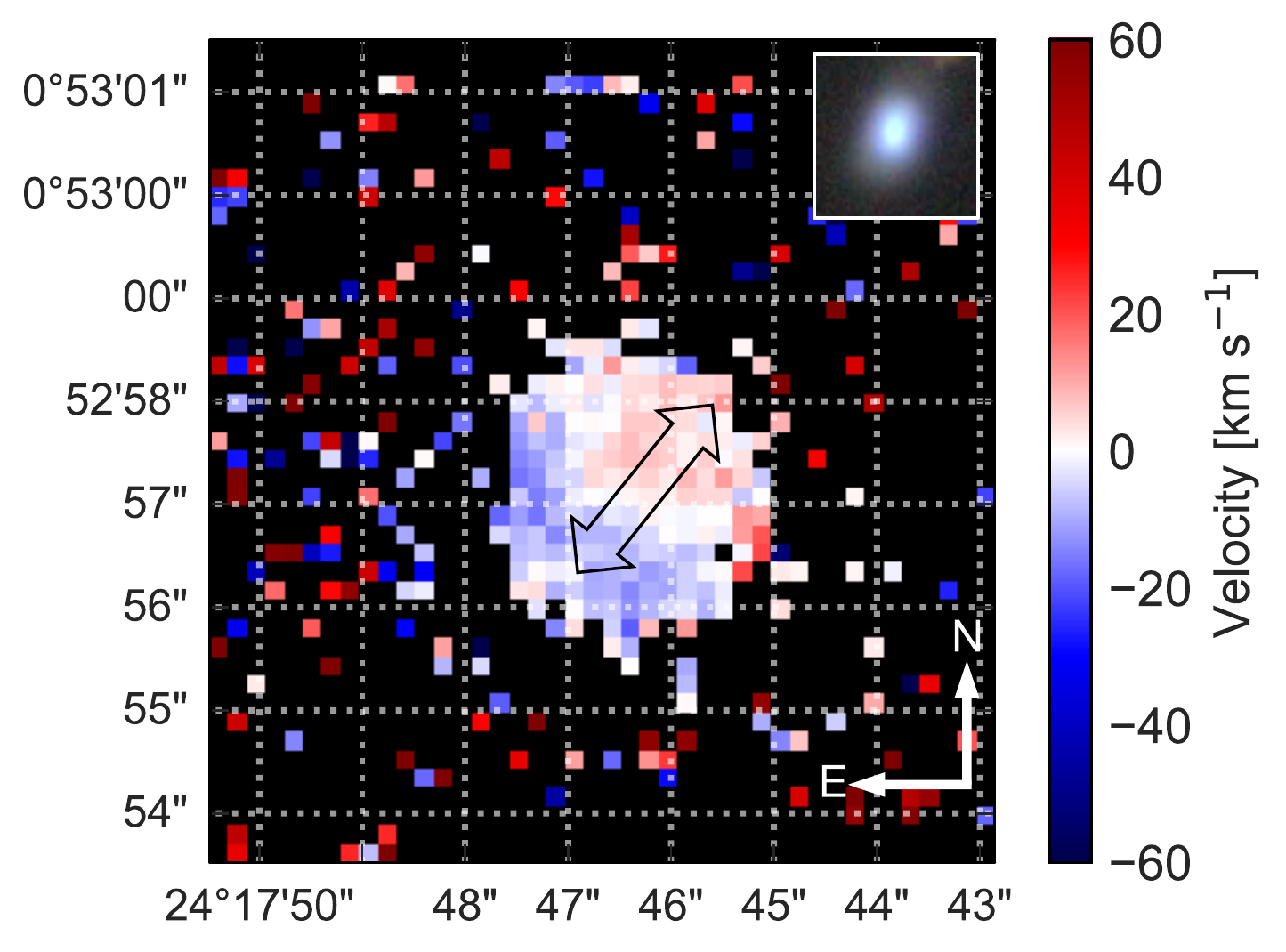}}
   \subfloat[SAM FP: rf0250 continuum]{\includegraphics[clip,width=0.7\columnwidth]{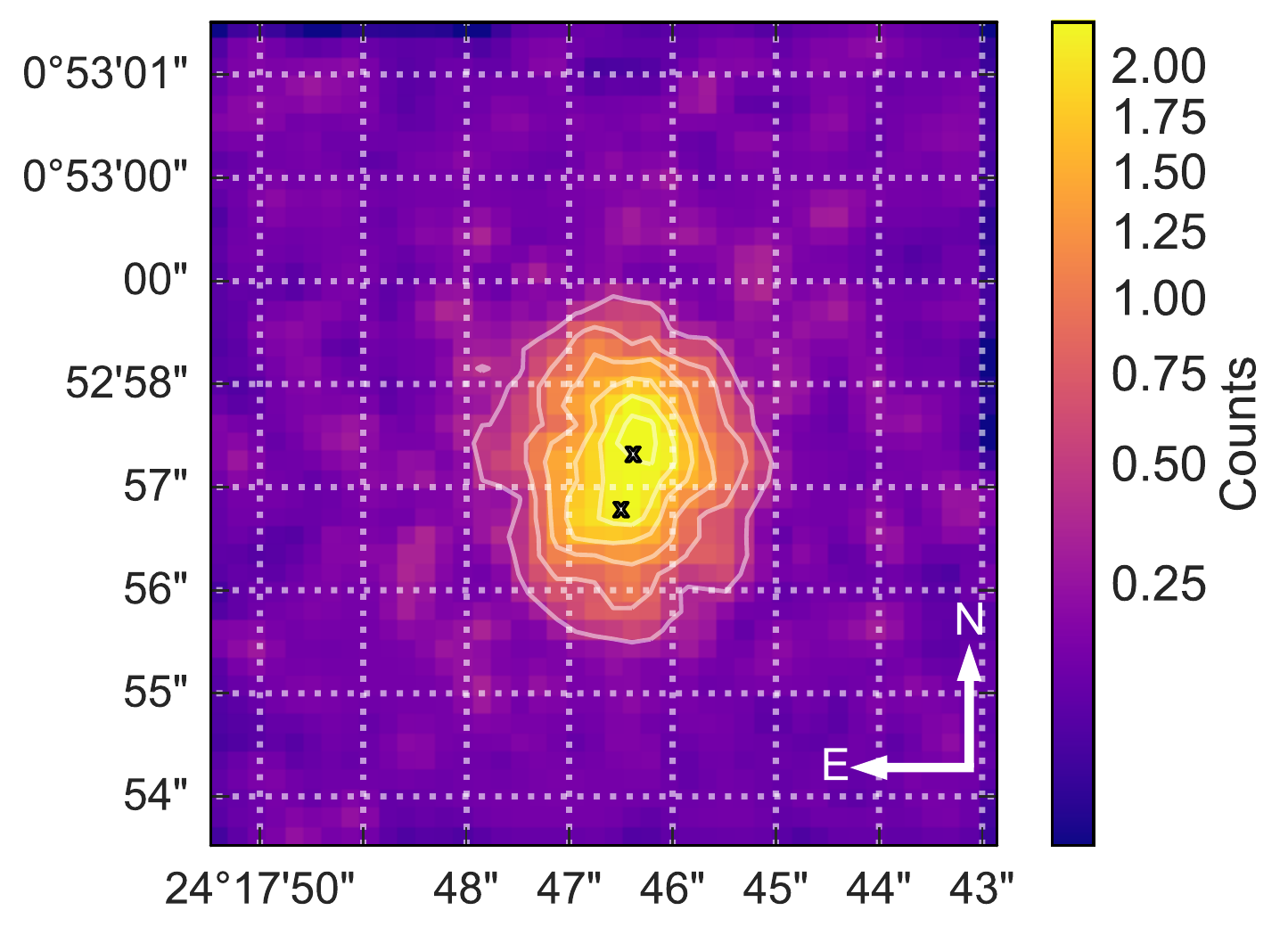}}
   \subfloat[SAM FP: rf0250 \halpha flux]{\includegraphics[clip,width=0.7\columnwidth]{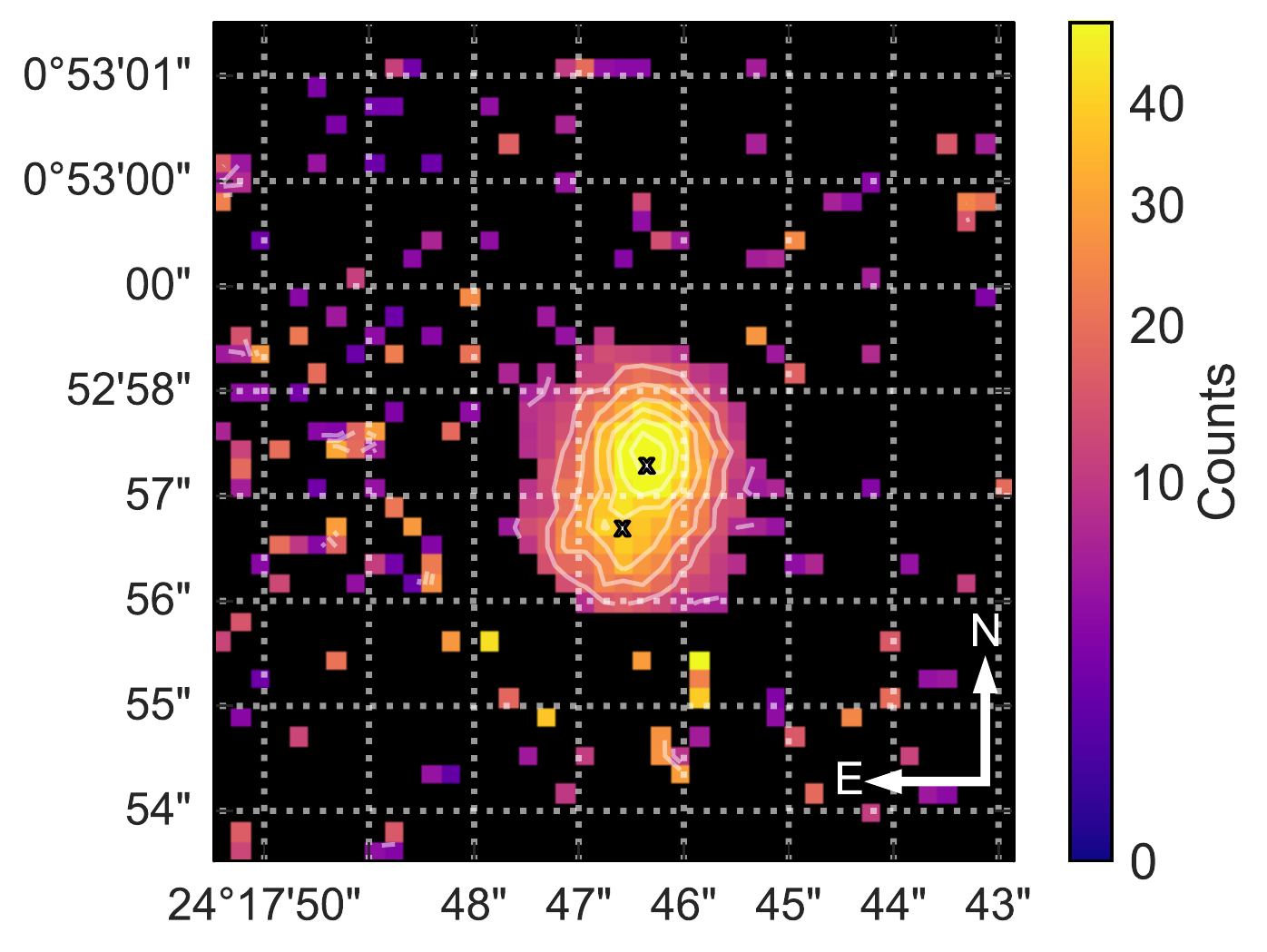}
}\\
\caption{Velocity fields (left column), continuum maps (middle column), and \halpha flux maps (right column) derived from 3D spectroscopic observations. Nuclei are denoted with an ``x" and rotation patterns with arrows (see \S\ref{subsec:vfield}), where appropriate. \cedit{For visual reference, photometry from DECaLS is inset in each velocity field image. The red object in the image for rf0266 is a background object at $z\sim 0.25$.}}
\label{fig:3dspec1}
\end{figure*}

\setcounter{figure}{12}
\begin{figure*}
\captionsetup[subfigure]{labelformat=empty}
  \subfloat[SAM FP: rs0804 velocity field]{\includegraphics[clip,width=0.7\columnwidth]{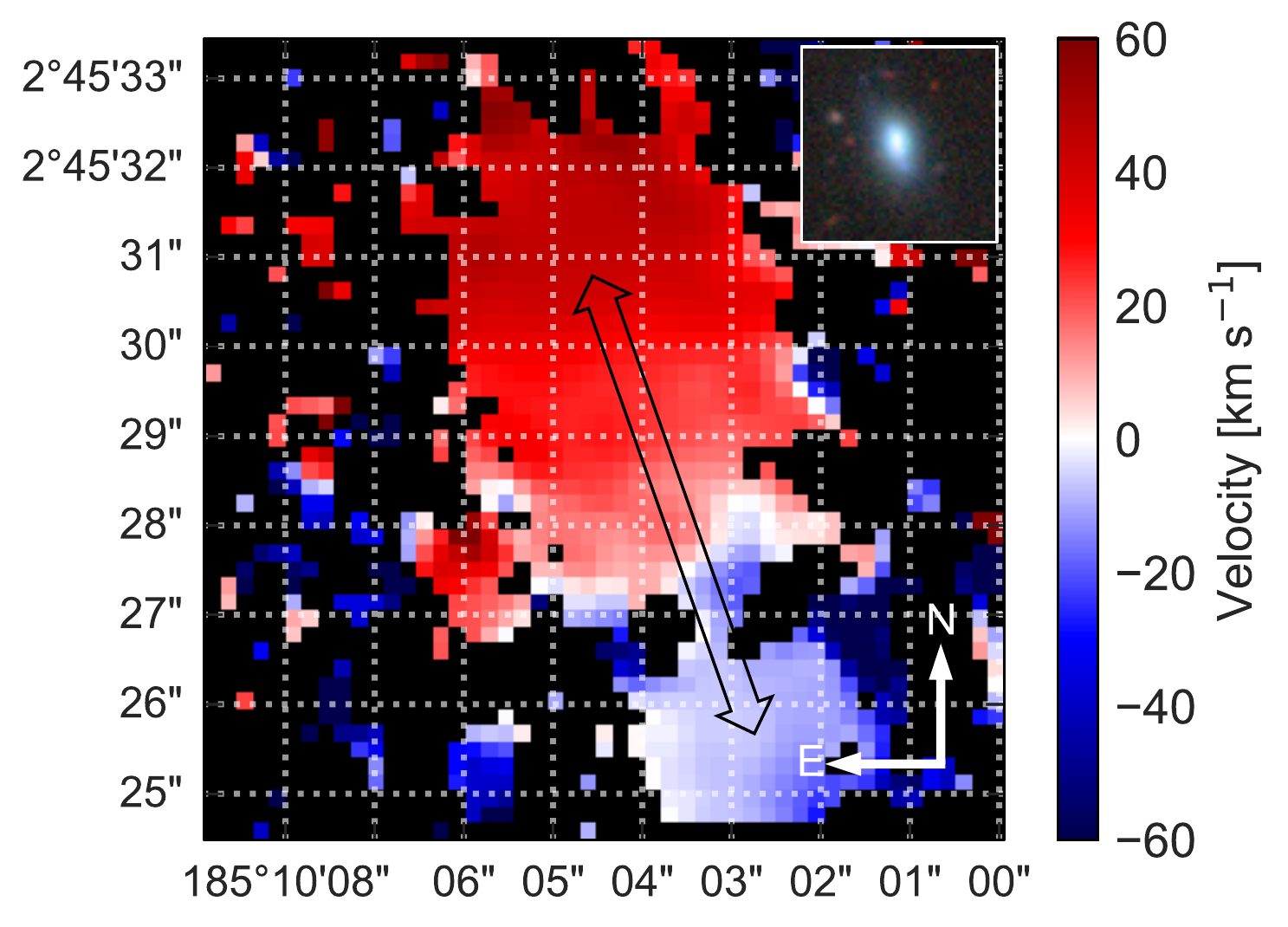}}
  \subfloat[SAM FP: rs0804 continuum]{ \includegraphics[clip,width=0.7\columnwidth]{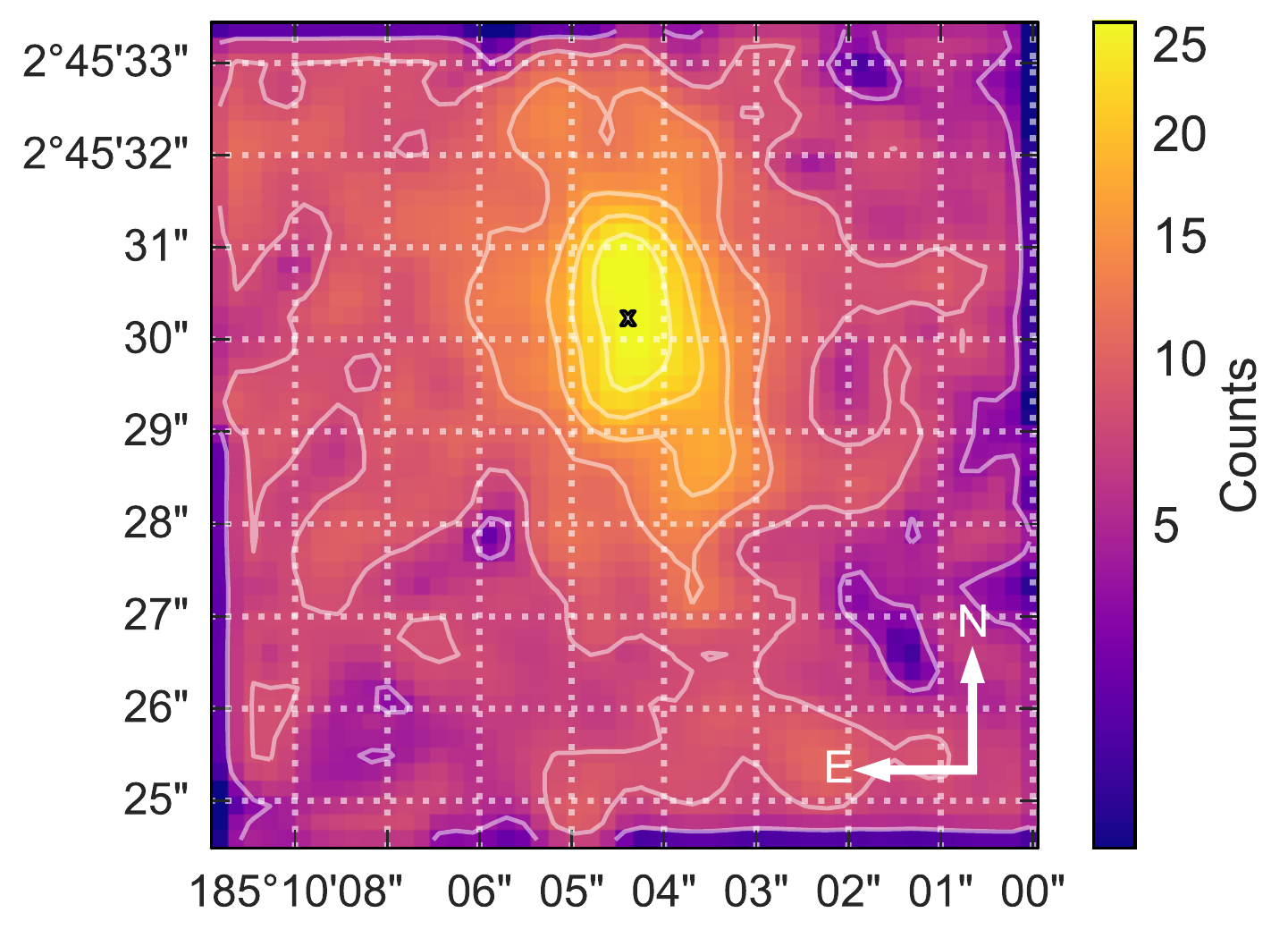}}
  \subfloat[SAM FP: rs0804 \halpha flux]{\includegraphics[clip,width=0.7\columnwidth]{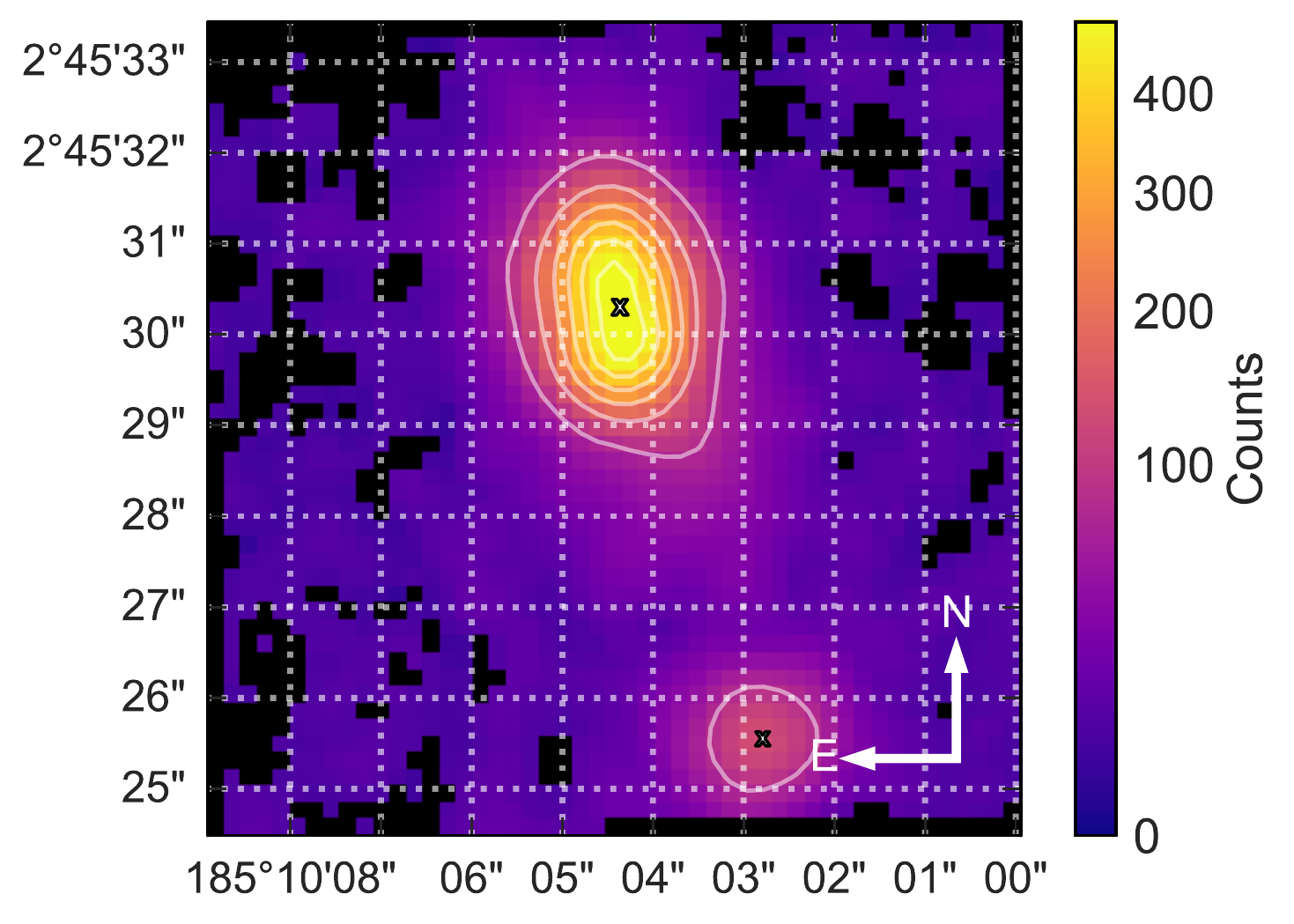}
}\\
   \subfloat[SAM FP: rs0804 main velocity field]{\includegraphics[clip,width=0.7\columnwidth]{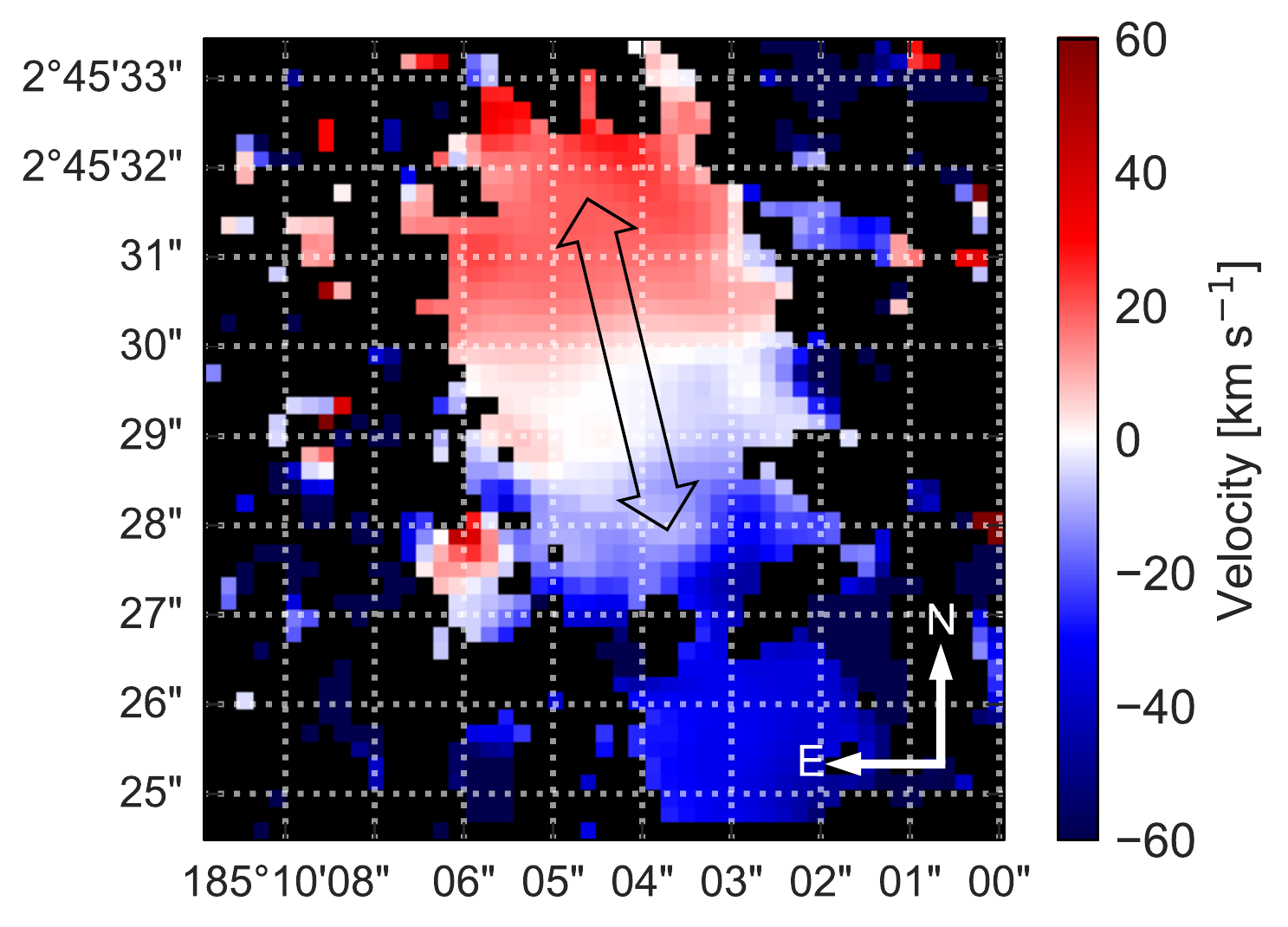}}
   \subfloat[SAM FP: rs0804 companion velocity field]{\includegraphics[clip,width=0.7\columnwidth]{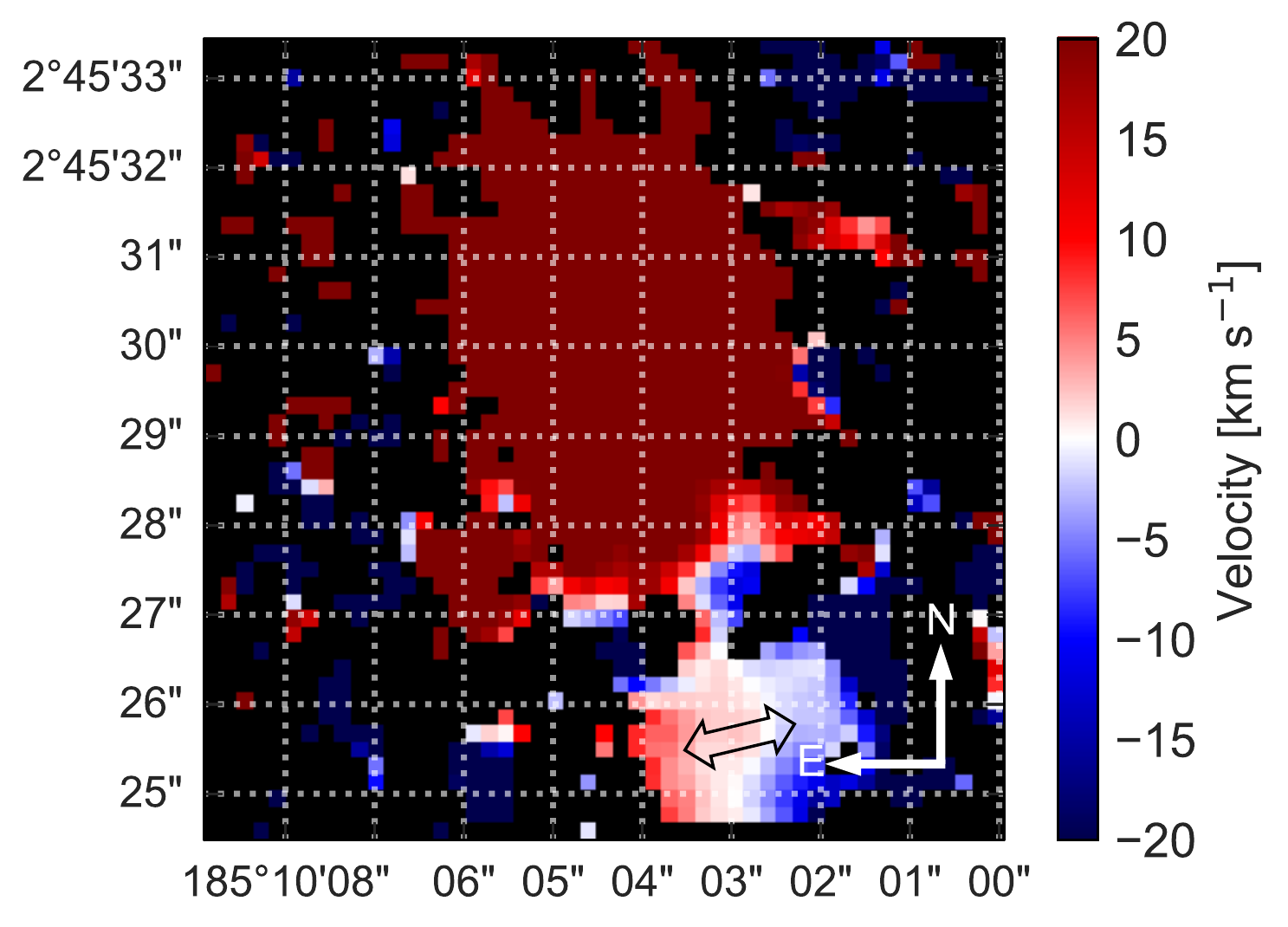}
}\\
   \subfloat[SIFS: rs1103 velocity field]{\includegraphics[clip,width=0.7\columnwidth]{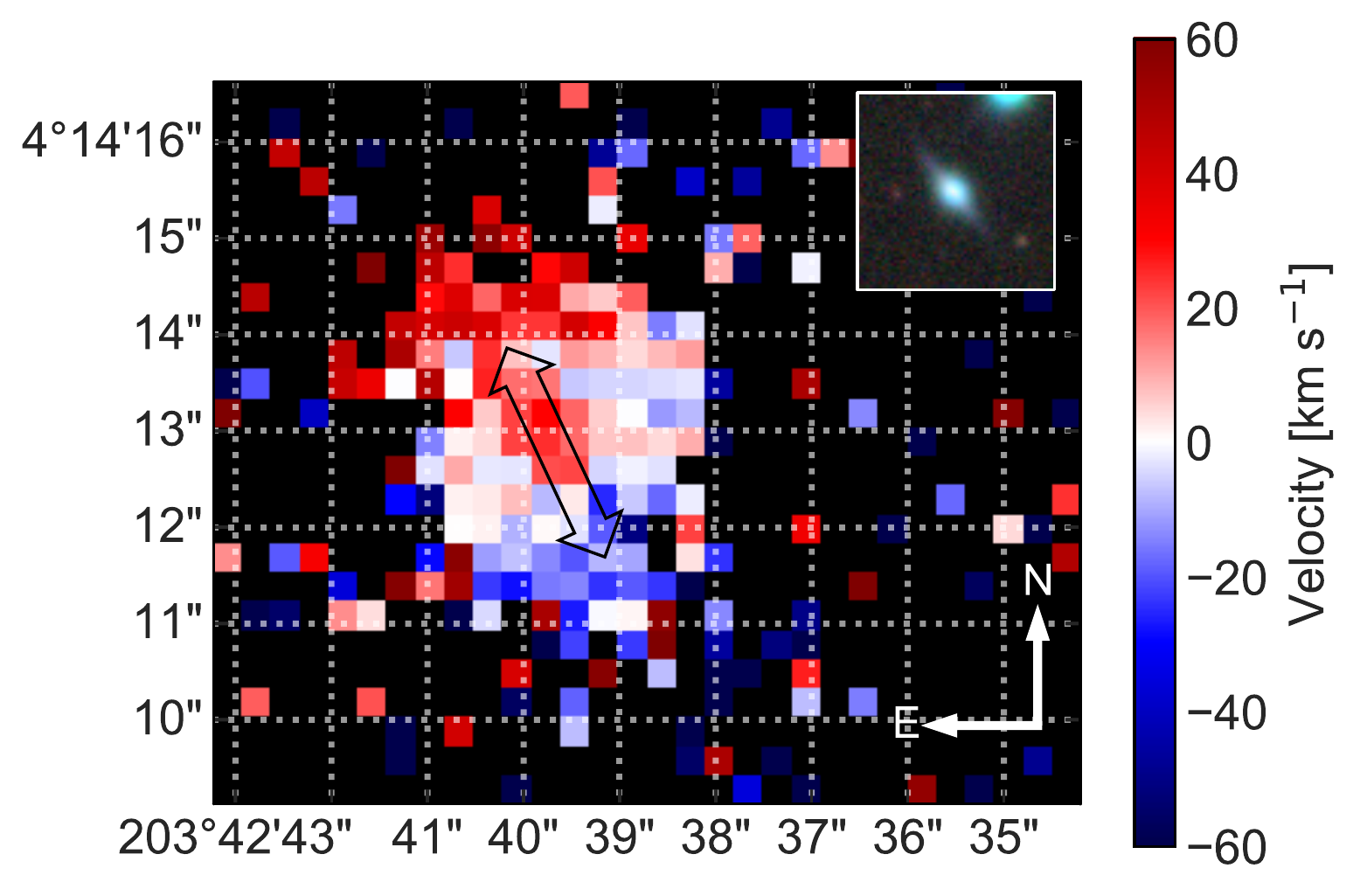}}
   \subfloat[SIFS: rs1103 \halpha flux]{\includegraphics[clip,width=0.7\columnwidth]{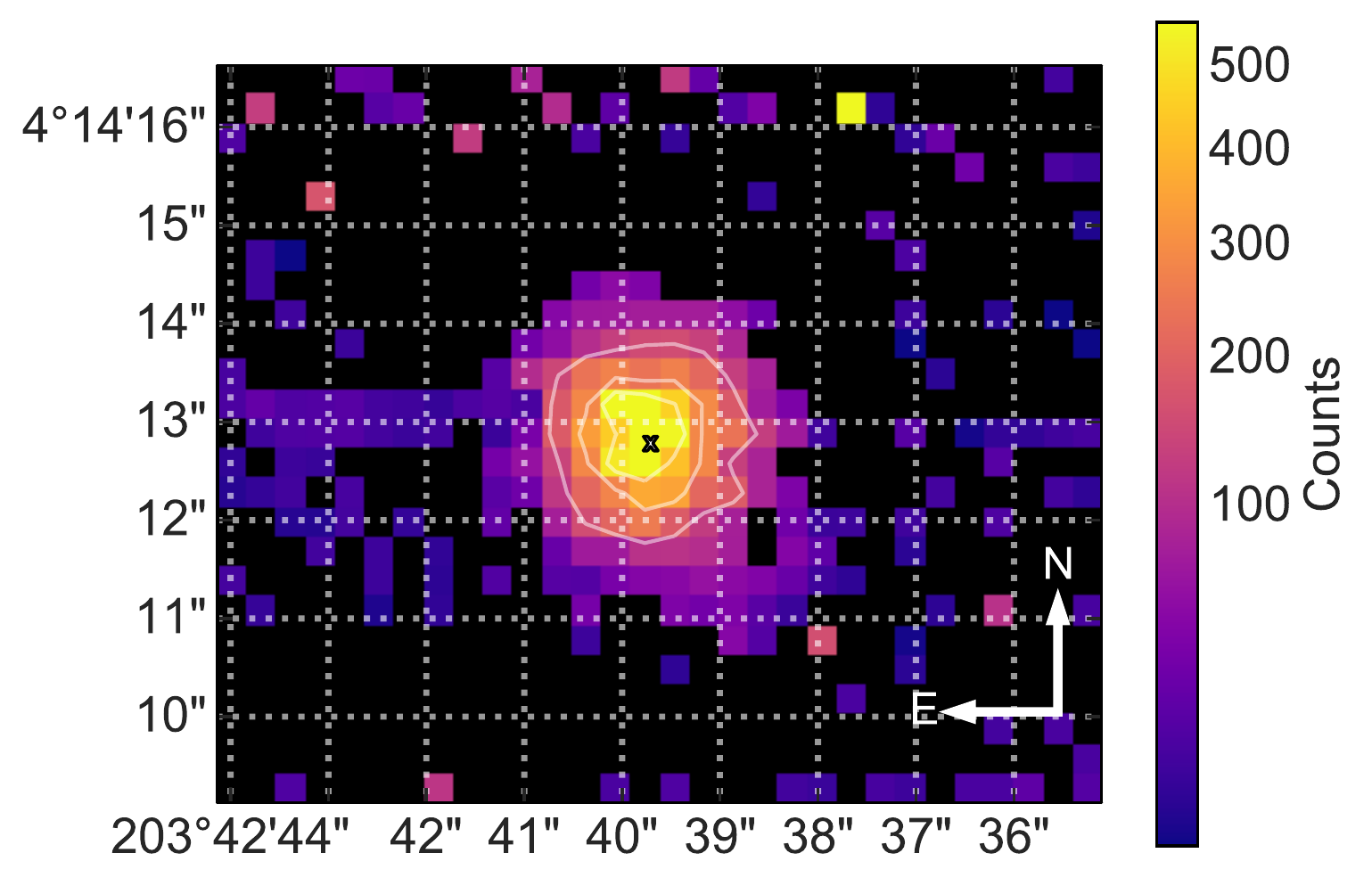}
}\\
   \subfloat[SIFS: rf0363 velocity field]{\includegraphics[clip,width=0.7\columnwidth]{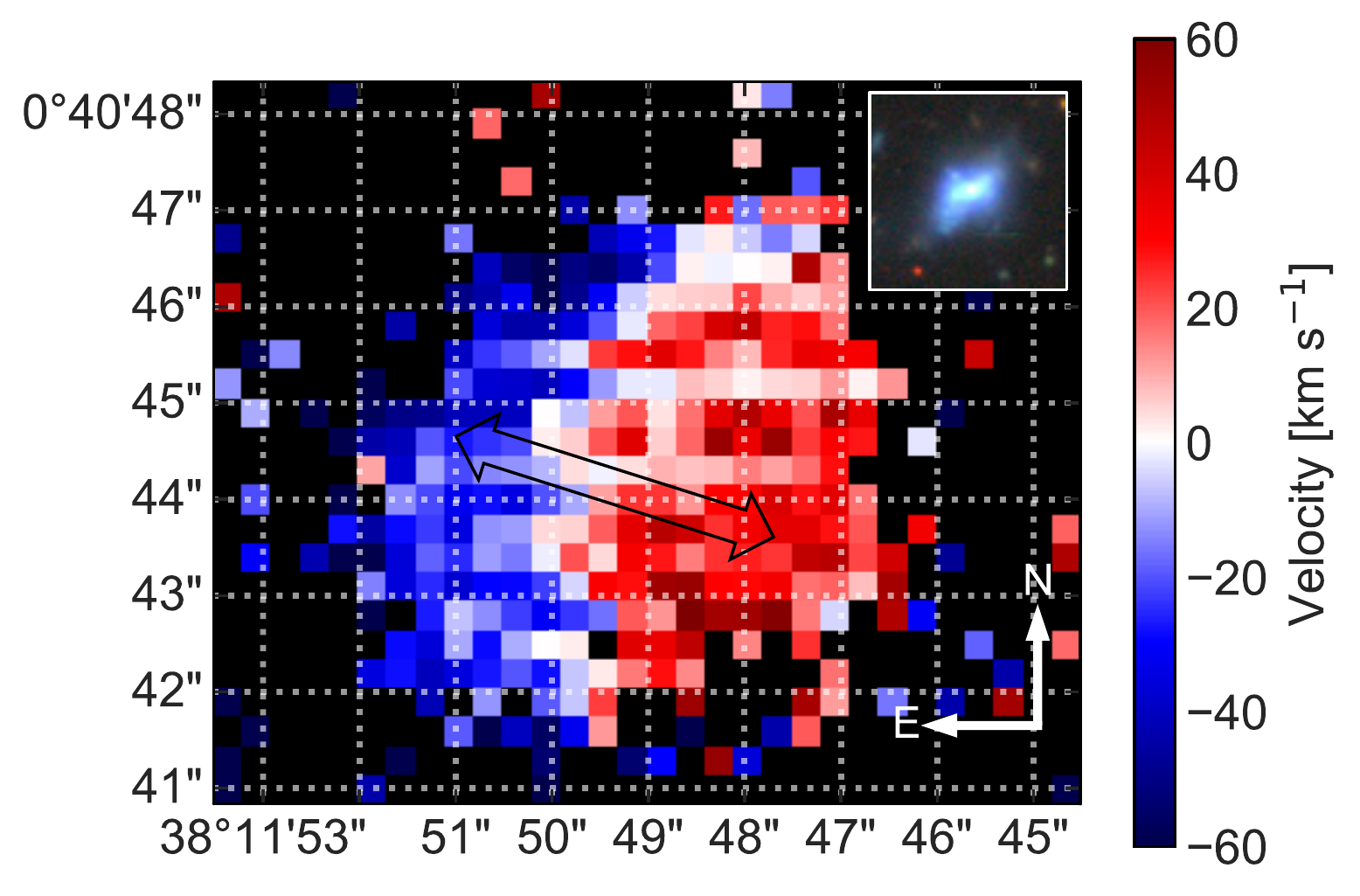}}
   \subfloat[SIFS: rf0363 continuum]{\includegraphics[clip,width=0.7\columnwidth]{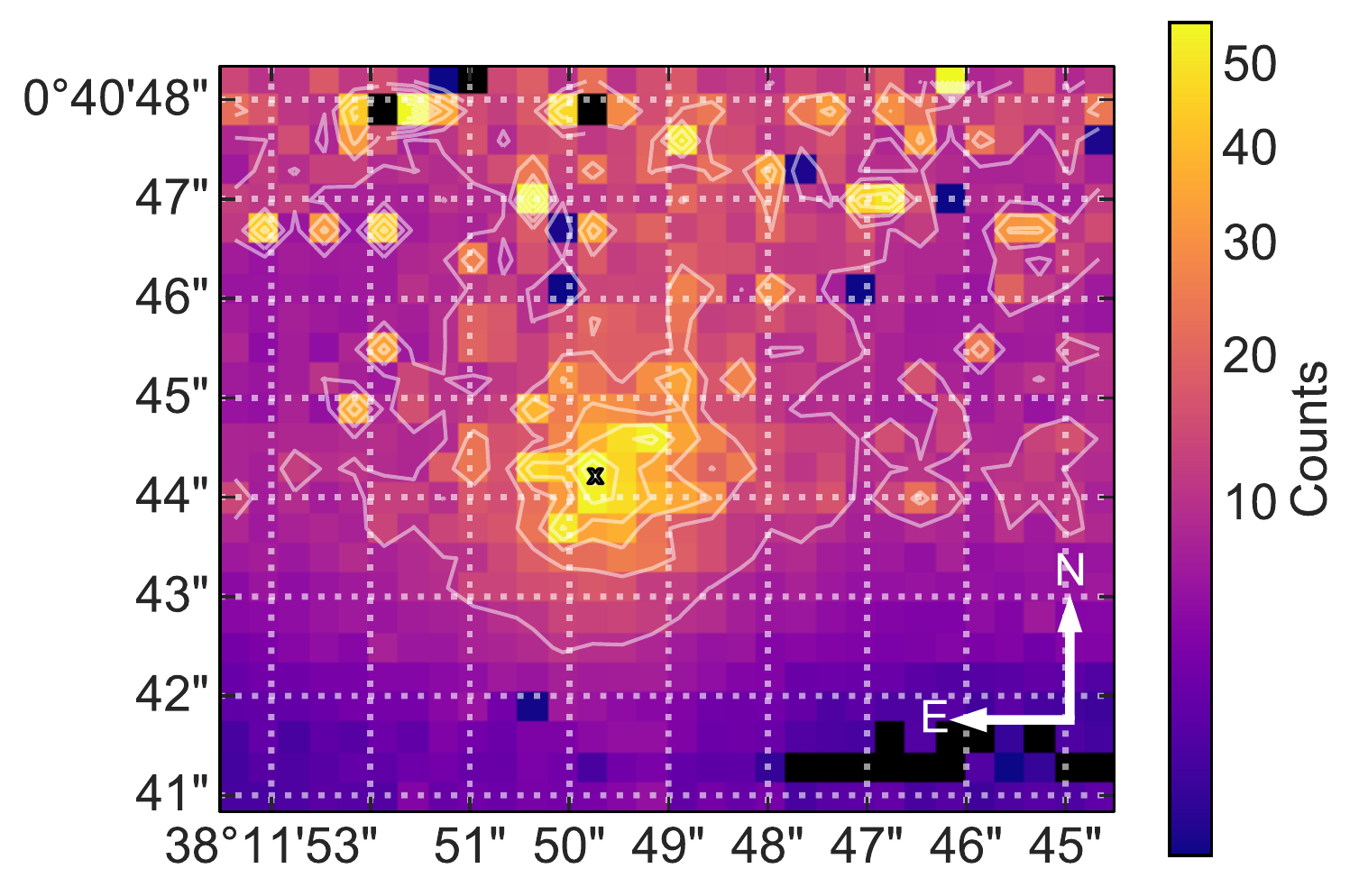}}
   \subfloat[SIFS: rf0363 \halpha flux]{\includegraphics[clip,width=0.7\columnwidth]{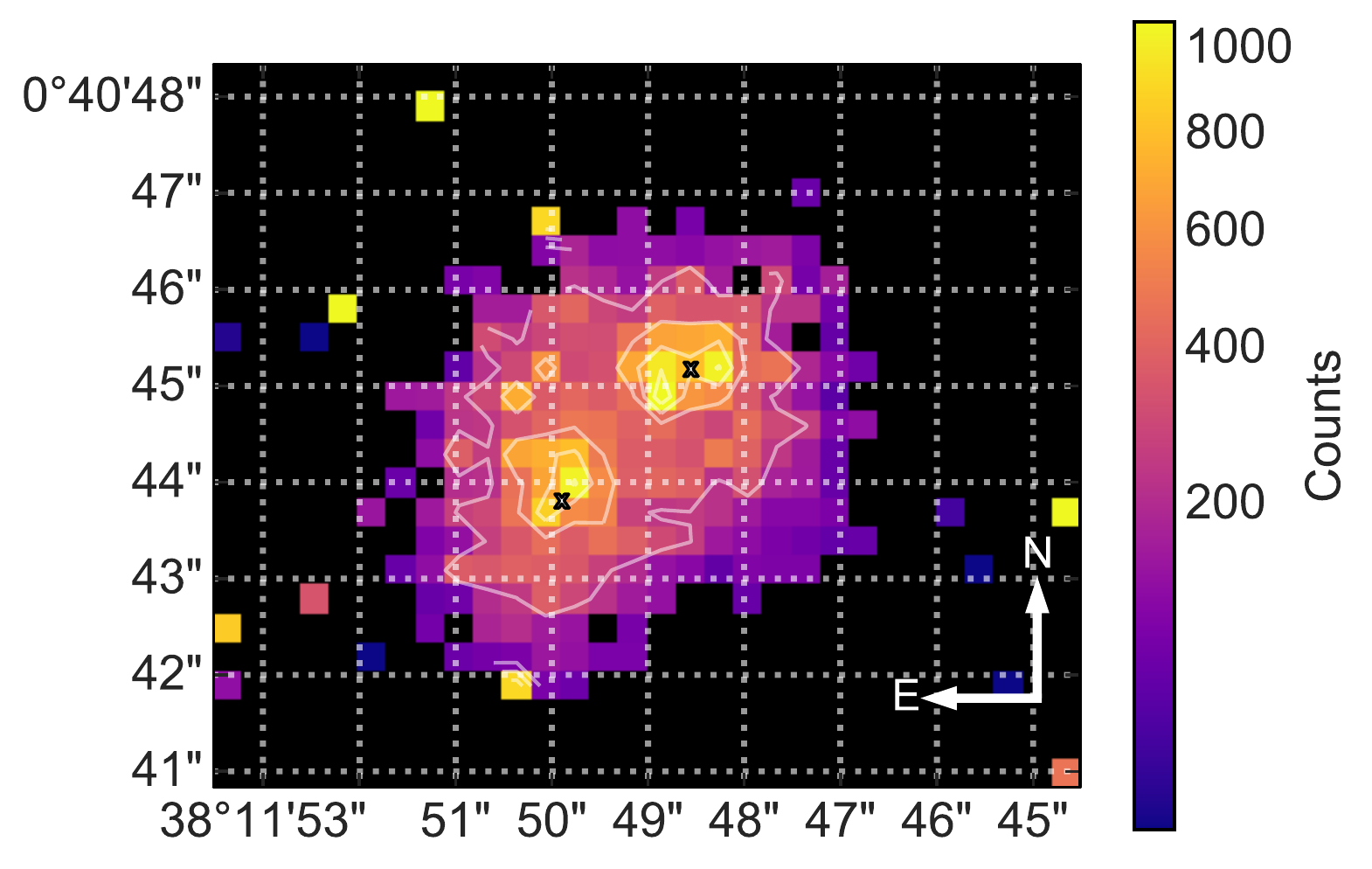}
}\\
\caption{Same as previous page. Additional velocity fields are shown for the main object and the companion in the case of rs0804.}
\vspace*{3in}
\label{fig:3dspec2}
\end{figure*}

\setcounter{figure}{12}
\begin{figure*}
\captionsetup[subfigure]{labelformat=empty}
   \subfloat[SIFS: rs0463 velocity field]{\includegraphics[clip,width=0.7\columnwidth]{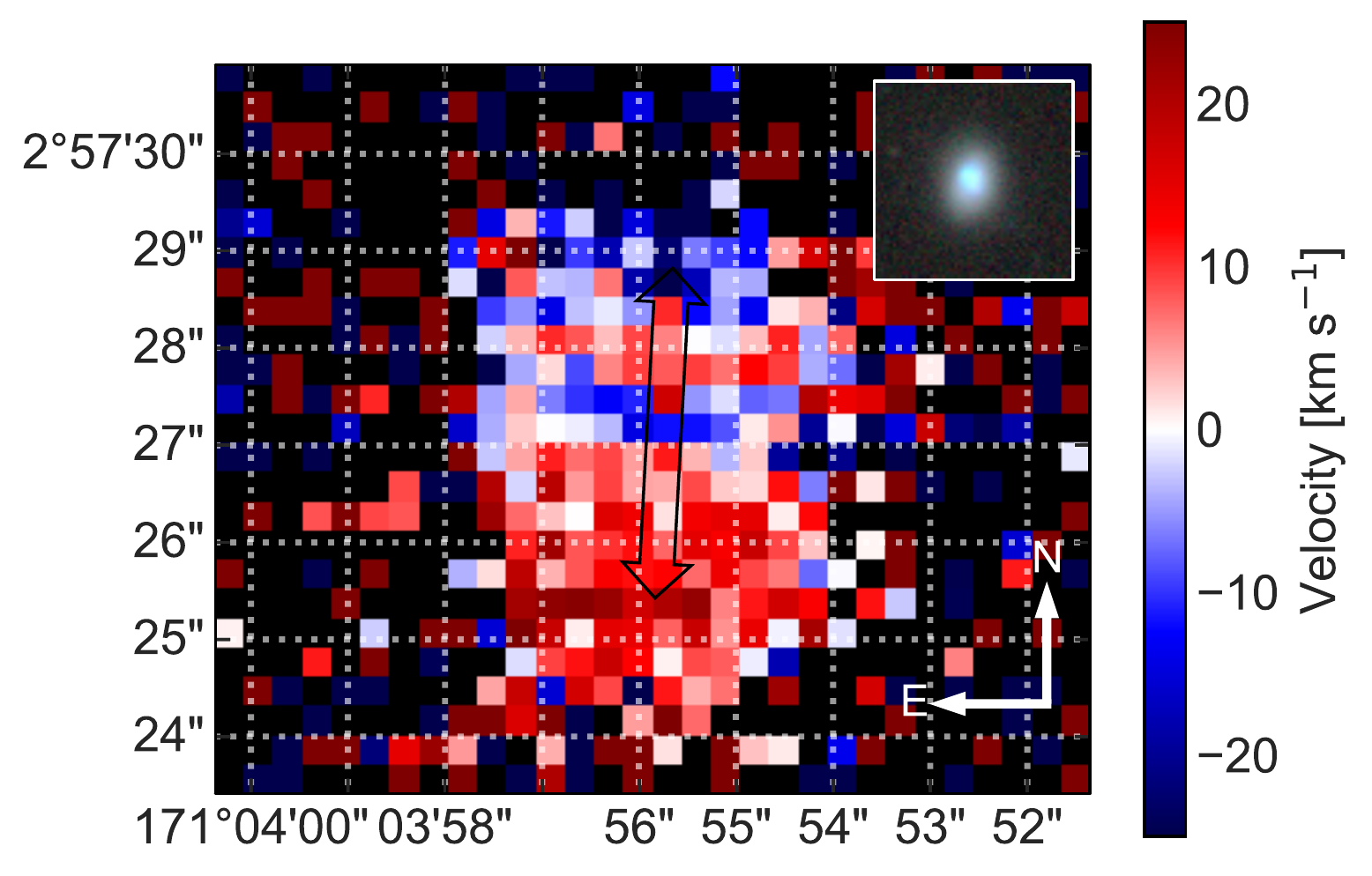}}
   \subfloat[SIFS: rs0463 continuum]{\includegraphics[clip,width=0.7\columnwidth]{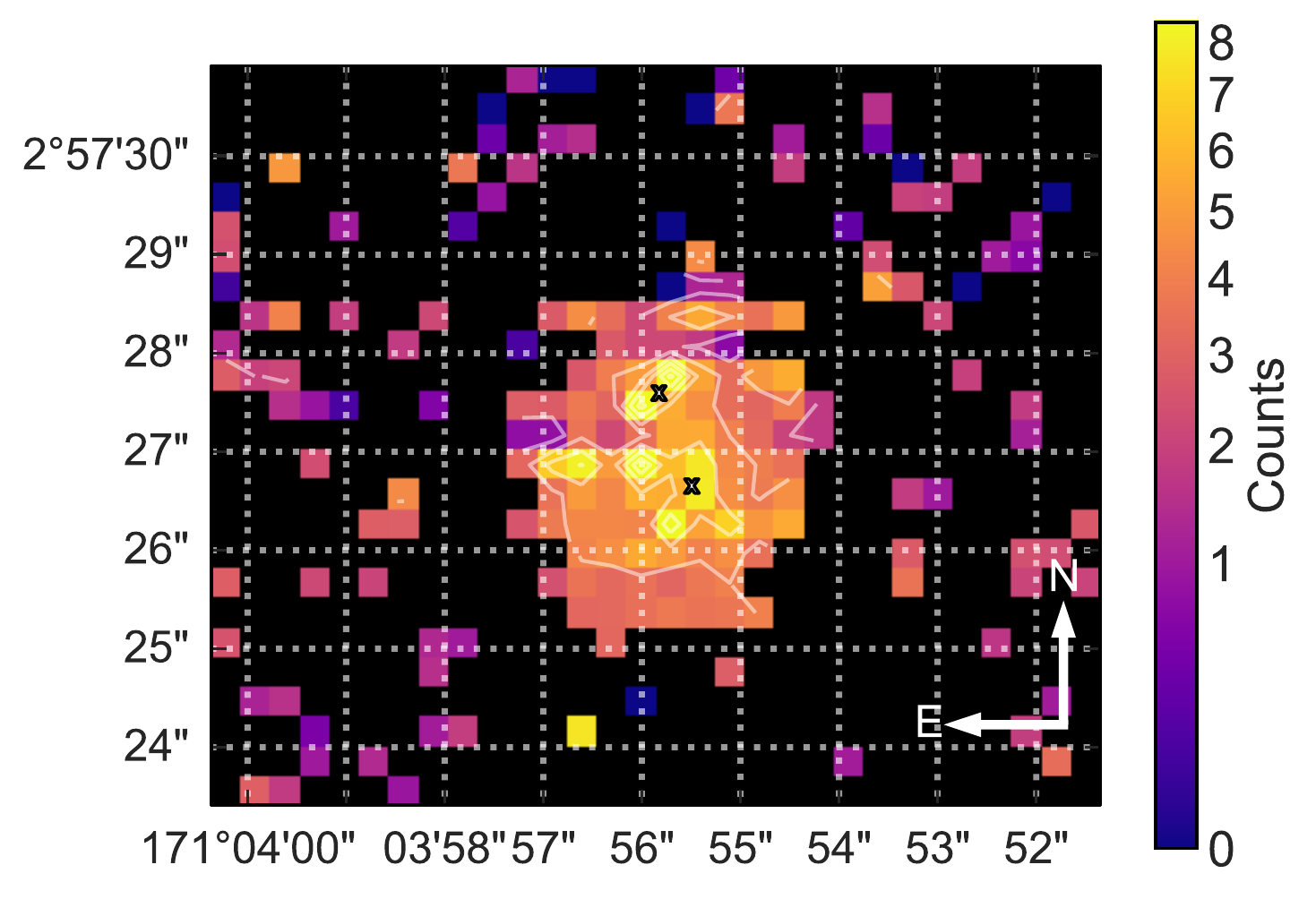}}
   \subfloat[SIFS: rs0463 \halpha flux]{\includegraphics[clip,width=0.7\columnwidth]{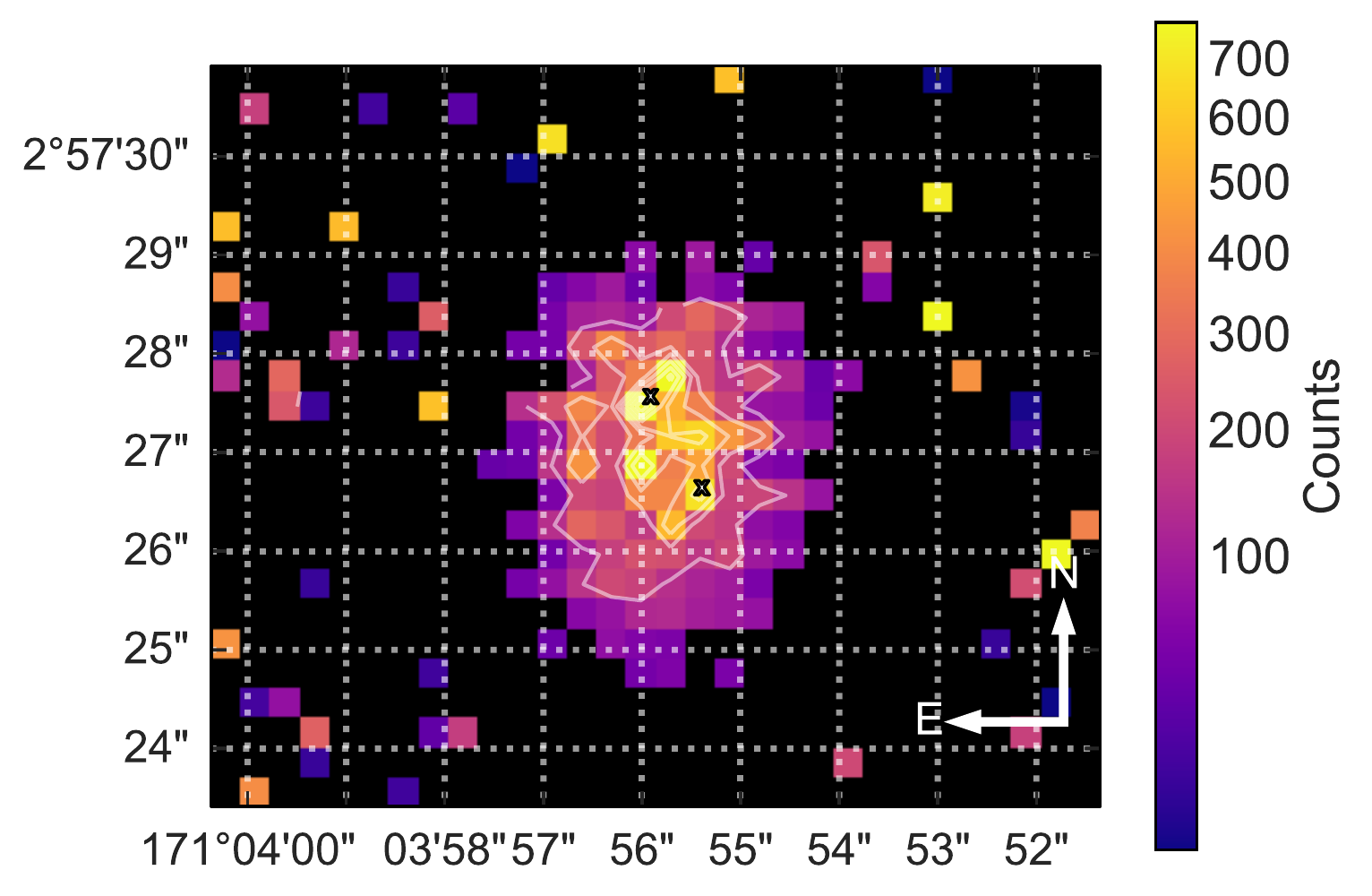}
}\\
   \subfloat[SIFS: rs1259 velocity field]{\includegraphics[clip,width=0.7\columnwidth]{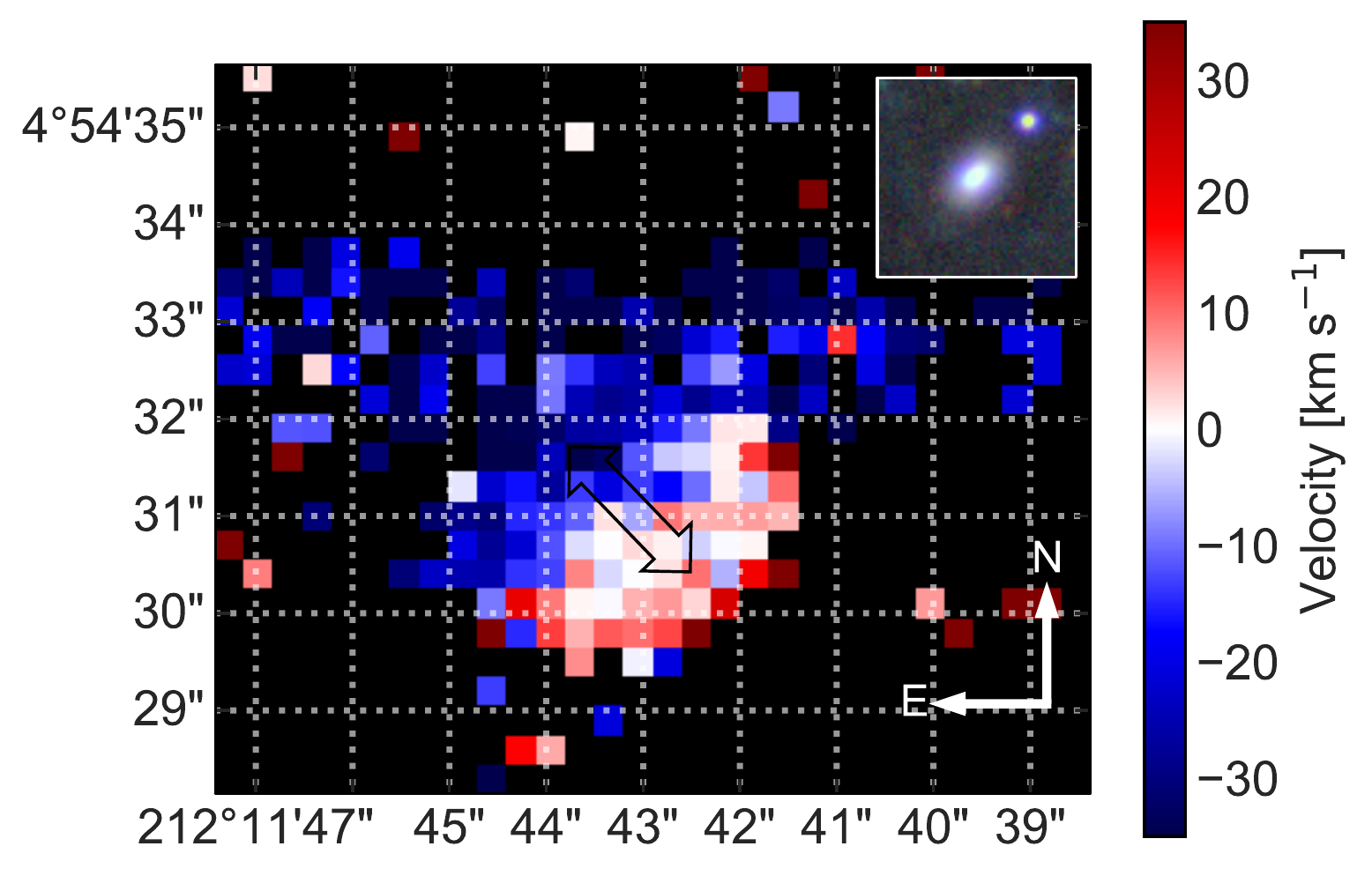}}
   \subfloat[SIFS: rs1259 continuum]{\includegraphics[clip,width=0.7\columnwidth]{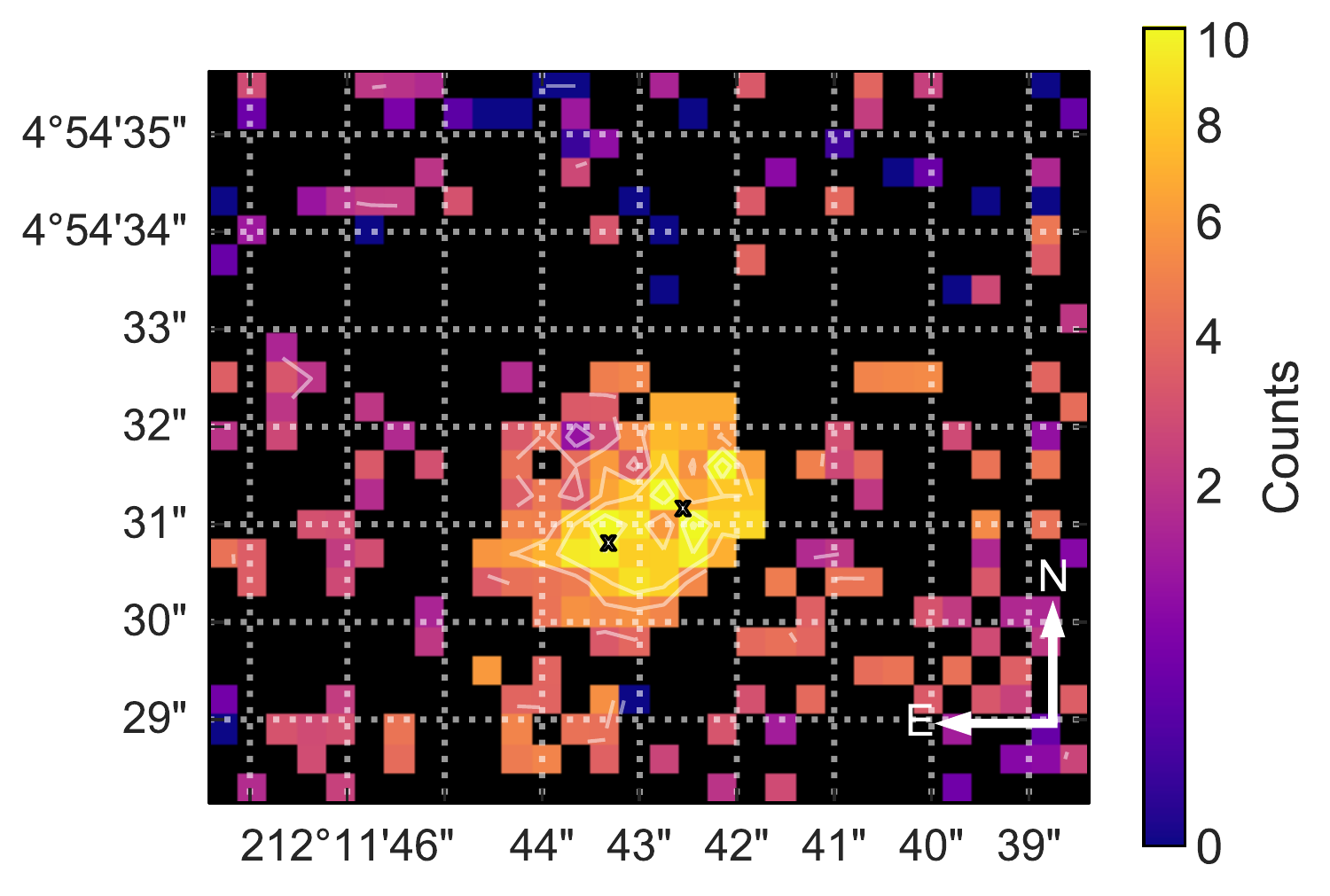}}
   \subfloat[SIFS: rs1259 \halpha flux]{\includegraphics[clip,width=0.7\columnwidth]{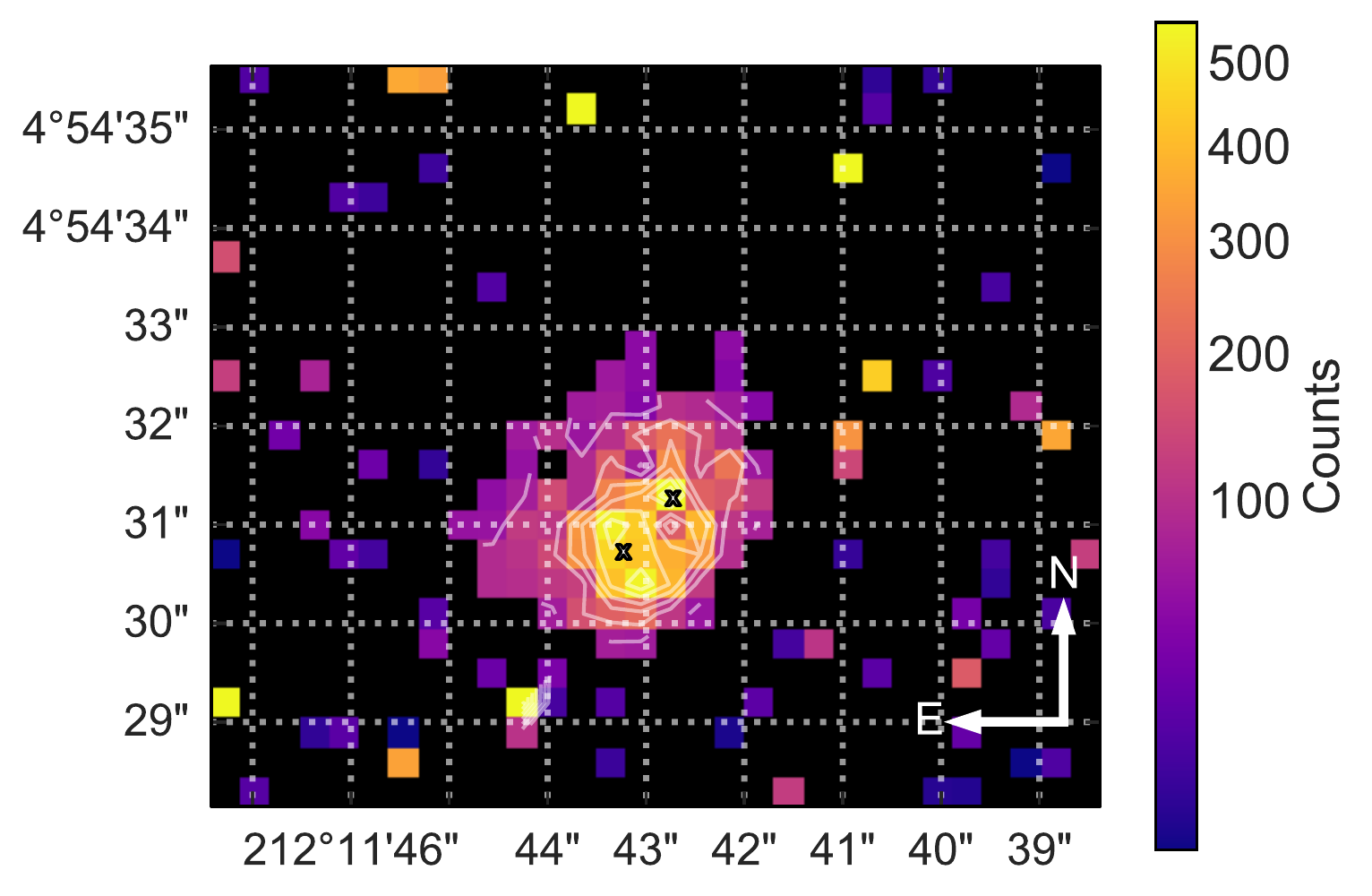}
}\\
\caption{Same as previous page. The effects of CCD ghosting are evident as horizontal smearing in the continuum images for these galaxies, which were observed in the 2018A semester.}
\vspace*{3in}
\label{fig:3dspec3}
\end{figure*}

\section{Discussion} \label{sec:discussion}

To review, within the volume-limited RESOLVE survey, we have identified a population of \cedit{50} low-redshift compact dwarf starburst (CDS) galaxies that have the same minimum SSFR used to identify high-$z$ blue nuggets, compactness equivalent to spheroid or bulged disk morphologies, and masses below  $\mstar < 10^{9.5}\, \msun$, \cedit{in the regime where rapid halo gas cooling and prolate galaxies are expected to persist to $z \sim 0$ }  \citep[e.g.,\ ][]{2006MNRAS.368....2D,2013ApJ...777...42K,2014MNRAS.438.1870D,2014ApJ...792L...6V}. We have argued that the numerical frequency, low halo masses $<10^{11.5}\, \msun$, gas-dominated composition, relative compactness in the mass-size relation, and SSFR and surface density distributions of RESOLVE CDS galaxies all match predicted and/or observed expectations for \cedit{low-$z$} blue nuggets (\S\ref{subsec:link}).

Since a defining feature of blue nuggets is their formation via gas-rich compaction events, we have also examined evidence for prolate morphology, \cedit{disturbed features in photometry}, and unusual kinematic structure to assess whether CDS galaxies form by either of the two main channels of compaction, gas-rich mergers or colliding cosmic streams (\S\ref{subsec:result_formation}). Based on DECaLS imaging \cedit{alone}, \cedit{$\sim$36\% (or $\sim$10\%) of RESOLVE CDS galaxies are "possible" (or "likely") mergers, having average classification scores $\geq$2.5 (or 3.5), compared to $\sim$15.2\% (or $\sim$4.8\%) of control galaxies matched in mass and compactness but having lower SSFRs.} \cedit{These visual merger classifications clearly represent a lower limit, as only two of seven CDS galaxies with follow-up 3D spectroscopy are classified as possible mergers in DECaLS photometry, but the follow-up 3D spectroscopy suggest mergers in at least 4 additional cases, for a total of at least 6 out of 7 galaxies with some evidence of past or present disturbance.}
All of these results are consistent with a gas-rich dwarf merger origin for the majority of CDS galaxies. We have also seen that CDS galaxies live in similar environments to the general dwarf population, if anything showing greater isolation (Figure~\ref{fig:nn} and \S\ref{subsubsec:environ}), consistent with merger remnant status. \par

\cedit{We find no clear evidence for the colliding gas streams formation scenario. RESOLVE CDS galaxies occupy halos in the mass regime of rapid cold gas accretion at $z = 0$ ($M_{\rm halo} \le 10^{11.5} \msun$, e.g.,\ \citealt{2005MNRAS.363....2K}; \citealt{2006MNRAS.368....2D}; \citealt{2011MNRAS.416..660L}; \citealt{2013MNRAS.429.3353N}),  but whether that accretion might take filamentary form remains to be seen. The colliding streams mechanism would need to mimic merger-like kinematic and photometric features to constitute a major formation channel; this possibility has not yet been explored with simulations. Moreover, the CDS galaxy axial ratio distribution is not obviously consistent with the prolateness expected for colliding stream formation, as discussed in \S\ref{subsubsec:prolate}. On the other hand, evidence for pure minor-axis rotation possibly linked to prolateness is seen in two of the seven CDS galaxies with 3D spectroscopy (\S\ref{subsubsec:kin}), with the odd caveat that they also show double nuclei perpendicular to the rotation direction, possibly but not convincingly suggesting merger activity. This ambiguous evidence for prolateness, combined with the gas richness and isolated environments of our CDS galaxies, leaves open the possibility that the colliding stream channel of fast-track growth may persist at a low level. This channel is expected to be suppressed at late epochs with the decreasing cosmic gas density of the universe as well as the thickening of cosmic filaments after $z \sim 2-3$ (\S\ref{subsubsec:environ}; \citealt{2006MNRAS.368....2D}). However, \citet{2014ApJ...792L...6V} find that galaxies in the cyclic refueling regime, which have high prolate fractions at high $z$, have residual nonzero prolate fractions all the way to $z\sim0$.} \par

The likelihood that the fast-track channel itself evolves over cosmic time is suggested by other considerations as well. Present-day gas-rich mergers do not create the level of compactness typical of high-$z$ blue and red nuggets (\S\ref{subsubsec:ssmdcomp}). Our CDS galaxies are less compact and more modestly star-forming than more massive high-$z$ blue nuggets, in line with the redshift and mass trends predicted in \citet{2015MNRAS.450.2327Z} and observed in \citet{2013ApJ...765..104B} and \citet{2013ApJ...776...63F}. Yet even within the context of this finding, our CDS galaxies cannot be characterized as examples of slow-track evolution. Their compact structures, gas-dominated composition, and abundant merger signatures clearly suggest formation by compaction. They occupy the $z\sim0$ halo mass regime of rapid gas accretion, and their specific star formation rates are consistent with the lower end of SSFRs observed for high-$z$ blue nuggets (\S\ref{subsubsec:sfr}). We propose that their reduced-intensity evolution may be usefully thought of as moderated fast-track evolution. The shift from fast-track to moderated fast-track growth is likely the result of the lower cold gas inventories and lower merger rates of the $z\sim0$ universe, along with the likely suppression of the colliding streams mode of compaction as cosmic filaments grow too large to penetrate hot halos at low $z$. \par

Blue nuggets forming via low-$z$ moderated fast-track growth may ultimately evolve differently than high-$z$ blue nuggets, even in the low-mass regime of cyclic refueling probed by our CDS sample. Whereas high-$z$ blue nuggets likely cycle through compaction episodes so quickly that they do not have time to build ``normal'' disk structure prior to quenching (consistent with the greater prevalence of disks in red rather than blue nuggets at high $z$; \citealt{2011ApJ...730...38V,2013ApJ...765..104B}), low-$z$ blue nuggets may be able to evolve into unquenched disk galaxies. \par

Precisely such evolution is seen for blue E/S0s and BCDs in the ``Fueling Diagram'' of \citet{2013ApJ...769...82S}. Stark et al.\ show that galaxies in the low-$z$ universe occupy three branches of a triangle-shaped locus in the parameter space of global molecular-to-atomic gas ratio, \htwohone, vs.\ blue-centeredness, a mass-corrected metric of color gradient. The authors argue that galaxies on the right branch of the Fueling Diagram, which is populated by low-mass blue-sequence E/S0s and BCDs, represent the late stages of gas-rich dwarf-dwarf mergers. These objects follow a track of increasing blue-centeredness paired with decreasing \htwohone, consistent with depletion of molecular gas by star formation consumption or feedback. Galaxies on the lower branch of the Fueling Diagram begin with properties similar to those on the right branch (e.g., many are low-mass blue E/S0s) and maintain depressed \htwohone\ ratios moving leftward toward increased red-centeredness in the diagram.  However, at the same time these galaxies \textit{increase} in total gas-to-stellar-mass ratio. \citet{2013ApJ...769...82S} ultimately conclude that this trend is consistent with outer-disk rebuilding and show it coincides with a transition to spiral-type morphology. \par

The moderated fast-track evolution we propose for \cedit{low-$z$} blue nuggets likely parallels the story of right-branch galaxies in the Fueling Diagram. We predict that follow-up molecular gas observations will reveal that present-day CDS galaxies fall on the right branch of this diagram, and the handful that do not appear to be gas-dominated in Figure~\ref{fig:nn} may in fact prove to be gas-dominated once molecular gas is included. \cedit{Although we lack molecular hydrogen data, we do find that CDS galaxies are, on average, more blue centered than the general RESOLVE dwarf population.} CDS galaxies, blue E/S0s, and BCDs mutually overlap in parameter space (e.g., as seen in the mass-radius relation, Figure~\ref{fig:mass_size}; see also \citealt{2009AJ....138..579K} and references therein). Recent literature \citep[e.g.,][]{2001A&A...373...24P, 2008MNRAS.388L..10B, 2009AJ....138..579K, 2013ApJ...769...82S} has pointed to minor and/or gas-rich mergers as a primary driver of blue E/S0 and BCD formation and evolution. Blue E/S0s preferentially occupy low-density environments and low-mass halos in the rapid accretion, gas-dominated regime \citep{2009AJ....138..579K, 2010ApJ...708..841W, 2015ApJ...812...89M}. Moreover, low-mass blue E/S0s occupy a ``sweet spot" for disk rebuilding, with both abundant gas and high surface mass density promoting efficient star formation \citep{2009AJ....138..579K, 2010ApJ...708..841W, 2010ApJ...725L..62W}. In their environmental study of BCDs, \citet{2001A&A...373...24P} show that the majority of their BCD sample ($\sim 80\%$) exist in the vicinity of a tidally interacting neighbor or exhibit merger morphology, suggesting that BCD evolution and starbursts are primarily driven by external factors in the majority of cases. \par

This last result highlights a key caveat in identifying CDS galaxies with blue E/S0s or BCDs --- the latter are more broadly defined, without our extreme SSFR requirement, so can include objects on their way to quenching via tidal interactions with large neighbors or cluster environments \citep[e.g.,][]{2001AJ....121..793R, 2001A&A...373...24P, 2006AJ....132.2432L}. In contrast, our CDS sample is unquenched and starbursting by design (\S\ref{subsubsec:sample}), resulting in an indirect selection for isolated environments and low-mass halos (Figure~\ref{fig:nn}, \S\ref{subsubsec:environ}). Only 7 of the \cedit{50} CDS galaxies in Figure~\ref{fig:nn} exist near massive companions. Thus their complex \cedit{photometric and} kinematic structures are strong evidence of merger-driven formation, whereas tidal interactions may play a larger role in the formation and/or quenching of blue E/S0s and BCDs in dense environments. In general, blue E/S0s follow a looser mass-size relation than CDS galaxies (Figure~\ref{fig:mass_size}), likely because blue E/S0s encompass a wider range of evolutionary states, including the lower branch of the fueling diagram associated with post-merger disk regrowth. Given that RESOLVE's CDS galaxies fall almost entirely within the locus of blue E/S0s in the mass-size plot, it is likely that low-mass, merger-formed blue E/S0s evolve through an initial moderated fast-track growth phase as transient blue nuggets before continuing on to build disks and evolve into ``normal'' galaxies. \par

We note however that our CDS sample is more narrowly defined in mass than both blue E/S0s and the higher-mass extension of BCDs, luminous blue compact galaxies (BCGs), which are rare at $z\sim0$. Blue E/S0s and luminous BCGs can have masses well above the gas-richness threshold scale at $\mstar \sim 10^{9.5} \,\msun$ \citep[e.g.,][]{2009MNRAS.396..818S,2009AJ....138..579K,2015A&A...583A..55O,2017MNRAS.470.4382R}. Detailed studies \citep{2004A&A...419L..43O, 2007A&A...474L...9M, 2008A&A...479..725C, 2015A&A...583A..55O} of BCGs in the local universe have shown that these objects regularly exhibit irregular kinematics and secondary dynamical components akin to those we see in \S\ref{subsubsec:kin}, consistent with merger formation, although some BCGs are also observed to have companions \citep{2004A&A...419L..43O, 2008A&A...479..725C}. For galaxies above $\mstar \sim 10^{9.5} \,\msun$ however, merger-driven compaction in $z\sim0$ galaxies appears closely tied to quenching by gas depletion \citep{2018ApJ...865...49W}. Consistent with this picture, low-$z$ blue E/S0s show declining evidence for disk rebuilding in the mass range $\mstar \sim 10^{9.5-10.5} \,\msun$ and seem to form almost entirely via quenching mergers at higher masses \citep{2009AJ....138..579K, 2009MNRAS.396..818S}. \par

At lower masses, green peas --- dense star-forming galaxies with unusually powerful and broad [OIII] 5007 \AA \ lines at $0.112 \lesssim z \lesssim 0.360$ first published by \citet{2009MNRAS.399.1191C} --- are another galaxy class with features potentially evocative of higher-$z$ blue nuggets. In a population study, \citet{2009MNRAS.399.1191C} calculate a median stellar mass of $\sim$$10^{9.5} \msun$ for green peas, with a range of $10^{8.5} - 10^{10.5} \msun$. Additionally, \citet{2009MNRAS.399.1191C} find that green pea mass-doubling times span $\sim 0.1$--1 Gyr. Our median SSFR of $\sim$$0.530$ Gyr$^{-1}$ (see Figure~\ref{fig:ssfr_dist}) is equivalent to a mass-doubling time of $\sim$$1.88$ Gyr, and the shortest mass-doubling times we measure approach only $\sim$0.5 Gyr. Given the universal decrease in star formation activity between the epoch of green peas and today, CDS galaxies may simply represent \cedit{lower-$z$} analogues of green peas \cedit{that reflect star-formation downsizing}. However, 3D spectroscopy of individual green pea galaxies reveals examples both with and without merger signatures, leaving open the possibility that some green pea starbursts may be fueled by colliding gas streams \citep{2017MNRAS.471.2311L}. \par

Overall, the above comparisons to blue E/S0s, BCDs/BCGs, and green peas suggest that the likely evolutionary trajectory of our candidate low-$z$ blue nuggets may be rather different from the trajectory of high-$z$ blue nuggets, even in the gas-rich regime below the threshold scale, $M_{\rm halo} \sim 10^{11.5}\, \msun$. As presented in both simulations and observations, it would appear that the extreme compaction inherent in high-$z$ fast-track evolution entails rapid, cyclic compaction-driven starburst activity until a galaxy's halo grows to the point of enabling quiescence and red nugget formation above the threshold scale. In contrast, the milder moderated fast-track evolution we see at low $z$ may allow CDS galaxies to rebuild disks with less interruption, so they have the potential to evolve into the typical star-forming disk galaxies we see in the local universe. \par

\section{Conclusions} \label{sec:conclusions}

In this work, we have explored the existence, properties, and formation of low-$z$ blue nuggets within a sample of compact dwarf starburst (CDS) galaxies in the RESOLVE survey, a volume-limited census of $z\sim0$ galaxies complete down to baryonic (stellar+cold gas) mass $\sim 10^{9.2} \, \msun$. We have identified \cedit{50} CDS galaxies at $z\sim0$ by criteria that should select \cedit{residual} blue nuggets in the regime where repeated regeneration of blue nuggets is expected (\S\ref{subsubsec:sample} and Figure~\ref{fig:selection}): all have dwarf stellar mass ($\mstar < 10^{9.5} \, \msun$), compact morphology ($\mu_{\Delta} > 8.6$, consistent with spheroid or bulged disk), and starburst activity above the lower cutoff defining high-$z$ blue nuggets ($\log(\textnormal{SSFR [{\rm Gyr}$^{-1}$}]) > -0.5$). If these galaxies indeed constitute a \cedit{residual} tail of the blue nugget phenomenon, three properties should characterize them -- compactness relative to contemporaneous galaxies, an upper halo mass limit of $M_{\rm vir} \sim 10^{11.5} \msun$ implying abundant gas, and formation via gas-rich mergers or other intense gas-compaction channels (\S\ref{intro}). \par

\begin{itemize}[leftmargin=*]
\item Consistent with their selection and potential identity as blue nuggets, CDS galaxies have $0.7$ dex higher surface mass densities than other RESOLVE dwarfs \cedit{\S\ref{subsubsec:ssmdcomp}}. They lie on a tight ridge in the mass-size relation (Figure~\ref{fig:mass_size}).

\item Although not selected on halo mass or gas content, $\sim$86\% of our CDS galaxies are found to reside in halos below $M_{\rm halo} \sim 10^{11.5}\, \msun$. As expected for blue nuggets in this halo mass regime, $\sim$68\% of CDS galaxies have atomic gas-to-stellar mass ratios $M_{\rm gas}/\mstar > 1$ (Figure~\ref{fig:nn}, left), with only $\sim$10\% appearing gas poor. The seven CDS galaxies that friends-of-friends group-finding identifies as satellites in larger halos also have $M_{\rm gas}/\mstar \gtrsim 1$, suggesting either recent infall\cedit{, as indicated by their high atomic-gas fractions,} or group-finding errors (\S\ref{subsubsec:environ}).

\item \cedit{A search for merger evidence in high-resolution DECaLS imaging reveals $\sim$$2\times$ as many mergers in the CDS galaxy sample vs.\ in a control sample selected identically in mass and morphology but with lower SSFRs, specifically  $\sim$$36.0\%$ vs. $\sim$$15.2\%$ "possible" mergers and $\sim$$10.0\%$ vs. $\sim$$4.8\%$ "likely" mergers, respectively (see \S\ref{subsubsec:doublenuc}).}

\item We have performed follow-up 3D spectroscopy with the GMOS IFU, SAM FP, and SIFS for seven low-$z$ CDS galaxies. Consistent with formation via gas-rich dwarf mergers, the velocity fields and continuum maps reveal double nuclei and/or disturbed kinematics (such as multi-component rotation and an ongoing kinematically-detected merger) in at least six of the seven (Figure~\ref{fig:3dspec1}). \cedit{The DECaLS data reveal merger signatures in only two of these seven galaxies, suggesting that the DECaLS classifications constitute a lower limit on the frequency of recent mergers in the CDS sample.}

\item The distribution of nearest-neighbor distances for CDS galaxies follows that of the general RESOLVE survey, with if anything slightly greater isolation (Figure~\ref{fig:nn}, right). Along with the aforementioned results, this isolation supports the idea that the majority of CDS galaxies are gas-rich dwarf merger remnants.

\item Our results neither favor nor rule out the persistence of a compaction channel involving colliding gas streams in rare cases. The CDS sample shows a distribution of projected axial ratios suggestive of oblate (rather than prolate) morphology (\S\ref{subsubsec:prolate} and Figure~\ref{fig:ba}). Our 3D spectroscopy reveals minor-axis rotation in four of seven CDS galaxies, but it is combined with major axis rotation in multi-component velocity fields and/or double nuclei in all cases (\S\ref{subsubsec:kin} and Figure~\ref{fig:3dspec1}). These results are inconclusive and leave open the possibility of a residual population of prolate blue nuggets formed by colliding streams at $z\sim0$, consistent with \citet{2014ApJ...792L...6V}. 

\item Our $z\sim0$ CDS galaxies exhibit surface densities (\S\ref{subsubsec:ssmdcomp}), approximate numerical frequency (\S\ref{subsubsec:freq}), and SSFRs (\S\ref{subsubsec:sfr},\ Figure~\ref{fig:ssfr_dist}) consistent with prior predictions and observations of \cedit{redshift evolution} in blue nuggets \citep{2013ApJ...765..104B,2013ApJ...776...63F,2014MNRAS.438.1870D,2015MNRAS.450.2327Z}. With the SSFR and compactness cuts used in this paper, CDS galaxies have $\sim$0.4 dex lower median SSFR and $\sim$1 dex lower median surface mass density compared to the high-$z$ blue nuggets of \citet{2013ApJ...765..104B}. 

\item We show that lowering our SSFR restriction to compensate for evolution in the star-forming main sequence from high $z$ \citep[see][]{2013ApJ...770...57B} introduces an additional 60 galaxies to the sample. This enlarged sample has median SSFR $\sim$0.5 dex lower than the Barro et al. sample. However, about a third of the newly added galaxies are satellites in massive halos with substantially lower gas-to-stellar mass ratios than our original CDS sample. Since we are interested in starburst activity driven by gas compaction, as opposed to other mechanisms such as group infall or tidal harassment, our original higher SSFR cut is more useful for isolating the blue nugget phenomenon (\S\ref{subsubsec:sfr}).

\end{itemize}

In keeping with the less extreme compactness and star formation rates observed in low-$z$ CDS galaxies relative to high-$z$ blue nuggets, we have suggested that the blue nugget formation channel has itself evolved over time (\S\ref{sec:discussion}). We call this channel moderated fast-track evolution, to distinguish it both from the qualitatively more violent fast track inferred at high $z$ and from the more gradual slow track observed for most present-day galaxies. We argue that this moderated fast-track has been observed in the Fueling Diagram of \citet{2013ApJ...769...82S}, which plots molecular-to-atomic gas ratio vs.\ blue-centeredness. Stark et al.\ argue that blue E/S0s and BCDs occupy the right branch of the Fueling Diagram during the late starburst stages of gas-rich dwarf mergers, then evolve along the lower branch during the post-starburst stages of gas accretion and disk regrowth. Based on their likely gas-rich dwarf merger origin, CDS galaxies should occupy the right branch of the Fueling Diagram. Intriguingly, Figure~\ref{fig:nn} shows that a small minority of CDS galaxies are \textit{not} gas-dominated based on atomic gas data alone, which may reflect either temporary depletion or significant unmeasured molecular gas as expected for upper right branch galaxies in the Fueling Diagram. We also find that CDS galaxies fall within the locus of blue E/S0s in the mass-size relation (Figure~\ref{fig:mass_size}), defining a tighter relation as expected for a narrower range of evolutionary states. This analogy suggests that low-$z$, low-mass blue nuggets will not become red nuggets, but instead follow the path of blue E/S0s in rebuilding disks and becoming ``normal" disk galaxies \citep[see also][]{2009AJ....138..579K, 2015ApJ...812...89M}. \par

In brief, we have demonstrated that the blue nugget evolutionary story has a \cedit{residual} tail that extends even into the present-day universe. The moderated fast-track evolution of these low-$z$ objects suggests that the violent compaction events that drove rapid galaxy evolution and star formation in the early universe still exist to some extent at the present epoch, particularly in the form of gas-rich dwarf mergers. However, the resulting \cedit{low-$z$} blue nuggets cannot evolve to the densities of high-$z$ red nuggets and indeed seem more likely to be early progenitors of normal disk galaxies. \par

\section*{Acknowledgements}
We are grateful to the anonymous referee whose comments have improved this work. We thank Vianney Lebouteiller for insightful comments and suggestions that greatly contributed to this work. MP acknowledges the financial support of the National Space Grant College and Fellowship Program and the North Carolina Space Grant Consortium. LF acknowledges the support from CNPq-Brazil. This work is in part based on observations obtained at the Southern Astrophysical Research (SOAR) telescope with the SIFS \citep{2003SPIE.4841.1086L} and SAM FP \citep{2017MNRAS.469.3424M} instruments. SOAR is a joint project of the Minist\'{e}rio da Ci\^{e}ncia, Tecnologia, Inova\c{c}\~{o}es e Comunica\c{c}\~{o}es (MCTIC) do Brasil, the U.S. National Optical Astronomy Observatory (NOAO), the University of North Carolina at Chapel Hill (UNC), and Michigan State University (MSU). This work is in part based on observations obtained at the Gemini Observatory acquired through the Gemini Observatory Archive and processed using the Gemini IRAF package, which is operated by the Association of Universities for Research in Astronomy, Inc., under a cooperative agreement with the NSF on behalf of the Gemini partnership: the National Science Foundation (United States), the National Research Council (Canada), CONICYT (Chile), Ministerio de Ciencia, Tecnolog\'{i}a e Innovaci\'{o}n Productiva (Argentina), and Minist\'{e}rio da Ci\^{e}ncia, Tecnologia e Inova\c{c}\~{a}o (Brazil). We acknowledge the use of DECam Legacy Survey data, which has made use of the resources described at \url{http://legacysurvey.org/acknowledgment/}. \par

\bibliographystyle{mnras}
\bibliography{Palumbo20} 

\appendix

\section{The Gemini Reduction Pipeline} \label{ap:gem}

The ultimate goal of the RESOLVE Gemini reduction pipeline is to transform raw 2D scientific and calibration exposures into a 3D data cube that contains spatial data on the $xy$-plane and spectral data on the $z$-axis. To do so, the pipeline uses a Python script to call tasks in the GMOS IRAF package\footnote{\url{http://www.gemini.edu/sciops/data-and-results/processing-software}}. \par

The pipeline is designed to run in a working directory initially containing the raw data, obtained from the Gemini Observatory Archive, and a handful of calibration files (e.g.\ a bias frame and a file containing a list of strong lamp lines used in the wavelength calibration step). The reduction is largely performed in the standard manner, including steps such as: bias and overscan subtraction, fiber identification, flat-fielding, bad pixel masking, arc extraction, wavelength solution creation, quantum efficiency correction, cosmic ray removal, sky subtraction, flux calibration, and data cube mosaicking. \par

Rather than the more traditional implementation of L.A. Cosmic \citep{2001PASP..113.1420V} in the cosmic ray removal step, we instead use PyCosmic\footnote{\url{http://www.bhusemann-astro.org/?q=pycosmic}} \citep{2012A&A...545A.137H}. PyCosmic uses the same Laplacian edge detection method as L.A. Cosmic, but with extensions to the algorithm specifically tailored for use with fiber-fed spectrograph data. Whereas L.A. Cosmic frequently falsely construes the bead-like structure of the fiber data as cosmic ray hits, PyCosmic is better able to differentiate the scientific data and therefore avoids destruction of strong emission lines, such as \halpha. \par

To mosaic the data cubes produced by the pipeline, we use the PyFu package\footnote{\url{http://drforum.gemini.edu/topic/pyfu-datacube-mosaicking-package/}} written by Gemini staff member James Turner. PyFy performs the merging of the cubes in a two-step process. First, the task \textit{pyfalign} makes a centroid fit over the brightest feature in each wavelength-summed cube and calculates the spatial offsets of the centroid in each region. With the spatial offsets quantified, the \textit{pyfmosaic} task resamples the cubes onto a common grid and then co-adds them. \par

The co-added cubes are assigned a World Coordinate Solution (WCS) based on the references coordinates and pixel scales written to the FITS header by the telescope. Differences in the real and assigned telescope pointing lead to slight inconsistencies in the absolute position of the galaxy. However, without any field stars to match, it is not easy (or necessary) to further refine the WCS. These discrepancies are somewhat apparent in Figure~\ref{fig:3dspec1}, where the center coordinates of the SAM FP and GMOS IFU observations of rf0250 and rf0266 do not precisely agree. \par

Unfortunately, observations in the Gemini program GS-2013B-Q-51 consistently lack twilight flats in one of the wavelength dithers. Therefore, we have been unable to create a response function to properly flat-field any observations in the redder wavelength dither. Moreover, without proper flat-fielding, the differential throughput of the pseudoslits is unaccounted for. Unfortunately, this ``two-slit issue" can create a false bifurcation in the final scientific data cube which may persist in the velocity fields, continuum, and line flux maps described in \S\ref{subsec:vfield} and \S\ref{subsec:cont}. Thus far, we have only managed to free ourselves from the two-slit issue by excluding the red dither exposures in the 2013B semester from the data cube creation step. Consequently, these observations have SNR reduced by a factor of $\,\sqrt[]{2}$. In \S\ref{subsec:vfield}, we show that consistency between GMOS IFU red dither data and SAM FP data suggest that any double nuclei and velocity field misalignment features are genuine physical features, as opposed to calibration effects. \par

\section{Notes on 3D Spectroscopy for Individual Galaxies} \label{ap:notes}
\subsection{rf0266} \label{subsubsec:rf0266}
The GMOS IFU observations of rf0266 reveal a complex kinematic structure in the velocity field. We observe two distinct rotation components, misaligned with respect to each other by about 90 degrees, as identified by eye. The SAM FP velocity field does not show two components clearly, likely due to the lower SNR observations toward the outskirts of the galaxy and worse seeing conditions. Nevertheless, the continuum and \halpha flux maps from both the GMOS IFU and SAM FP reveal a single nucleus. \par

\subsection{rf0250} \label{subsubsec:rf0250}
As in the case of rf0266, the GMOS IFU observations of rf0250 reveal multiple misaligned rotation components. The slightly lower fidelity SAM FP observations have primarily revealed the inner component of rotation in the higher surface-brightness region of the galaxy. Moreover, both the SAM FP and GMOS IFU \halpha flux maps reveal a double nucleus, also seen in the SAM FP continuum. The positions of the two distinct nuclear peaks align with the peaks of the inner velocity curve, suggesting not only that the observed misalignment is a real, physical feature but also that rf0250 likely formed as the result of recent merger activity. \par

\subsection{rs0804} \label{subsubsec:rs0804}
The low-$z$ CDS galaxy rs0804 provides a unique and informative example, as it is in the beginning stages of a minor merger. The presence of a smaller companion is immediately evident in the velocity field and H$\alpha$ flux map. The companion is just barely visible in the continuum image, as the narrow spectral window of SAM FP prevents adequate sampling of a wide section of continuum. The velocity maps for this galaxy, which are re-zeroed for both the main galaxy and its companion, reveal distinct and approximately orthogonal rotation patterns in each object, supporting the idea that the two objects are distinct galaxies, rather than the smaller being a stripped off gas cloud or structural anomaly. \par

\subsection{rs1103} \label{subsubsec:rs1103}
The low-$z$ CDS galaxy rs1103 is quite small in spatial extent in comparison to the other CDS galaxies. The low SNR of these observations (since these data were taken during bright time) has allowed us to capture the velocity field within only the high surface brightness nuclear region of the galaxy. As such, while we do observe a possibly anomalous redshifted peak near the center of the galaxy, potentially indicative of recent accretion, we do not observe multiple misaligned rotation components in the velocity field, nor evidence for a double nucleus in either the continuum or \halpha flux maps. (However, the continuum map is not shown due to extremely low SNR.) \par 

\subsection{rf0363} \label{subsubsec:rf0363}
SIFS observations of the low-$z$ CDS galaxy rf0363 reveal that the galaxy is experience particularly strong bursts of star formation in two distinct knots near the nucleus of the galaxy, as revealed by the \halpha flux map. The presence of only a single concentration of light in the continuum image might suggest that these \halpha flux concentrations may not be a true double nucleus; however the continuum DECaLS image for this galaxy shows three knots, two of which are aligned in the same sense as seen in the H$\alpha$ imaging. The two knots seen in both DECaLS and H$\alpha$ are aligned along the major axis, whereas the third knot is in the minor axis direction. The direction of the velocity gradient is also along the minor axis, consistent with possible prolate structure. These observations leave open the possibility that the starburst may be driven by either cosmic gas accretion or a recent gas-rich merger. \par

\subsection{rs0463}\label{subsubsec:rs0463}
Distinct double nuclei features appear in both the continuum and \halpha maps for rs0463. The southwesterly nucleus is also evident as a separate concentration in both SDSS and DECaLS imaging. The velocity map for this galaxy shows evidence of perturbation, particularly in the northerly, blue-shifted region. This region appears to have a redshifted outflow which approximately corresponds in position with the northerly double nucleus. \par

\subsection{rs1259}\label{subsubsec:rs1259}
This galaxy also exhibits evidence for a double nucleus. As in rf0363, it has a pair of knots aligned with the major axis, whereas the rotation is along the minor axis. However in rs1259 there are only two knots, which are seen in both the SIFS continuum and H$\alpha$ maps. DECaLS imaging appears to have insufficient resolution to separate these knots. Again, the juxtaposition of double nuclei along the major axis with minor-axis rotation leaves open both gas compaction scenarios. \par

\section{Calculation of $\Delta$ Quantities} \label{app:delta}
\cedit{In \S\ref{subsubsec:ssmdcomp}, we calculate three ``$\Delta$'' quantities as defined in \citet{2017ApJ...840...47B}:}

\begin{align}
	\Delta \Sigma_{1.5} &= \log \Sigma_{1.5} - \log \Sigma_{1.5}^{\rm Q} \\
	\Delta \Sigma_{\rm e} &= \log \Sigma_{\rm e} - \log \Sigma_{\rm e}^{\rm Q} \\
	\Delta \mu_\Delta &= \mu_\Delta - \mu_\Delta^{\rm Q} \\
\end{align}

\noindent \cedit{The superscript ``Q" on the subtracted term in each equation denotes the median density at a given mass for quiescent galaxies. We obtain this quantity by fitting a power-law relating each density metric to stellar mass for quiescent galaxies, as in \citet{2017ApJ...840...47B}:}

\begin{equation}
	\log \Sigma^{\rm Q}=\alpha\left[\log \left(\frac{M_{\star}}{M_{\odot}}\right)-10.5\right]+\log A
\end{equation}

\noindent \cedit{As described in \citet{2017ApJ...840...47B}, we select quiescent galaxies by fitting the SFR-MS and iteratively removing galaxies more than 0.7 dex below the fit:}

\begin{equation}
\log \mathrm{SFR}=\mu\left[\log \left(\frac{M_{\star}}{M_{\odot}}\right)-10.5\right]+\log C
\end{equation}

\noindent \cedit{Lastly, we can calculate $\Delta  {\rm SFR}$ as:}

\begin{equation}
	\Delta {\rm SFR} = \log{\rm SFR} - \log {\rm SFR}^{\rm MS}.
\end{equation}

\bsp
\label{lastpage}
\end{document}